\documentclass{lmcs}

\usepackage{hyperref}
\usepackage{microtype}
\usepackage[utf8]{inputenc} 
\usepackage{mathtools,amssymb,stmaryrd,amsmath}
\usepackage{thmtools,thm-restate}
\usepackage[inference,ligature]{semantic}
\usepackage[dvipsnames]{xcolor}
\usepackage{color}
\usepackage{xspace} 
\usepackage{listings}
\usepackage{enumerate}
\usepackage{url}
\usepackage{macros}
\usepackage{hyperref}

\AtBeginDocument{}

\title[A Theory of Available-by-Design Communication Systems]{A Theory of Available-by-Design Communicating Systems}

\author[H. López, F. Nielson and H. R. Nielson]{Hugo A. López}	
\address{Technical University of Denmark\\Kongens Lyngby, Denmark}	
\email{hulo@dtu.dk}  

\author[]{Flemming Nielson}	
\address{Technical University of Denmark\\Kongens Lyngby, Denmark}	
\email{fnie@dtu.dk}  

\author[]{Hanne Riis Nielson}	
\address{Technical University of Denmark\\Kongens Lyngby, Denmark}	
\email{hrni@dtu.dk}  


\begin{document}

\begin{abstract}
  Choreographic programming is a programming-language design approach
  that drives error-safe protocol development in distributed
  systems. Starting from a global specification (\emph{choreography})
  one can generate distributed implementations.  The advantages of
  this top-down approach lie in the correctness-by-design principle,
  where implementations (\emph{endpoints}) generated from a
  choreography behave according to the strict control flow described
  in the choreography, and do not deadlock. 
  Motivated by challenging scenarios in Cyber-Physical Systems (CPS),
  we study how choreographic programming can cater for dynamic
  infrastructures where not all endpoints are always available.  We introduce the Global Quality Calculus (\GCQ),  a
  variant of choreographic programming for the description of communication systems where \emph{some}
  of the components involved in a communication might fail. \GCQ features 
  novel
  operators for multiparty,
  partial and collective
  communications. 
  This paper studies the nature of failure-aware communication: First,
  we introduce 
  \GCQ  syntax, semantics and examples of its use. The interplay
  between failures and collective communications in a choreography can
  lead to choreographies that cannot progress due to absence of
  resources. In our second contribution, we provide a type
  system that ensures that choreographies can be realized despite
  changing availability conditions.
  A
  specification in 
  \GCQ guides the implementation of distributed endpoints when paired
  with global (session) types. Our third contribution provides an
  endpoint-projection based methodology for the generation of
  failure-aware distributed processes.  We show the correctness of the
  projection, and that well-typed choreographies with availability
  considerations enjoy progress.
 \end{abstract}


\maketitle




\section{Introduction}

\emph{Choreographies} are a well-established formalism in concurrent
programming, with the purpose of providing a \emph{correct-by-construction}
framework for distributed systems
\cite{carbone7scc,Carbone:2013:DMA:2429069.2429101}. Using Alice-Bob’s
style protocol narrations, they provide the structure of
interactions among components in a distributed system. Combined with a
behavioral type system, choreographies are capable of deriving
distributed (endpoint) implementations. Endpoints generated from a
choreography ascribe all and only the behaviors defined by
it. Additionally, interactions among endpoints exhibit correctness
properties, such as liveness and deadlock-freedom. In practice,
choreographies guide the implementation of a system, either by
automating the generation of correct deadlock-free code for each
component involved, or by monitoring that the
execution of a distributed system behaves according to a protocol
\cite{carbone7scc,DBLP:journals/corr/NeykovaBY14,DBLP:conf/forte/BocchiCDHY13}.

In this paper we study the role of \emph{availability} when building
communication protocols. In short, availability describes the ability
of a component to engage in a communication. Insofar, the study of
communications using choreographies assumed that components were
always available. We challenge this assumption on the light of new
scenarios. The case of Cyber-Physical Systems (CPS) is one of them. In
CPS, components become unavailable due to faults or because of changes
in the environment. Even simple choreographies may fail when including
availability considerations. Thus, a rigorous analysis of availability
conditions in communication protocols becomes necessary, before
studying more advanced properties, such as deadlock-freedom or
protocol fidelity.

Practitioners in CPS take availability into consideration, programming
applications in a \emph{failure-aware} fashion. First, application-based QoS
policies replace old node-based ones. Second, one-to-many and many-to-one
communication patterns replace peer-to-peer communications. Still,
programming a CPS from a component viewpoint such that it respects an
application-based QoS is difficult, because there is no centralized
way to ensure its enforcement.

This work reports initial steps towards a methodology for the
development of  failure-aware communication protocols, as
exemplified by CPS. We depart from choreographic programming as a
reference model, extending it in order to cope with the intrinsic
characteristics present in communication protocols for CPS. This
resulted in a novel language for choreographies, the Global Quality
Calculus (\GCQ). The novel characteristics of \GCQ include a multiparty,
asynchronous model of communication, including collective
message-passing operators, such as broadcast, collective message aggregators,
and collective method selections. Communication is rarely perfect in
CPS, and successful communications depend on the availability of
components.  \GCQ plays important consideration on this
aspect, by including component availability as a first-class
consideration to deem a communication successful. 

We present the following contributions:

\paragraph{\bf First: A formal model for Failure-aware Choreographies:}
We present the
Global Quality Calculus (\GCQ), a process calculus aimed at capturing the most
important aspects of CPS, such as variable availability
conditions and multicast communications.
It is a generalization of the Global
Calculus \cite{Carbone:2013:DMA:2429069.2429101}, enriched with
collective communication primitives and explicit availability
considerations. Central to \GCQ is the inclusion 
\emph{quality predicates} \cite{nielson2013calculus}  and 
optional datatypes, whose role is to allow
for communications where only a subset of the original participants is
available. Models in \GCQ can accommodate in this way 
application-based QoS policies, instead of a node-centric
approach. 

\paragraph{\bf Second: A Type system to control progress in
failure-aware choreographies.}
Our second contribution relates to the verification of failure-aware
protocols. We focus on \emph{progress}. As an application-based QoS, a
progress property requires that at least a minimum set of components
is available before firing a communication action.  
Changing availability conditions 
may leave collective communications without enough required
components, forbidding the completion of a protocol.  We introduce a
type system, orthogonal to session types, that ensures that well-typed
protocols with variable availability conditions do not get stuck,
preserving progress.

\paragraph{\bf Third: A Choreographic programming methodology for
  available-by-design distributed systems.} In our
third contribution, we propose a methodology to generate distributed
implementations from failure-aware choreographies.  To do so, we resort
on previous works on \emph{choreographic programming}
\cite{Carbone:2012:SCP:2220365.2220367,Carbone:2013:DMA:2429069.2429101}. Starting
from a specification in \GCQ, one can generate the distributed
implementation in terms of interacting processes. The language of
endpoints used is an extension of standard session $\pi$ calculi with
quality-based input/output processes, asynchronous and queue-based
communication. Quality choreographies are paired with a
session-type system, taking inspiration on previous works on session
types for collective communications \cite{Lopez2015OOPSLA}. As such,
session types for quality choreographies guarantee that the
specification can follow a given protocol. Moreover, they are
important in that session types guide the projection to correct
behavior. In this paper, we are interested in
\emph{availability-by-design}, a variant of deadlock-freedom that
ensures communication under minimal set of available components.



This paper is a revised and extended version of
\cite{Lopez2016Quality}. In particular, the syntax, semantics and type
system controlling progress capabilities in
\GCQ have been revised and simplified. In addition, in this paper we
provide novel sections detailing the development based on session types and endpoint
projection (c.f. \S \ref{sec:session-types}, \S \ref{sec:epc}), that
was absent in the original presentation.

\paragraph{\bf Document Structure:} In Section \S\ref{sec:langDesign} we
introduce the design considerations for a calculus with
variable availability conditions and we 
present a minimal working example to illustrate the calculus in
action. Section \S\ref{sec:language} introduces
syntax and semantics of \GCQ. 
The progress-enforcing type system is presented in
Section \S\ref{sec:type-sys}.  Section \S \ref{sec:session-types}
presents the session type methodology for quality choreographies. The
Endpoint model, and the projection from quality choreographies is
presented in Section \S \ref{sec:epc}.  Section
\S\ref{sec:relwork} discusses related work. Finally, Section \S\ref{sec:conclusions} 
concludes. 
Appendix \ref{appendix:proofs} includes additional lemmata and proofs
of the main results in the paper. 


\section{Towards a Language for CPS Communications} \label{sec:langDesign}

The design of a language for CPS requires a \emph{technology-driven}
approach, that answers to requirements regarding the nature of
communications and devices involved in CPS.  Similar approaches have
been successfully used for Web-Services
\cite{carbone2006theoretical,yoshida2013scribble,DBLP:conf/ecows/MontesiGZ07},
and Multicore Programming
\cite{Lopez2015OOPSLA,DBLP:conf/coordination/CogumbreiroMV13}.
The considerations on CPS used in this work come
from well-established sources
\cite{CPS-book,Perillo2005Wireless-sensor}. We will proceed by
describing their main differences with respect to traditional
networks.

\subsection{Unique Features in CPS Communications}

Before defining a language for communication protocols in CPS, it is
important to understand the taxonomy of networks where they
operate. CPS are composed by \emph{sensor networks} (SN) that perceive important measures
of a system, and \emph{actuator networks} that change it.  Some of the most important
characteristics in these networks include asynchronous operation,
sensor mobility, energy-awareness, application-based protocol
fidelity, data-centric protocol development, and multicast
communication patterns. We will discuss each of them.

\subsection*{Asynchrony.}
  Depending on the application, deployed sensors in a
network have less accessible mobile access points, for instance,
sensors deployed in harsh environmental conditions, such as arctic or
marine networks. Environment may also affect the lifespan of a sensor,
or increase its probability of failure. To maximize the lifespan of some
sensors, one might expect an \emph{asynchronous operation}, letting sensors
remain in a standby state, collecting data periodically. 

\subsection*{Sensor Mobility.}
The implementation of sensors in autonomic devices brings about
important considerations on \emph{mobility}. A sensor can move away from the
base station, making their interactions energy-intensive. In contrast,
it might be energy-savvy to start a new session with a different base
station closer to the new location.  

\subsection*{Energy-Awareness.}
Limited by finite energetic resources, SN must
optimize their energy consumption, both from node and application
perspectives. From a node-specific perspective, a node in a sensor
network can optimize its life by turning parts of the node off, such
as the RF receiver. From a application-specific perspective, a
protocol can optimize it energy usage by reducing its traffic. SN
cover areas with dense node deployment, thus it is unnecessary that
all nodes are operational to guarantee coverage. Additionally, SN must
provide \emph{self-configuration} capabilities, adapting its behavior to
changing availability conditions. Finally, it is expected that some of
the nodes deployed become permanently unavailable, as energetic
resources ran out. It might be more expensive to recharge the nodes
than to deploy new ones. The SN must be ready to cope with a decrease
in some of the available nodes.

\subsection*{Data-Centric Protocols.} One of the most striking
differences to traditional networks is the \emph{collaborative}
behavior expected in SN. Nodes aim at accomplishing a similar,
universal goal, typically related to maintaining an application-level
quality of service (QoS). Protocols are thus data-centric rather than
node-centric. Moreover, decisions in SN are made from the aggregate
data from sensing nodes, rather than the specific data of any of them
\cite{pattem2008impact}. Collective decision-making based in
aggregates is common in SN, for instance, in protocols suites such as
SPIN \cite{heinzelman1999adaptive} and Directed Diffusion
\cite{intanagonwiwat2000directed}. Shifting from node-level to
application-level QoS implies that \emph{node fairness} is
considerably less important than in traditional networks. In
consequence, the analysis of \emph{protocol fidelity}
\cite{honda1998lpa} requires a shift from node-based guarantees
towards application-based ones.

\subsection*{Multicast Communication.}
Rather than peer-to-peer message passing, one-to-many and many-to-one
communications are better solutions for energy-efficient SN, as
reported in
\cite{heinzelman2002application,deng2005balanced}. However, as the
number of sensor nodes in a SN scales to large numbers, communications
between a base and sensing nodes can become a limiting
factor. Many-to-one traffic patterns can be combined with data
aggregation services (e.g.: TAG \cite{madden2002tag} or TinyDB
\cite{madden2003design}), minimizing the amount and the
size of messages between nodes.

\subsection{Model Preview} \label{sec:preview}

We will
illustrate how the requirements for CPS communications have been
assembled in the our calculus
through a minimal example in Sensor Networks (SN).   
The syntax of
our language is inspired on the Global Calculus
\cite{carbone7scc,Carbone:2013:DMA:2429069.2429101} extended with collective
  communication operations \cite{Lopez2015OOPSLA}.

\begin{example}\label{example-sensors}
\begin{figure}[t!]
{
\begin{lstlisting}[
  backgroundcolor = \color{light-gray},frame=tlbr,framesep=4pt,framerule=0pt,xleftmargin=5pt,framexleftmargin=0.1em,
  mathescape,
  columns=fullflexible,
  basicstyle=\fontfamily{lmvtt}\selectfont,
  escapeinside={(*@}{@*)},
  numbers=left, numberstyle=\tiny, numbersep=5pt
]
$\initA{\thread_1[S_1]
  \Blue{\{\qpredpp[ Acc_1]\}}
  ,\thread_2[S_2]
  \Blue{\{\qpredpp[ Acc_2]\}}
  , \thread_3[S_3]
  \Blue{\{\qpredpp[ Acc_3]\}}
  }{\thread_0[M]
    \Blue{\{\qpredpp[ Acc_0]\}}
  }{temperature}{k}
    \pfx$ (*@\label{ex1-line0}@*)
$\choice{\thread_0
  \Blue{\{\qpredpp[Acc_0];\qpredpp[Ms_0]\}}
}{\thread_1
  \Blue{\{\qpredpp[Acc_1];\qpredpp[ Ms_1]\}}
  , \thread_2
  \Blue{\{\qpredpp[Acc_2]; \qpredpp[Ms_2]\}}
  , \thread_3
  \Blue{\{\qpredpp[Acc_3]; \qpredpp[Ms_3]\}}
 }{\green{q_1}}{k}{measure} \pfx $ (*@\label{ex1-line1}@*)
$\reduceA{\thread_1
  \Blue{\{\qpredpp[Ms_1];\qpredpp[E_1]\}}
  .``1", \thread_2
  \Blue{\{\qpredpp[Ms_2];
    \qpredpp[E_2]\}}
  .``{-2}",
  \thread_3
  \Blue{\{\qpredpp[Ms_3]; \qpredpp[E_3]\}}
  .``5"}{\thread_0
  \Blue{\{\qpredpp[Ms_0]; \qpredpp[E_0]\}}
  :x_m}{\green{q_2}}{k}{avg} \pfx \INACT $(*@\label{ex1-line2}@*)
\end{lstlisting}
\caption{Example: Sensor network choreography}
\label{fig:example}
}
\end{figure}
  Figure \ref{fig:example} portrays a simple SN choreography for
  temperature measurement.  Line \ref{ex1-line0} models a
  \emph{session establishment} phase between sensors
  $\thread_1, \thread_2, \thread_3$ (each of them implementing the
  indexed role
  $S_i$) and a monitor $\thread_m$ with role $M$. In Line
  \ref{ex1-line1}, $\thread_m$ executes a \emph{collective selection}
  of method $measure$ at each node. In Line \ref{ex1-line2}, an asynchronous
  many-to-one communication (e.g. \emph{reduce}) is performed between
  sensors and the monitor. Quality predicates $\green{q_1, q_2}$ model
  application-based QoS, established in terms of availability
  requirements for each of the nodes. For instance,
  $\green{q_1} = \green{q_2} = \forall$ only allows communications
  with all sensors in place, and
  $\green{q_1} = \forall, \green{q_2} =2/3$ tolerates the absence of
  one of the  sensors in data harvesting. 
  Once nodes satisfy applications' QoS requirements, an
  aggregation operation will be applied to the messages received, in
  this case computing the average value. 

  Considerations regarding the impact of available components in a
  communication must be tracked explicitly. Annotations $\Blue{\{X;Y\}}$ (in blue
  font) define \emph{capabilities}, that is, control
  points achieved in the system. The $\Blue{X}$ in
  $\thread\Blue{\{X;Y\}}$ denotes the \emph{required}
  capability for $\thread$ to act, and $\Blue{Y}$ describes the
  capability \emph{offered} after $\thread$ has engaged in an
  interaction.  No preconditions are necessary for establishing a new session, so no required capabilities are necessary in Line 1. After a session has
  been established, capabilities $\Blue{(Acc_i)_{\irange i03}}$ are available in the
  system.  Lines 
  \ref{ex1-line1} and  \ref{ex1-line2} will modify which capabilities
  are present in  the
  system depending on the number of available threads. For example, a
  model with $\green{q_1} = 2/3$ can advance from 
  $\Blue{Acc_0, Acc_1, Acc_2, Acc_3}$ to   $\Blue{Ms_0, Acc_1, Ms_2,
    Ms_3}$. 
  There may be cases in which an execution of
  the protocol will not generate necessary capabilities for a
  communication operation to be engaged, leaving a protocol stuck. One
  case will be if $\green{q_1} = 2/3, \green{q_2} = \forall$, since not
  enough $\Blue{Ms_i}$ capabilities can be provided. We will defer the discussion
  about the interplay of capabilities and quality predicates to
  Section \ref{sec:type-sys}.

\end{example}


\section{The Global Quality Calculus (\GCQ)} \label{sec:language}
  In the following, $\chor$ denotes a choreography;  $p$ denotes an
  annotated thread $\thread[A]\{\Blue{ X};\Blue{ Y}\}$, where   $\thread$ is a
thread,
$\Blue{X,Y}$ are atomic formulae and $A$ is a role
annotation. We will use $\VEC{\thread}$ to denote  $\{\thread_1,
\ldots, \thread_j\}$ for a finite $j$.
Variable 
$a$ ranges over {\em service channels},
intuitively denoting the public identifier of a service, and $k \in \Names$
ranges over a finite, countable set of  session (names), created at runtime. 
Variable $x$ ranges over variables local
to a thread.
We use terms $t$ to denote data
and expressions $e$ to denote optional data, much like the use of option data types in programming languages
like Standard ML \cite{Harper-SMLBook}. Expressions include arithmetic and other first-order expressions excluding service and
session channels. In particular, the expression
$\some{t}$ signals the presence of some data $t$ and $\none$ the absence of
data. In our model, terms denote closed values $v$. 
Names $m,n$ range over threads and
session channels. For simplicity of presentation, all models in the paper are
finite. 

\begin{definition}[\GCQ syntax]\label{globalcalculussyntax}\rm
  \begin{align*}
&\text{\emph{(Choreographies)}} &    \chor::=\ & \phantom{{}\mid{}}
   \chorAct \pfx \chor  
          ~\mid~ \ifthenelsek {e@p} {\chor} {\chor} 
        ~\mid~ \INACT         
        ~\mid~ \highlightBox{\new{r} \chor}
  \end{align*}
  \begin{align*}
  &\text{\emph{(Annotated threads)}}   & p ::=
                                          &\phantom{{}\mid{}}
                                            \thread[A]\Blue{\{X;Y\}} \\                                
\\ 
  &\text{\emph{(Interactions)}}   & \chorAct ::=
                                          &\phantom{{}\mid{}}
                                            \initA{\VEC{p_r}}{\VEC{p_{s}}}{a}{k} &&(init) \qquad \qquad \qquad \\ 
&    & &         ~ \mid ~  \bcastA{p_r}{\VEC{p_s:x_s}}{q}{e}{k}         &&(bcast)\\
&    & &         ~ \mid ~ \reduceA{\VEC{p_r.e_r}}{p_{s}:x}{q}{k}{op} &&(reduce)\\
&    & &         ~ \mid ~ \choiceA{p_r}{\VEC{p_s}}{q}{k}{l}  &&(select)
  \end{align*} 
\end{definition}

A novelty in this variant of the Global calculus is the addition of \emph{quality
predicates} $\green{q}$ binding vectors in a multiparty communication.
Essentially, $\green{q}$ determines when sufficient
inputs/outputs are available. For example, $\green{q}$ can be $\exists$,
meaning that one sender/receiver is required in the interaction, it
can be $\forall$ meaning that all of them are needed, or it can be
$m/n$, describing that $m$ out of $n$ components are needed.  The syntax of
$\green{q}$ and other examples can be summarised in Figure
\ref{fig:syntax-quality-predicate}.
We require 
$\green{q}$ to be monotonic (in the sense that
$\green{q}(\VEC{\thread_r})$ implies $\green{q}(\VEC{\thread_s})$ for all
$\VEC{\thread_s} \subseteq \VEC{\thread_r}$) and satisfiable.

\begin{figure}[t]
{\footnotesize
  \begin{align*}
    \lencod{\green{\forall}}(\VEC{\thread_r}) &= ( |\{\thread_i \in
    \VEC{\thread_r} \mid \thread_i = \true \} | = n ) & n & = |\VEC{\thread_r}| \\
    \lencod{\green{\exists}}(\VEC{\thread_r}) & = ( |\{\thread_i \in
    \VEC{\thread_r} \mid \thread_i = \true \} | \ge 1 ) & n & = |\VEC{\thread_r}| \\
    \lencod{\green{ m/n}}(\VEC{\thread_r}) & = ( |\{\thread_i \in
    \VEC{\thread_r} \mid \thread_i = \true \} | \ge m ) & n & = |\VEC{\thread_r}| 
  \end{align*}
  \caption{Quality predicates: syntax $\green{q}$ and semantics $\lencod{\green{q}}$.}
  \label{fig:syntax-quality-predicate}
}
\end{figure}

Choreographies are composed by standard operators of restriction,
  if-then choice and inaction, as
standard in the literature. 
 We will focus our
discussion on the novel interactions. 
%
First, \!\startK defines a (multiparty) \emph{session initiation} between
active annotated threads $\VEC{p_r}$ and annotated service threads
$\VEC{p_{s}}$ over a (shared) service channel $a$.
Each active thread
(resp. service thread) implements the behaviour of one of the roles in
$\VEC{A_r}$ (resp. $\VEC{A_s}$), sharing a new session
name $k$. We assume that a session is established with at least two
participating processes, therefore $2 \le |\VEC{p_r}|+|\VEC{p_s}|$,
and that threads in $\VEC{p_r}\cup \VEC{p_s}$ are pairwise different.

The language features broadcast, reduce and selection as collective
interactions. 
A \emph{broadcast} implements a one-to-many communication pattern,  
where a session channel
$k$ is used to transfer the evaluation of expression $e$ (located at
$p_r$)  
to threads in  $\VEC{p_s}$, 
with the resulting binding of variable $x_i$ at $p_i$, for each $p_i
\in \VEC{p_s}$. 
A \emph{reduce} combines one-to-many communicationsas well as
aggregation functions.   In
$\reduceA{\VEC{p_r.e_r}}{p_{s}:x}{q}{k}{op}$, each annotated thread
$p_i$ in $\VEC{p_r}$ evaluates an expression $e_i$, and the aggregate of all receptions
is evaluated using   
$\mathsf{op}$ (an operator defined on multisets 
such as $\operator{max}, \operator{min}$, etc.)
Interaction $\choiceA{p_r}{\VEC{p_s}}{q}{k}{l}$ describes a collective \emph{label selection}: $p_r$
communicates the selection of label $l$ to peers in $\VEC{p_s}$
through session $k$. In order to simplify the technical development, we will
assume that $\green{q} = \forall$ in
$\choiceA{p_r}{\VEC{p_s}}{q}{k}{l}$ (that is, we require all
receiving participants to perform a collective selection). 

Central to our language are \emph{progress capabilities}. Pairs of
atomic formulae
$\{\Blue{X};\Blue{Y}\}$ at each annotated thread state 
the necessary preconditions for a thread to engage ($\Blue{X}$), and
the capabilities provided after its interaction ($\Blue{Y}$). As we
will see in the semantics, there are no associated preconditions for
session initiation (i.e. threads are created at runtime), so we
normally omit them.  Explicit $x@p$ (resp. $e@p$)
indicate the variable/boolean expression $x$ (resp. $e$) is located at
$p$. 
Term $\new r \chor$ represents the restriction of a name $r$ in
$\chor$, and it will be only used at runtime. The same notation
standard will be used for all the terms written surrounded by boxes (as in
$\highlightBox{\new r \chor}$). We often omit $\INACT$, empty
vectors  and atomic formulae $\{\Blue{ X};\Blue{ Y}\}$ from annotated threads when unnecessary. 

The set of free term variables  $\fv \chor$, free names $\fn(\chor)$,
free threads $\ft{\chor}$, service channels
$\fsc{\chor}$  and roles $\roles{\chor}$  are inductively defined as
usual for $\chor$ and for $\chorAct$.
 An interaction $\chorAct$ in  $\chorAct \pfx \chor$ can bind session
channels, choreographies and variables. In $(init)$, variables $\{\VEC{p_r}, a\}$ are free while variables $\{\VEC{p_s}, k\}$ are bound
  (since they are freshly created). 
  In $(bcast)$, variables $\VEC{x_s}$ are
  bound.  The $(reduce)$ interaction binds $\{x\}$. 
\subsection{Expressivity}
The
importance of roles 
is only crucial in a \!\!$\startK$\!\! interaction. Technically, one
can infer the role of a given thread $\thread$ used in an interaction
$\chorAct$  by looking at the \!$\startK$\!\! interactions
preceding it in the abstract syntax tree. 
\GCQ can still represent unicast message-passing
patterns as in \cite{carbone7scc}. Unicast communications
$\unicastA{p_1}{p_2:x}{e}{k} $ can be encoded in multiple
ways using broadcast/reduce operators. For instance,  $
\bcastA{p_1}{p_2:x}{\forall}{e}{k}$ and
$\reduceA{p_1.e}{p_2:x}{\forall}{id}{k}$ are just a couple of possible
implementations. Similar considerations apply also for unicast selection
$\unicastChoiceA ABlk$.
\subsection{Semantics} \label{sec:global:semantics}
Choreographies are considered modulo standard structural and swapping congruence
relations (resp. $\equiv$, $\swaps$). Structural congruence $\equiv$
is 
defined as the least congruence relation on $\chor$ supporting
$\alpha-$renaming, such that rules $\new x\new y \chor \equiv \new
y\new x  \chor$ and $\new x \INACT \equiv \INACT$ hold.

The swap congruence  \cite{Carbone:2013:DMA:2429069.2429101}
provides a way to reorder non-conflicting interactions, allowing for a restricted form of
asynchronous behavior. Non-conflicting interactions are those
involving sender-receiver actions that do 
not conform a control-flow dependency.
For instance,
$\bcastA{\thread_A}{\thread_B:x_B}{q_1}{e_A}{k_1} \pfx
\bcastA{\thread_C}{\thread_D:x_D}{q_2}{e_C}{k_2} \swaps
\bcastA{\thread_C}{\thread_D:x_D}{q_2}{e_C}{k_2} \pfx
\bcastA{\thread_A}{\thread_B:x_B}{q_1}{e_A}{k_1}$.  
Formally, let $\threads{\chor}$ be the
set of threads in $\chor$, defined inductively as $\threads{\chorAct
  \pfx C} \eqDef \threads{\chorAct} \cup \threads{C}$, and
$\threads{\chorAct} \eqDef \bigcup_{i = \{1..j\}} \thread_i$ if
$\chorAct = \bcastA{\thread_1[A_1]}{\thread_2[A_2]:x_2, \ldots,
  \thread_j[A_j]:x_j}{q}{e}{k}$ (similarly for $(init)$, $(reduce)$ and
$(select)$, and standardly for the other process constructs in
$\chor$).  
The swapping congruence rules are 
presented in Figure \ref{fig:swap-rel}, where we use the shorthand notation
$A ~\#~ B$ to denote set disjointness,  $A \cap B = \emptyset$.  
\begin{figure}[t!]
  \begin{gather*}
    \frac{ 
      \threads{\chorAct} ~\#~       \threads{\chorAct'} 
    }{
      \chorAct \pfx (\chorAct' \pfx \chor) \swaps \chorAct' \pfx (\chorAct \pfx \chor)
    }
\quad 
    \frac{ p \notin \threads{\chorAct}}{
      \ifthenelsek {e@p} { \chorAct \pfx \chor_1}
      { \chorAct \pfx \chor_2} \swaps
      \chorAct \pfx\ifthenelsek {e@p} { \chor_1} { \chor_2} }
   \\[\ruleskip]    
    \frac{
      p \neq r
    }{ \begin{array}{l}
      \ifthenelsek {e@p} {(\ifthenelsek {e'@r} {\chor_1} {\chor_2}) \\}
      { (\ifthenelsek {e'@r} {\chor'_1} {\chor'_2})}
       \end{array}
       \swaps \begin{array}{l}
      \ifthenelsek {e'@r} {(\ifthenelsek {e@p} {\chor_1} {\chor'_1})\\} 
      { (\ifthenelsek {e@p} {\chor_2} {\chor'_2})\end{array}}}
  \end{gather*}
  \caption{Swap congruence relation, $\swaps$}
  \label{fig:swap-rel}
\end{figure}

A state $\sigma$ keeps track of the capabilities achieved by a
thread in a session, and it is formally defined as set of maps $(\thread, k
) |-> X$. 
The rules in Figure \ref{fig:store-update} define state manipulation
operations, including update ($\updateState{\sigma'}$), and lookup ($\sigma(\thread, k)$).
\begin{figure}[t]
  \begin{align*}
    & \Rule{
    \Blue{ Y} =
         \Blue{ X} \text{ if } (\thread, k, X ) \in \sigma  \quad
    \Blue{ Y} =     \Blue{\emptyset} ~~ \text{o.w.}
    } 
    {  \sigma(\thread, k) = \Blue{ Y}} 
\quad
     \Rule{
    \delta = 
      \left\{ 
      (\thread, k, X) ~|~ \begin{array}{l} (\thread, k,
                                           X) \in \sigma \\
                                           \land  (\thread, k, Y) \in \sigma' 
                                         \end{array}
    \right\}
    } 
    {  \updateState{\sigma'} = (\sigma\backslash \delta), \sigma'}     
  \end{align*}
\caption{State lookup and update rules}
\label{fig:store-update}
\end{figure}

A substitution  $\theta = 
\subst{(p_1,\some{v_1}),  \ldots,  (p_n,\some{v_n}) }{x_1@p_1, \ldots, x_n@p_n }$
maps each
variable  $x_i$ at $p_i$ to optional data $\some{v_i}$ for $1 \le i \le n$.
A composition $\theta_1 \circ \theta_2(x)$ is defined as $\theta_1 \circ \theta_2(x)
::= \theta_1(\theta_2(x))$, and $q(t_1, \ldots, t_n) = \bigwedge_{i
\in 1 \le i \le n} t_i$ if $q =  \forall$, $q(t_1, \ldots, t_n) = \bigvee_{i
\in 1 \le i \le n} t_i$ if $q =  \exists$, and possible combinations
therein. As for process terms,  $\theta(\chor)$ denotes the
application of substitution $\theta$ to a term $\chor$ (and similarly
for $\chorAct$).
\begin{figure*}[t!]
{\small
  \begin{align*}
    &    
      \qquad
      \inferenceg{^G|Init}
      {
      \begin{array}{c}
        \chorAct =\initA{ \VEC{\thread_r[A_r]\{\Blue{Y_r}\}}
        }{ \VEC{\thread_s[B_s]\{\Blue{ Y_s}\}}}{a}{k}  
        \quad
        \sigma' = [(\thread_i,k)  |-> \Blue{ Y_i} ]_{i = 1}^{|\VEC{\thread_r}|} 
        \quad
        \sigma'' = [(\thread_i,k)  |-> \Blue{ Y_i} ]_{i = 1}^{|\VEC{\thread_s}|} 
      \end{array}
      }
    {
     \conf{\sigma}{\initA{ \VEC{\thread_r[A_r]\{\Blue{Y_r}\}}}{ \VEC{\thread_s[B_s]\{\Blue{ Y_s}\}}}{a}{k}  \pfx \chor}
    \action{\chorAct} 
    \conf{\sigma[\sigma'[\sigma'']]}{\new{\VEC{\thread_s},k}\chor} 
    }
    \\\\    
    & \quad
\inferenceg{^G|Bcast}
    {       
      \begin{array}{c}
        J \subseteq \VEC{\thread_r}
        \quad
        q(J)
        \quad
        \forall {\iset i{\{\thread_A\} \cup J}}: \Blue{ X_i} \subseteq
        \sigma(\thread_i, k) \land
        \sigma'(t_i, k) = \exchange{\Blue{ X_i}}{\Blue{ Y_i}}(\sigma(t_i,k))
        \quad
        \evalOp{e@\thread_A}{ v }
        \\
        \forall {\irange {i}{1,}{,|\VEC{\thread_r}|}}: \theta(x_{i}) =
        \left\{
        \begin{array}{l} 
          \some{v} ~~ \thread_i \in J  \\
          \none \qquad \text{o.w.}
        \end{array}
        \right.
        \\
        \chorAct =
        \bcastA{
        \thread_A[A]\{\Blue{ X_A;Y_A}\}
        }{\VEC{\thread_r[B_r]\{\Blue{X_r;Y_r}\}:x_r}}{q}{v}{k}
      \end{array}
    } {
    \conf{\sigma}{    \left(
    \bcastA{
    \thread_A[A]\{\Blue{ X_A;\!Y_A}\}
    }{\VEC{\thread_r[B_r]\{\Blue{ X_r;\!Y_r}\}\!:\! x_r}}{q}{e}{k} \right) \!\pfx\! \chor
    } 
    \action{    
    \theta(\chorAct)
    }\ 
   \conf{\sigma[\sigma']}{\theta(\chor) }}
    \\\\ & \qquad
                   \inferenceg{^G|Sel}
                   {    
                   \begin{array}{c}
                     J \subseteq \VEC{\thread_r} 
                     \quad
                     q(J)
                     \quad
                     \forall {\iset i{\{\thread_A\} \cup J}}: \Blue{ X_i} \subseteq
                     \sigma(\thread_i, k) \land
                     \sigma'(\thread_i, k) =
                     \exchange{\Blue{ X_i}}{\Blue{ Y_i}}(\sigma(\thread_i, k))
                     \\
                   \chorAct =
                   \choiceA{\thread_A[A]\{\Blue{ X_A;Y_A}\}}{\VEC{\thread_r[B_r]\{\Blue{ X_r;Y_r}\}}}{q}{k}{l_h}
                   \end{array}
                   } {
                   \conf{\sigma}{ 
                   \left(
                   \choiceA{\thread_A[A]\{\Blue{ X_A;Y_A}\}}{\VEC{\thread_r[B_r]\{\Blue{ X_r;Y_r}\}}}{q}{k}{l_h}
                   \right)                \pfx \chor} 
                   \action{
                   \chorAct 
                   } 
                       \conf{\sigma[\sigma']}{\chor }}
\\\\ & 
               \inferenceg{^G|Red}
               {
               \begin{array}{c}
                 J \subseteq \VEC{t_r} 
                 \quad
                 q(J)
                 \quad
                 (\evalOp{e_i@\thread_i}{ v_i })_{\iset{\thread_i}{J}}
                 \quad
                 \forall {\iset {\thread_i}{\{\thread_B\} \cup J}}: \Blue{ X_i} \subseteq
                 \sigma(\thread_i, k) \land
                 \sigma'(\thread_i, k) = \exchange{\Blue{ X_i}}{\Blue{
                 Y_i}}(\sigma(\thread_i,k))
                 \\
                 {\forall {\irange {i}{1,}{,|\VEC{\thread_r}|}}: \theta(x_i) =
                 \left\{
                 \begin{array}{l} 
                   \some{v_i} ~~ \thread_i \in J  \\
                   \none \qquad \text{o.w.}
                 \end{array}
                 \right.}
                 \quad
                 \evalOp{\mathsf{op}(\theta)}{ \some{v} }         
                 \quad
                 dom(\sigma') = J \cup \{\thread_B\}
                 \\
                 \chorAct = \reduceA{\VEC{\thread_r[A_r]\{\Blue{
                 X_r;Y_r}\}.v_r}}{\thread_B[B]\{\Blue{
                 X_B;Y_B}\}:x}{q}{k}{op} 
               \end{array}
       } {
    \begin{array}{l}
    \conf{\sigma}{  \reduceA{\VEC{\thread_r[A_r]\{\Blue{
    X_r;Y_r}\}.e_r}}{\thread_B[B]\{\Blue{
    X_B;Y_B}\}:x}{q}{k}{op} \pfx \chor\ }
    \action{\chorAct \subst{\some{v}}{x@\thread_B}}\ \\ \qquad \qquad \qquad \qquad \qquad \qquad \qquad \qquad \qquad \qquad \qquad \qquad \qquad 
    \conf{\sigma[\sigma']}{\chor
    \subst{\some{v}}{x@\thread_{B}}}
    \end{array}
    }
\\\\ &
                       \inferenceg{^G|Res}
                       { 
                       \conf{\sigma}{\chor}
                       \action{\lambda}
                       \conf{\sigma'}{\chor'}
                       }{ 
                       \conf{\sigma}{\new{r} \chor} 
                       \action{\lambda}
                       \conf{\sigma'}{\new{r} \chor'} }
                     \quad 
                     \inferenceg{^G|Cong}
                    {
                    \begin{array}{c}
                      \chor \,\mathcal R\, \chor' \quad
                       \conf{\sigma}{\chor'} \action{\lambda}
                       \conf{\sigma'}{\chor''} 
                      \quad
                      \chor'' \,\mathcal R\, \chor''' 
                      \quad
                      \mathcal R \in \{ \equiv, \swaps\} 
                    \end{array}
                     }  { \conf{\sigma}{\chor}
                     \action{\lambda}
                     \conf{\sigma'}{\chor'''}}
                     \\\\ &\qquad \qquad \qquad \qquad                             \qquad \qquad 
                     \inferenceg{^G|If}
                     {\begin{array}{c}
                   i=1 \text{ if } \evalOp{e@\thread}{\true}, 
                   \quad
                   i=2 ~\text{ o.w. }
                 \end{array}
                            } { 
                            \conf{\sigma}{\ifthenelsek{e@\thread}{\chor_1}{\chor_2}} 
                            \action{\tau} 
                            \conf{\sigma}{\chor_i }}
  \end{align*}
}
  \caption{\GCQ : Operational Semantics}
  \label{fig:global:semantics}
\end{figure*}


We now have all the ingredients to understand the semantics of
\GCQ. The set of transition rules in $\action{\lambda}$ is defined
as the minimum relation on names, states, and choreographies satisfying the
rules in Figure~\ref{fig:global:semantics}. The operational semantics
is given in terms of labelled transition rules. Intuitively, 
 a transition
$\ltransition{\conf{\sigma}{\chor}}{\lambda}{\conf{\sigma'}{\chor'}}$
expresses that a 
configuration $<. \sigma, \chor .>$ fires an action 
$\lambda$ and evolves into $<. \sigma', \chor' .>$.
The exchange function $\exchange{X}{Y}Z$
returns $(Z\backslash X)\cup Y$ if $ X \subseteq Z$ and $Z$ otherwise. 
Actions
are defined as
$\lambda::=\{\tau, \chorAct
\}$, where
$\chorAct$ denotes interactions, and $\tau$ represents an internal
computation.
Relation $\evalOp{e@p}{v}$ describes the evaluation of
a expression $e$ (in $p$) to a value $v$. 

We now give intuitions on the most representative operational rules.
Rule
$\Did{^G|Init}$ models initial interactions: state $\sigma$ is updated
to account for the new threads in the session, updating the
set of used names in the reductum.
%
Rule $\Did{^G|Bcast}$ models broadcast: 
given an expression evaluated at the sender, one needs to check that
there are
enough receivers ready to get a message. Such a check is performed by
evaluating $q(J)$. In case of a positive evaluation, the execution of
the rule will:
(1) update the current state with the new states of each
participant engaged in the broadcast, and (2) apply the partial
substitution $\theta$ to the continuation $\chor$.  Rule
$\Did{^G|Sel}$ behaves in a similar way. 
%
The behaviour of a reduce operation is described via rule
$\Did{^G|Red}$. 
If all required threads are present, one can
proceed by evaluating the operator to the set of received
values, binding variable $x$ to its results. Rule $\Did{^G|Cong}$ allows
choreographies to evolve up to swap and structural
congruence. Finally, rule 
$\Did{^G|If}$ represent
standard 
if-then-else constructs in sequential languages.

In contrast to
previous works in multiparty sessions
(e.g. \cite{castagna2011global}), we present an \emph{early} semantics: it
allows for transitions to match with distinct  moves, depending on
which participants are available first. We opted to favor an
application-based QoS rather than a node-based QoS, as described in
Section \ref{sec:langDesign}. 
\begin{remark}[Broadcast vs. Selection] 
The inclusion of separate language constructs for communication and
selection takes origin in early works of structured communications
\cite{honda1998lpa}. Analogous to method invocation in
object-oriented programming, selections play an important role in
making choreographies projectable to distributed implementations. We
illustrate their role with an example. Assume a session key $k$ shared
among threads $p, r, s$, and an evaluation of $e@p$ of boolean
type. The choreography $\unicastA{p}{r:x}{e}{k} \pfx   \ifK\, (x@r) \,\thenK\,
\left(\unicastA{r}{s:y}{d}{k}\right) \,\elseK\, \\ \left(\unicastA{s}{r:z}{f}{k}\right)$
branches into two different communication flows: one from $r$ to $s$ if the
evaluation of $x@r$ is true, and one from $s$ to $r$ otherwise.
Although the evaluation of the guard in the  $\ifK$ refers only to
$r$, the projection of such choreography to a distributed system
requires $s$ to behave differently based on the decisions made by
$r$. The use of a selection operator permits $s$ to be
notified by $r$
about which behavior to implement:
$\unicastA{p}{r:x}{e}{k} \pfx   \ifK\, (x@r) 
\,\thenK\, \\ \left(  \unicastChoiceA{p}{r}{k}{l_1} \pfx
  \unicastA{r}{s:y}{d}{k}\right)
\,\elseK\, 
\left(\unicastChoiceA{p}{r}{k}{l_2} \pfx
  \unicastA{s}{r:z}{f}{k}\right)
$
\end{remark}
\begin{remark}[Broadcast vs. Reduce]
  We opted in
  favor of an application-based QoS instead of a classical node-based QoS, as
  described in Section \ref{sec:langDesign}.  This  consideration
  motivates the asymmetry of broadcast and reduce commands: both
  operations are blocked unless enough receivers are available,
  however, we give precedence to senders over receivers.  In a
  broadcast, only one sender needs to be available, and provided
  availability constraints for receivers are satisfied, its evolution will be
  immediate. In a reduce, we will allow a delay of the transition,
  capturing in this way the fact that senders can become active in
  different instants.
\end{remark}

The reader familiar with the Global Calculus may
have noticed the absence of a general asynchronous behaviour in our
setting. In particular,  rule:
\begin{align*}
                    \inferenceg{^G|Asynch}
                    {
                    \begin{array}{c}
                      \conf{\sigma}{\chor}
                      \action{\lambda} 
                      \conf{\sigma'}{\new{\VEC{r}} \chor'}
                      \quad
                      \chorAct \neq \startK
                      \quad
                      \senders{\chorAct} \subseteq \fn(\lambda)
                      \\
                      \receivers{\chorAct} ~\#~ \fn(\lambda)
                      \quad
                      \forall_{r \in \VEC r}\ ( 
                      r \in \bn(\lambda)
                      \quad
                      r \notin \fn(\chorAct)
                      )
    \end{array}
    }
    {
  \conf{\sigma}{\chorAct \pfx \chor}
  \action{\lambda} 
  \conf{\sigma'}{\new{\VEC{r}}(\chorAct \pfx \chor')}
    }
\end{align*}
corresponding to the extension of rule 
{\small$\Did{^C|_{ASYNCH}}$} in
\cite{Carbone:2013:DMA:2429069.2429101} with collective
communications, is absent in our
semantics.  The reason behind it lies in the energy considerations of our
application: consecutive communications may have different
energetic costs, affecting the availability of sender nodes. Consider
for example the configuration 
{\small
\[    
\begin{array}{l}
  \langle \sigma,     (
    \bcastA{
    \thread_A[A]\{\Blue{ X;Y}\}
    }{\VEC{\thread_r[B_r]:x_r}}{\exists}{e}{k}) 
  \pfx \bcastA{
    \thread_A[A]\{\Blue{ X;Y}\}
    }{\VEC{\thread_s[B_s]:x_s}}{\forall}{e}{k}
     \rangle
\end{array}
\] 
}
with $\VEC{\thread_r} ~\#~ \VEC{\thread_s}$ and $\sigma(\thread_A,k) = X 
$. If the order of the broadcasts is shuffled, the
second broadcast may consume all energy resources for $\thread_A$,
making it unavailable later. Formally, the execution of a broadcast  update the capabilities
offered in $\sigma$ for $\thread_A, k$ to $Y$, inhibiting two
communication actions with same capabilities to be reordered.   We will refrain the use rule $\Did{^G|Asynch}$ in our semantics.


We say that a choreography $\chor$ is restriction-free if $\chor$ does
not contain any subterm $\new x\chor$. We can proceed to define our
notion of progress.

\begin{definition}[Progress: Choreographies]
  Choreography 
  $\chor$ progresses if there exists
  $ \chor', \sigma',$ 
  $\lambda$ such that  
  $\conf{\sigma}{\chor}
  \action{\lambda} \conf{\sigma'}{\chor'}$,
  for all $\sigma$. 
\end{definition}






\section{Type-checking Progress} \label{sec:type-sys}
One of the challenges regarding the use of partial collective
operations concerns the possibility of getting into runs with
locking states. 
Consider a variant of Example \ref{example-sensors} 
with $\green{q_1} = \exists$ and $\green{q_2} = \forall$.
This choice leads to a blocked configuration. The system blocks
since the collective selection in Line (\ref{ex1-line1}) continues after a subset of the receivers in
$\thread_1,\thread_2,\thread_3,$ have executed the command. Line
(\ref{ex1-line2}) requires all senders to be ready, which will not be
the most general case. The system will additionally block if participant dependencies among communications
are not preserved. The choreography in Figure \ref{example2} blocks for $\green{q_1} = \exists$, since the
selection operator in Line (\ref{ex1-line1}) can execute a
communication over $\thread_2$, blocking the reduce operation  in Line
\ref{example:causality}.
\begin{figure}[!h]
{
\begin{lstlisting}[
  backgroundcolor = \color{light-gray},frame=tlbr,framesep=0pt,framerule=0pt,xleftmargin=5pt,framexleftmargin=0.1em,
  mathescape,
  columns=fullflexible,
  basicstyle=\fontfamily{lmvtt}\selectfont,
  escapeinside={(*@}{@*)},
  numbers=left, numberstyle=\tiny, numbersep=5pt,
  firstnumber=2
]
  ... Lines 1,2 in Figure $\ref{fig:example}$.
  $\reduceA{\thread_1 \Blue{\{\qpredpp[Ms_1]; \qpredpp[E_1]\}}.1,
    \thread_3 \Blue{\{\qpredpp[Ms_3]; \qpredpp[E_3]\}}.5}{\thread_m \Blue{\{\qpredpp[Ms_0];
    \qpredpp[E_0]\}}:x_0}{\exists}{k}{avg}\pfx \INACT $(*@\label{example:causality}@*)
\end{lstlisting}
}
\caption{Variant of Example \ref{example-sensors} with locking states}
\label{example2}
\end{figure}

We introduce a type system to ensure progress on variable availability
conditions. A judgment is written as $\isEntailed{\chor}$, where
$\ltypeEnv$ is a list of formulae in
Intuitionistic Linear Logic (ILL) \cite{Girard:1987}. 
Intuitively,
$\isEntailed{\chor}$ is read as \emph{the formulae
  in $\ltypeEnv$ describe the program point immediately before}
$\chor$. Formulae  $\psi \in \ltypeEnv$ take the form of the constant
$\true$, ownership types of the form  $p: \ownerTP{k}{A}{X}$, and the
linear logic version of conjunction, disjunction and implication
$(\otimes, \oplus, \lolli)$.
Here $p: \ownerTP{k}{A}{X}$ is an \emph{ownership type}, asserting that $p$ behaves as
the role $A$ in session $k$ with atomic formula $X$. 
Moreover, we require $\ltypeEnv$ to
contain formulae free of linear implications in    $\isEntailed{\chor}$\footnote{We
do, however, use the full set of operators when performing proof search}.

Figure \ref{fig:global:types-causality} presents selected rules for the type system for
\GCQ. 
Since the rules for
inaction, restriction, conditionals and non-determinism are standard, we focus
our explanation on the typing rules for communications. 
Rule $\Did{TInit}$ types new sessions: $\ltypeEnv$ is extended with 
function $\isInit{ \widetilde{\thread_p[A]\{X\} }, k}$, that returns a list of
   ownership types $\VEC{ \thread_p \colon \ownerTP kAX }$.
   Conditions $    \VEC{\thread_s} \not\subseteq \threads{\ltypeEnv}$
   and $ k \notin \keys{\ltypeEnv} $ 
    ensure that new names do not exist neither in the threads
     nor in the used keys 
    in $\ltypeEnv$.

The typing rules for broadcast, reduce and selection are analogous, so
we focus our explanation in  $\Did{TBcast}$. Here we abuse of the
notation, writing $\isEntailed{ \chor}$ to denote type checking, and
$\Psi \vdash \psi$ to denote formula entailment.
The semantics of $\forallplus J$ s.t. $\mathbf{C}:D$ is given by
$\forall J$ s.t.  $\mathbf{C}:D \land \exists J$ s.t. $\mathbf{C}$.  The judgment
\[\isEntailed{(\bcastA{\thread_A[A]\Blue{\{X_A;Y_A\}}}{\VEC{\thread_r[B_r]\Blue{\{X_r;Y_r\}}:x_r}}{q}{e}{k}) \pfx \chor}\]
succeeds if environment $\ltypeEnv$ can provide capabilities for
sender $\thread_A [A]$ and for a valid subset $J$
of the  receivers in $\VEC{\thread_r[B_r]}$.  
$J$ is a valid subset if it contains enough threads to render the quality
predicate true ($q(J)$), and judgment   
$    \typerule{
  \psi_A,   (\psi_j)_{j \in J},  \phi \lolli \phi'
}{}{
  \phi'
}$ 
is provable.  This proof succeeds if  $
\psi_A $ and $ (\psi_j
) _{j \in J}$ contain ownership types for the
sender and available receivers with corresponding
capabilities. 
Finally, the type of the  continuation $\chor$ will consume the
resources used in the sender and all involved receivers, updating them with new
capabilities for  the threads engaged.

 \begin{figure*}[!ht]
{\small 
  \begin{align*}
& \text{Choregraphy Formation }    (\isEntailed{ \chor}), \\
&     \qquad \qquad \quad
\inferenceg{Tinit}
{ 
  \begin{array}{c}
    \isEntailed[\ltypeEnv, \isInit{
    \VEC{\thread_r[A_r]\Blue{\{Y_r\}}}, \VEC{\thread_s[B_s]\Blue{\{Y_s\}}},
    k}]{\chor}
    \quad
    \VEC{\thread_s} \not\subseteq \threads{\ltypeEnv} \quad  k \notin \keys{\ltypeEnv} 
  \end{array}
  }
    { 
    \isEntailed{ \initA{\widetilde{\thread_r[A_r]\Blue{\{Y_r\}}}}{
    \VEC{\thread_s[B_s]\Blue{\{Y_s\}}}}{a}{k} \pfx \chor } 
    } 
    \\\\ &
                      \hspace{\horizontalskip}
                      \inferenceg{Tbcast}
                    { \begin{array}{c}
                        \forall^{\ge 1} J.\ s.t. \left(
                        \begin{array}{c}
                          J \subseteq \VEC{\thread_r}
                          ~\land~
                          q(J) 
                          ~\land~
                          \ltypeEnv = \psi_A, (\psi_j)_{j \in
                          J},  \ltypeEnv' 
                          \\[\ruleskip] 
                          ~\land~
                          \typerule
                          {
                          \psi_A, (\psi_j)_{j \in
                          J}
                          }{}{
                          \thread_A \colon
                          \ownerTP{k}{A}{X_A} \bigotimes_{j \in J} (\thread_j \colon
                          \ownerTP{k}{B_j}{X_j}  ) 
                          } 
                        \end{array}
                        \right):
                        \\
                        \isEntailed[      \thread_A \colon
                        \ownerTP{k}{A}{Y_A},  \left(\thread_j 
                        \colon \ownerTP{k}{B_j}{Y_j} \right)_{j \in
                        J},
                        \ltypeEnv']{\chor} 
                        ~~
                        \isOptData[e@\thread_A]
                        ~~
                        (\isOptData[x_i@\thread_i]) _{i =1}^{|\VEC{\thread_r}|}
                      \end{array}
    }
    {
    \isEntailed{
    \left( \bcastA{\thread_A[A]\Blue{\{X_A;Y_A\}}}{\VEC{\thread_r[B_r]\Blue{\{X_r;Y_r\}}:x_r}}{q}{e}{k} \right) \pfx \chor }
    } 
    \\\\ &                    \hspace{\horizontalskip}                    
                    \inferenceg{Tred}
                    { \begin{array}{c}
                        \forall^{\ge 1} J.\ s.t. \left( 
                        \begin{array}{c}
                          J \subseteq \VEC{\thread_r}
                          ~\land~ 
                          q(J)
                        ~\land~
                          \ltypeEnv = \psi_B, (\psi_j)_{j \in J}, \ltypeEnv' 
                          \\
                          ~\land~
                          \typerule
                          {
                          \psi_B,  (\psi_j)_{j \in J}
                          }{}{
                          \thread_B \colon
                          \ownerTP{k}{B}{X_B} \bigotimes_{j \in J} (\thread_j \colon
                          \ownerTP{k}{A_j}{X_j}  ) 
                          } 
                        \end{array}
                        \right): \\[\ruleskip]
                        \isEntailed[
                        \thread_B \colon  \ownerTP{k}{B}{Y_B},
                        \left(\thread_j \colon \ownerTP{k}{A_j}{Y_j}
                        \right) _{j \in J}, 
                        \ltypeEnv']{\chor}
                        \quad
                        (\isOptData[e_i@\thread_i])_{i =1}^{|\VEC{\thread_r}|}
                        \quad
                        \isOptData[x@\thread_B]
                      \end{array}
    } 
    {
    \isEntailed[\ltypeEnv]{\left(
    \reduceA{\VEC{\thread_r[A_r]\Blue{\{X_r;Y_r\}}.e_r}}{\thread_B[B]\Blue{\{X_B;Y_B\}}:x}{q}{k}{op}
    \right) \pfx \chor }
    } 
    \\\\ & 
                    \hspace{\horizontalskip}
                    \inferenceg{Tsel}
                    { \begin{array}{c}
                        \forall^{\ge 1} J.\ s.t. \left(
                        \begin{array}{c}
                          J \subseteq \VEC{\thread_r}
                          ~\land~
                          q(J) 
                          ~\land~
                          \ltypeEnv = \psi_A, (\psi_j)_{j \in
                          J},  \ltypeEnv' 
                          \\[\ruleskip] 
                          ~\land~
                          \typerule
                          {
                          \psi_A, (\psi_j)_{j \in
                          J}
                          }{}{
                          \thread_A \colon
                          \ownerTP{k}{A}{X_A} \bigotimes_{j \in J} (\thread_j \colon
                          \ownerTP{k}{B_j}{X_j}  ) 
                          } 
                        \end{array}
                        \right):
                        \\
                        \isEntailed[      \thread_A \colon
                        \ownerTP{k}{A}{Y_A},  \left(\thread_j 
                        \colon \ownerTP{k}{B_j}{Y_j} \right)_{j \in
                        J},
                        \ltypeEnv']{\chor} 
                      \end{array}
    }
    {
    \isEntailed[\ltypeEnv]{ \left( \choiceA{\thread_A[A]\Blue{\{X_A;Y_A\}}}{\VEC{\thread_r[B_r]\Blue{\{X_r;Y_r\}}}}{q}{k}{l_h} \right)                \pfx \chor}
    } 
    \quad 
    \inferenceg{Tinact}
    { 
    \quad 
    } 
    {
    \isEntailed[\ltypeEnv]{ \INACT }
    } 
    \\\\ & 
           \qquad \qquad \qquad \qquad
           \inferenceg{Tcond}
           { 
           \isEntailed[\ltypeEnv]{\chor_1}
           \qquad
           \isEntailed[\ltypeEnv]{\chor_2} 
           }  
           {
           \isEntailed[\ltypeEnv]{ \ifthenelsek {e@\thread} {\chor_1}
           {\chor_2}    }
           } 
           \qquad
           \inferenceg{Tres}
           { \isEntailed{\chor}} 
           {
           \isEntailed{\new{x}\chor}
           }  
    \\\\
& \text{Data Typing,  }  
\\[\ruleskip] & 
  \qquad   \qquad   \qquad   \qquad   \qquad  \inferenceg{TD1}{}{\isData[t@p]} 
  \qquad
  \inferenceg{TD2}{}{\isData[v@p]}
  \\ &
  \inferenceg{TOD1}{}{\isOptData[e@p]}
  \quad
  \inferenceg{TOD2}{\isData[v]}{\isOptData[\some{v}@p]}
  \quad
  \inferenceg{TOD3}{}{\isOptData[\none@p]}
\\
& \text{State Formation } (\isAState[\sigma]), \\
& \qquad \qquad \qquad\inferenceg{TS1}{}{\isAState[\emptyset]}
  \qquad \qquad
  \inferenceg{TS2}{\isAState[\sigma]\quad
  \sigma(\thread[A], k) = \emptyset \quad \Blue{ X} \in
  dom(\signature)}{\sigma,(\thread[A], k, \Blue{ X}) \colon \stateP}
    \\[\ruleskip] & \qquad \qquad 
  \inferenceg{TS3}{\isAState[\sigma]\quad
  (\thread[A], k, \Blue{ X}) \in \sigma \quad \Blue{ Y} \in
  dom(\signature)}{\exchange{X}{Y}(\sigma(t,k)) \colon \stateP}
\qquad 
  \inferenceg{TS4}{
  \isAState[\sigma]\quad 
  \isAState[\delta] 
  }{\sigma \backslash \delta \colon \stateP}\\
& \text{Formulae Formation }  (\isAFormula[\ltypeEnv]),\\
&   \quad   \quad
  \inferenceg{TF1}{}{\isAFormula[\cdot]}
  \quad
  \inferenceg{TF2}{\isAFormula[\psi] \quad \isAFormula}{\isAFormula[\psi, \ltypeEnv]}
  \quad
  \inferenceg{TF3}{}{\isAFormula[\true]}
  \quad
  \inferenceg{TF4}{}{\isAFormula[\thread \colon \ownerTP{k}{A}{X} ]}
\\ &\qquad \qquad 
    \inferenceg{TF5}{\isAFormula[ \psi] \quad \isAFormula[ \psi']
  \quad \circ \in \{\otimes, \oplus \}   }{\isAFormula[ \psi
  \circ \psi'  ]}
\qquad 
    \inferenceg{TF6}{\isAFormula[ \psi] \quad \isAState[\delta]
  }{\isAFormula[ \psi \backslash \delta
   ]}
  \end{align*}
}
  \caption{\GCQ: Type checking}
  \label{fig:global:types}
  \label{fig:global:types-causality}
\end{figure*}


\begin{example}
  In Example \ref{example-sensors}, $\isEntailed[\true]{ \chor}$ if
  $(\green{q_1} = \forall) \land (\green{q_2} = \{\forall,\exists\})$. In the case $\green{q_1} = \exists, \green{q_2} = \forall$, the same typing
  fails. Similarly, $\isNotEntailed[\true]{ \chor}$ if $\green{q_1} =
  \exists$, for the variant of Example \ref{example-sensors} in Figure
  \ref{example2}.
\end{example}
A type preservation theorem must consider the interplay between the
state and formulae in $\ltypeEnv$. We write $\isState{\sigma}{\ltypeEnv}$ to
say that the tuples in $\sigma$ entail the formulae in
$\ltypeEnv$. For instance,  $\isState{ \sigma}{\thread \colon
  \ownerTP{k}{A}{X}}$ iff $(\thread, k, X) \in \sigma$. 

\begin{definition}[State satisfaction] \label{app:state-satisfaction}
The entailment relation between a state $\sigma$ and  an environment
$\ltypeEnv$, and the entailment relation between a state $\sigma$ and
a formula $\psi$ are written $\isState{\sigma}{\ltypeEnv}$ and
$\isState{\sigma}{\psi}$, respectively. They are defined as follows:
\begin{align*}
&  \isState{\sigma}{\cdot} & \iff & \sigma \text{ is defined}\\
&  \isState{\sigma}{\psi, \ltypeEnv} & \iff & \isState{\sigma}{\psi}
                                              \text{ and } \isState{\sigma}{\ltypeEnv}\\
&  \isState{\sigma}{\true} & \iff & \sigma \text{ is defined}\\
&  \isState{\sigma}{\thread: \ownerTP{k}{A}{X}} & \iff & (\thread,k,X) \in
                                                  \sigma\\
&  \isState{\sigma}{\psi_1 \otimes \psi_2} & \iff & \sigma = \sigma',
                                                    \sigma'' ~ | ~
                                            \isState{\sigma'}{\psi_1}
                                                    ~\land~\isState{\sigma''}{\psi_2}\\
&  \isState{\sigma}{\psi \backslash \delta} & \iff & \exists \sigma'
                                                     \text{ s.t. }
                                                     \isState{\sigma'}{\psi}
                                                     ~\land~ \sigma =
                                                     \sigma'
                                                     \backslash \delta
\end{align*}

\end{definition}

\begin{restatable}[Type
  Preservation]{theorem}{theoremTypePreservation}
  \label{thm:type-preservation}
  If
  $\conf{\sigma}{\chor} \action{\lambda} 
  \conf{\sigma'}{\chor'}$,
  $\isState{\sigma}{\ltypeEnv}$, and $\isEntailed{\chor}$, then
  $\exists \ltypeEnv'.~ \isEntailed[\ltypeEnv']{\chor'}$ and
  $\isState{\sigma'}{\ltypeEnv'}$.
\end{restatable}
\begin{proof}
  It follows by rule induction on the transition relation   $\conf{\sigma}{\chor} \action{\lambda} 
  \conf{\sigma'}{\chor'}$. Details are
  presented in  Appendix \ref{lem-type-preservation-proof}.
\end{proof}

\begin{restatable}[Well-typed choreographies progress]{theorem}{theoremProgressChor}
\label{thm:progress}
  If $\isEntailed \chor$, $\isState{\sigma}{\ltypeEnv}$
  and $\chor \not \equiv \INACT$, then 
  $\chor$ progresses.
\end{restatable}
\begin{proof}
  Proof by contradiction. Let us assume that $\isEntailed \chor$, $\isState{\sigma}{\ltypeEnv}$
  and $\chor \not\equiv \INACT$ and $\conf{\sigma}{\chor}
  \not\action{\lambda}$. We proceed by
  induction on the structure of  $\chor$ to show that such $\chor$
  does not exists.
\end{proof}

The decidability of type checking depends on the provability of
formulae in our ILL fragment. Notice that the formulae used in type checking corresponds to the 
Multiplicative-Additive fragment of ILL, whose provability is
decidable \cite{lincoln1995deciding}. For typing collective
operations, the number of checks grows
according to the amount of  participants involved. Decidability
exploits the 
fact that for each
interaction the number of participants is bounded.

\begin{theorem}[Decidability of Typing]\label{thm:type-decidability}
  $\isEntailed{\chor}$ is decidable.
\end{theorem}


 \section{Session Types}
\label{sec:session-types}

We now present a type system which allows one to specify multiparty
protocols, allowing only specifications that respect causality
relations between interactions.

 \begin{definition}[Global Types: Syntax]\label{globalcalculustypes}\rm
{
  \begin{align*}
    &\text{\emph{(Sorts)}}    &    S ::= &\phantom{{}\mid{}}
                                            \boolT \quad \mid \intT \quad \mid
                  \stringT \quad \mid \ldots                                                                    
    \\
    &\text{\emph{(Global Types)}}     & G ::= & \phantom{{}\mid{}}  \bcastT {A}{\VEC{B}}S \pfx G && (broadcast) \\
    & & & ~ \mid ~ \redT {\VEC{A}}{B}S \pfx G && (reduce) \\
    & &  &~ \mid ~  \branchT {A}{\VEC
         B}lG && (branch) \\
    & &  &~ \mid ~  \endT && (end) 
  \end{align*}
}
\end{definition}

The syntax of global types describes the flow of interactions one can
have in \GCQ. Sorts $\boolT, \intT, \stringT, \ldots$ describe basic
value types. Type $\bcastT {A}{\VEC{B}}S \pfx G$ dictates the presence
of a one-to-many
communication from role $A$ to roles $\VEC{B}$ of sort $S$, followed by a
continuation of type $G$. The type $\redT {\VEC{A}}{B}S \pfx G$
describes a many-to-one communication from roles $\VEC{A}$ to role $B$
with sort $S$. In type $\branchT {A}{\VEC{B}}lG$, role $A$ will spawn
method identified with label $l_i$ collectively on threads
implementing roles $\VEC{B}$, following with a continuation of
type $G_i$ in each of the receivers. The type $\endT$ indicates termination and is often
omitted. 



The
labelled type transition relation $\ltransition{G}{\alpha}{G'}$
expresses the abstract execution of
protocols, where $\alpha = \{\bcastT{A}{\VEC{B}}{S}, \redT{\VEC{A}}BS,
\selT{A}{\VEC{B}}{l}\}$. $G \action{\alpha} G'$ is the smallest relation on 
global types satisfying the rules given in Figure
\ref{fig:global:types:semantics}. Intuitively, the transition 
$\ltransition{G}{\alpha}{G'}$ expresses in $\alpha$ the interaction
consumed. 
Rules $\Did{^G|Bcast}$ and $\Did{^G|Branch}$ track the one-to-many
communications performed in a protocol, and rule $\Did{^G|Red}$
records the many-to-one patterns.
Rule $\Did{^G|Swap}$ captures the swapping notion existing in
choreographies, and it is based on a swap relation for types $G
\swapsT G'$. The set of
rules documenting the behavior of $\swapsT$ is presented in Figure
\ref{fig:swap-rel-types}. 

\begin{figure}[t]
{\small
  \begin{align*}
    &  
      \quad
      \inferenceg{^G|Bcast}
      {\alpha \eqDef \bcastT{A}{\VEC{B}}{S}} 
      {\bcastT {A}{\VEC{B}}S \pfx G
      \action{\alpha} 
      G}
                \qquad                 \qquad                 \qquad                 
      \inferenceg{^G|Red}
      {\alpha \eqDef \redT {\VEC{A}}{B}S} 
      {\redT {\VEC{A}}{B}S \pfx G
      \action{\alpha} 
      G} 
\\[\ruleskip] &
      \inferenceg{^G|Branch}
      {\alpha \eqDef \selT{A}{\VEC{B}}{l_j}} 
      {\branchTP {A}{\VEC B}{l_i}{G_i}{i \in I\cup\{j\}}
      \action{\alpha} 
      G_j} 
                \qquad                 \quad                 
      \inferenceg{^G|Swap}
      {G_1 \swapsT G'_1 \action{\alpha} G'_2 \swapsT G_2 } 
      {G_1
      \action{\alpha} 
      G_2} 
  \end{align*}
}
  \caption{Type transitions for global types,  $\ltransition{G}{\alpha}{G'}$}
  \label{fig:global:types:semantics}
\end{figure}


\begin{figure}[t!]
{\small
  \begin{gather*}
    \inferenceg{^{GS}|^{Bc}_{Bc}}{ 
      (\{A\} \cup \roles{\VEC B}) ~\#~    (\{C\} \cup \roles{\VEC D})
    }{
      \begin{array}{c}
        \bcastT {A}{\VEC{B}}S \pfx \bcastT
        {C}{\VEC{D}}{S'} \swapsT 
      \bcastT {C}{\VEC{D}}{S'} \pfx
        \bcastT {A}{\VEC{B}}S
      \end{array}
    }
\\[\ruleskip]
    \inferenceg{^{GS}|^{Bc}_{Rd}}{ 
      (\{A\} \cup \roles{\VEC B}) ~\#~    (\roles{\VEC C} \cup \{D\})
    }{
      \begin{array}{c}
      \bcastT {A}{\VEC{B}}S \pfx \redT
        {\VEC{C}}{D}{S'} \swapsT 
      \redT {\VEC{C}}{D}{S'} \pfx
        \bcastT {A}{\VEC{B}}S
        \end{array}
    }
\\[\ruleskip]
    \inferenceg{^{GS}|^{Bc}_{Br}}{ 
      (\{A\} \cup \roles{\VEC B}) ~\#~    (\{C\} \cup \roles{\VEC D})
    }{
      \begin{array}{c}
        \bcastT {A}{\VEC{B}}S \pfx
        \branchT {C}{\VEC D}{l}{G} \swapsT 
        \branchT {C}{\VEC D}{l}{G}
        \pfx \bcastT {A}{\VEC{B}}S
      \end{array}
    }
\\[\ruleskip]
    \inferenceg{^{GS}|^{Rd}_{Br}}{ 
      (\roles{\VEC A} \cup \{B\}) ~\#~    (\{C\} \cup \roles{\VEC D})
    }{
      \begin{array}{c}
        \redT {\VEC{A}}{B}S \pfx \branchT
        {C}{\VEC D}{l}{G} \swapsT 
        \branchT {C}{\VEC D}{l}{G}
        \pfx \redT {\VEC{A}}{B}S
      \end{array}
    }
\\[\ruleskip]
    \inferenceg{^{GS}|^{Rd}_{Rd}}{ 
      (\roles{\VEC A} \cup \{B\}) ~\#~    (\roles{\VEC C} \cup \{D\})
    }{
      \begin{array}{c}
        \redT {\VEC{A}}{B}S \pfx \redT
        {\VEC C}{ D}{S'} 
        \swapsT 
        \redT {\VEC C}{ D}{S'}
        \pfx \redT {\VEC{A}}{B}{S}
      \end{array}
    }
\\[\ruleskip]
    \inferenceg{^{GS}|^{Br}_{Br}}{ 
      (\{A\}\cup \roles{\VEC B}) ~\#~    (\{C\} \cup \roles{\VEC D})
    }{
      \begin{array}{c}
        \branchTP {A}{\VEC{B}}{l_i}{\branchTP
        {C}{\VEC D}{l'_j}{G_{ij}}{j \in J}
        }{i \in I} \swapsT 
        \branchTP
        {C}{\VEC{D}}{l'_j}{\branchTP{A}{\VEC{B}}{l_i}{G_{ij}}{i
        \in I}}{j \in J}         
      \end{array}
    }
  \end{gather*}
}
  \caption{Swap relation for global types, $\swapsT$}
  \label{fig:swap-rel-types}
\end{figure}


A type judgment is written as $\isType{\ltypeEnv}{\chor}{\stypeEnv}$. 
 We commonly refer to $\typeEnv$ and $\stypeEnv$ as the \emph{service} and
\emph{session } environments, respectively. The unrestricted
environment $\typeEnv$ contains different types of information. First, it
contains maps from process variables to sorts, as in $x@p \colon S$. Second,
it contains maps from service channels to global types, as in $a
\colon \serviceT GAB$, where $\VEC A$ and $\VEC B$ represent the roles of the
active and service processes, respectively. Furthermore, we assume that
$\VEC A$ and $\VEC B$ are the only roles in $G$.  Third, it contains ownership types, as in $p:
\ownerT kA$, asserting that $p$ behaves as
the role $A$ in session $k$. Note that the ownership types used in this stage are a
variant of the ownership types used  in Section
\ref{sec:type-sys} with no capabilities.   The environment $\ltypeEnv$
denotes a set of linear logic formulae, and it is the same environment
described in Section \ref{sec:type-sys}. Finally, the linear
environment $\stypeEnv$ contains maps from session variables
to Global types, as in $k \colon G$. Its purpose is to track the state of each running
session with respect to the protocol. 
The set of typing rules defining the judgments for 
$\isType{\psi}{\chor}{\stypeEnv}$  is presented in
Figure \ref{fig:global:sessiontypes}.

\begin{figure}[t]
  \begin{align*}
    & \typeEnv ::= &&   \phantom{{}\mid{}}
                                            \emptyset     && 
    & \stypeEnv ::= & \phantom{{}\mid{}} \emptyset  \\
    & &&  ~ \mid ~  \typeEnv, a \colon \serviceT GAB &
    && &~ \mid ~  \stypeEnv, k : G
    \\
    & && ~ \mid ~  \typeEnv, x@p \colon S 
    & \\
    & &&  ~ \mid ~  \typeEnv, p: \ownerT kA 
  \end{align*}

\caption{Global Types: Typing environments}
\label{fig:global:environments}
\end{figure}

\begin{figure*}[t!]
  {\small
    \begin{align*}
      & \qquad \qquad
        \inferenceg{TGinit}
        { 
        \begin{array}{c}
          \isIndex{a}{\serviceT GAB} \quad 
          \isSType[\Gamma,\isInit{ \{\VEC{\thread_r[A_r] },
          \VEC{\thread_s[B_s]}\}, k}]
          {\chor}{\stypeEnv, k \colon G}  
          \quad
          \VEC{\thread_s}\ \#\ \typeEnv 
        \end{array}
      }
      { \isSType{ \initA{\VEC{\thread_r[A_r]\Blue{\{Y_r\}}}}{
      \VEC{\thread_s[B_s]\Blue{\{Y_s\}}}}{a}{k} \pfx \chor } {\stypeEnv}} 
      \\\\ &
             \inferenceg{TGbcast}
             { \begin{array}{c}
                 \isIndex[\typeEnv]{ \thread_A} { \ownerT{k}{A} } 
                 \quad
                 \isIndex[\typeEnv]{ \thread_i} { \ownerT {k}{B_i} } 
                 \quad 
                 \isIndex{e@\thread_A}{S}
                 \quad
                 \isSType[\typeEnv, \VEC{x_r@\thread_r[B_r]} \colon S]
                 {\chor}{\stypeEnv, k
                 \colon G }
                 \quad \irange{i}{1,}{,|\VEC{\thread_r}|}
               \end{array}
      }
      {\isSType{
      \left( \bcastA{\thread_A[A]\Blue{\{X_A;Y_A
      \}}}{\VEC{\thread_r[B_r]\Blue{\{X_r;Y_r\} }:x_r}}{q}{e}{k}
      \right) \pfx \chor  }{\stypeEnv, k \colon \left(
      \bcastT{A}{\VEC B}{S} \pfx G \right) }} 
      \\\\ &
             \inferenceg{TGred}
             { \begin{array}{c}
                 \isIndex[\typeEnv]{ \thread_i} { \ownerT{k}{A_i} } 
                 \quad
                 \isIndex[\typeEnv]{ \thread_B} { \ownerT{k}{B} } 
                 \quad 
                 \isIndex{e_i@\thread_i}{S}
                 \quad
                 \isSType[\typeEnv, x@t_B \colon S]
                 {\chor}{\stypeEnv, k
                 \colon G }     
                 \quad \irange{i}{1,}{,|\VEC{\thread_r}|}
               \end{array}
      } 
      {\isSType{\left(
      \reduceA{\VEC{\thread_r[A_r]\Blue{\{X_r;Y_r\}}.e_r}}{\thread_B[B]\Blue{\{X_B;
      Y_B\}}:x}{q}{k}{op}      \right) \pfx \chor }{\stypeEnv, k
      \colon  \left(       \bcastT{\VEC A}{B}{S} \pfx G\right) }}    
      \\\\ &
             \inferenceg{TGsel}
             { 
             \begin{array}{c}
               \isIndex[\typeEnv]{ \thread_A} { \ownerT{k}{A} } 
               \quad
               \isIndex[\typeEnv]{ \thread_i} { \ownerT{k}{B_i} } 
               \quad 
               \isSType
               {\chor}{\stypeEnv, k
               \colon G_h }
               \quad
               h \in I
               \quad \irange{i}{1,}{,|\VEC{\thread_r}|}
             \end{array}
      } 
      {\isSType{ \left( \choiceA{\thread_A[A]\Blue{\{X_A;Y_A\}}}{\VEC{\thread_r[B_r]\Blue{\{X_r;Y_r\}}}}{q}{k}{l_h} \right)
      \pfx \chor}{\stypeEnv, k \colon \branchT A{\VEC B}lG}} \\\\
      & 
        \qquad
        \qquad
        \inferenceg{TGcond}
        { \isSType{\chor_1}{\stypeEnv} \quad
        \isSType{\chor_2}{\stypeEnv} \quad
        \isProp{e@\thread} }  
        {\isSType{ \ifthenelsek {e@\thread} {\chor_1} {\chor_2}
        }{\stypeEnv}} 
                                                   \qquad 
                                                   \inferenceg{TGinact}
                                                   { \isEnd} 
                                                   {\isSType{ \INACT
                                                   }{\stypeEnv}} 
      \\\\ & 
                                                   \qquad \qquad                                                    \qquad \qquad 
             \inferenceg{TGres}
             { \isSType[\Gamma\backslash x]{ \chor}{\stypeEnv \backslash x}}
             {\isSType[\Gamma]{ \new{x} \chor}{\stypeEnv}}
             \qquad
             \inferenceg{TG}{
             \isEntailed{\chor} 
             \qquad
             \isSType{\chor}{\stypeEnv}
             }{
             \isType{\ltypeEnv}{\chor}{\stypeEnv}
             }\\
& \text{Optional Data Typing,  }  
\\[\ruleskip] & 
  \qquad   \qquad   \qquad   \qquad   \qquad  \inferenceg{TGopt_1}{\isIndex{v}{S}}{\isIndex{\some{v}}{S}} 
  \qquad
  \inferenceg{TGopt_2}{}{\isIndex{\none}{S}}
    \end{align*}
}
  \caption{\GCQ: Type checking - Global types}
  \label{fig:global:sessiontypes}
\end{figure*}


Intuitively, a choreography $\chor$ is well typed with respect to
$\typeEnv, \ltypeEnv$ and $\stypeEnv$  if its shared channels are
used, its processes behave according to the global types in
$\typeEnv$, and the capabilities for each collective communication in
$\ltypeEnv$ are respected.
We proceed to describe the typing rules in Figure
\ref{fig:global:sessiontypes}. 

An important observation to make is
that the information tracked by $\ltypeEnv$ is independent from
environments $\typeEnv$ and $\stypeEnv$. This allows us to divide the
analysis into two independent analyses, one for capabilities and one for global types
\cite{Nielson1999Type-and-Effect}. Rule $\Did{TG}$ represents this
fact, dividing the analysis of $
\isType{\ltypeEnv}{\chor}{\stypeEnv}$ into separate analysis for
causalities $           \isEntailed{\chor} $ and for global types $
\isSType{\chor}{\stypeEnv}$. All other rules in Figure
\ref{fig:global:sessiontypes} pertain only global typing (no
$\ltypeEnv$ is required).

In rule $\Did{TGinit}$ the sequence of interactions started with an
$(init)$ action is typed: each process in the term is given
ownership to the role declared for it in the created session $k$
through the function $\isInit{ \widetilde{\thread_p[A] }, k}$, which returns a set of
   ownership assignments $\{\thread_q : \ownerT kB \mid \thread_q[B]
   \in \widetilde{\thread_p[A]}\}$. The condition $\VEC{\thread_s}
   ~\#~ \typeEnv$ ensures that service threads are fresh. 

The typing rules for communications are grouped into $\Did{TGbcast}$, 
$\Did{TGred}$ and $\Did{TGsel}$. In each of them we must check that
the communication performed by the threads involved at both the sender
and receivers own their respective roles in the communication over the 
session in use. This is ensured by checking that the type environment
contains the ownership typings $\thread_p: \ownerT kA$ and $\thread_i:
\ownerT k{B_i}$ ($\irange i2j$) for a broadcast operation, and similarly
for reduce and selection. In $\Did{TGbcast}$, we additionally check
that the type expression sent by the sender corresponds to the carried
sort $<.S.>$, and that the continuation is typed according to the
initial type environment $\typeEnv$ extended with the assignment of
type $S$ to the variables used by the receiver threads in $\thread_2,
\ldots, \thread_j$. 
Rule $\Did{TGred}$ behaves complementary, ensuring that
each of the expressions used by the sender threads behave according
to the same sort $<.S.>$, and ensuring that the continuation is typed according to the
initial type environment $\typeEnv$ extended with the assignment of
type $S$ to the variable used by the receiver in the communication. 
Rule $\Did{TGsel}$ types labelled selections: A selection of a label
$l_h$ on session $k$ is well-typed if the label is in those allowed by
the protocol of session $k$ $(h \in I)$. The continuation must then
implement the selected continuation $G_h$ on session $k$.

Rule $\Did{TGcond}$ is the standard typing rule used to type
conditional blocks: we must check that the expression $e@\thread$  is
a proposition, and that each of the branches of the conditional are
typed according to the session environment $\stypeEnv$.
%
%
Rule $\Did{TGres}$ types name restriction in the standard way. Observe
that if $r$ is a process identifier, then restriction only affect
$\typeEnv$ since $\stypeEnv$ does not refer to processes.

 Rule $\Did{TGinact}$ types
   termination: $\INACT$ is well-typed under any unrestricted
   environment $\typeEnv$ and session environment $\stypeEnv$ if each
   session $k$ typed in $\stypeEnv$ has a type $\endT$, meaning that it
   has been successfully terminated. Predicate $\isEnd$ is
   formalized as $\{\true \mid \forall k : G \in \stypeEnv, G
   = \endT \}$ and $\false$ otherwise.


\begin{example}[The typing in practice]
Consider the variant of Example \ref{example-sensors}:
\begin{align*}
  &\initA{\thread_1 \Blue{\{\qpredpp[X_1]\}}[S_1], \thread_2 \Blue{\{\qpredpp[X_2]\}}[S_2], \thread_3 \Blue{\{\qpredpp[X_3]\}}[S_3]}{\thread_m \Blue{\{\qpredpp[X_m]\}}[M]}{temperature}{k}
    \pfx \\ 
  &\bcastA{\thread_m \Blue{\{\qpredpp[X_m]; \qpredpp[Y_m]\}}}{\thread_1 \Blue{\{\qpredpp[X_1]; \qpredpp[Y_1]\}}:x_1, \thread_2 \Blue{\{\qpredpp[X_2]; \qpredpp[Y_2]\}}:x_2, \thread_3 \Blue{\{\qpredpp[X_3]; \qpredpp[Y_3]\}}:x_3}{\forall}{{\bf today}}{k} \pfx \\
  &\reduceA{\thread_1 \Blue{\{\qpredpp[Y_1]; \qpredpp[Z_1]\}}.temp, \thread_2 \Blue{\{\qpredpp[Y_2]; \qpredpp[Z_2]\}}.temp,
    \thread_3 \Blue{\{\qpredpp[Y_3]; \qpredpp[Z_3]\}}.temp}{\thread_m \Blue{\{\qpredpp[Y_m]; \qpredpp[Z_m]\}}:x_m}{\forall}{k}{max} \pfx \INACT
    \numberthis \label{ex:typing}
\end{align*}

We can show that the choreography in \ref{ex:typing} is typable under
environments $\ltypeEnv = \emptyset$ and
$\typeEnv = temperature\colon \bcastT{M}{ <. S_1,S_2,S_3 .>
}{\dateT} \pfx \redT{ <. S_1,S_2,S_3 .>}{M}{\floatT} \pfx \endT$.


\end{example}

\begin{figure}[t]
  {
    \begin{align*}
      &\quad 
        \inferenceg{^L|Bcast}
        { 
        \begin{array}{c}
          \isIndex{\thread_A}{\ownerT kA}
          \qquad 
          \isIndex{\thread_{i}}{\ownerT k{B_i}}
          \qquad
          \isIndex{v@\thread_A}{S} 
          \qquad
          \irange{ i}{1}{|\VEC{B}|}
        \end{array}
      }
      { \isLType{
      \bcastA{
      \thread_A[A]
      }{\VEC{\thread_r[B_r]:sb_r}}{q}{v}{k}
      }
      {k \colon \bcastT {A}{\VEC{B}}S }} \\\\ &
                        \inferenceg{^L|Red}
                        { 
                        \begin{array}{c}
                          \isIndex{\thread_B}{\ownerT kB}
                          \qquad 
                          \isIndex{\thread_{i}}{\ownerT k{A_i}}
                          \qquad
                          \isIndex{v_i@\thread_i}{S} 
                          \qquad \irange{ i}{1}{ |\VEC{A}|}
                        \end{array}
                        }
                        { \isSType{\reduceA{\VEC{\thread_r[A_r].v_r}}{\thread_B[B]:\some{v}}{q}{k}{op}}
      {k \colon \redT {\VEC{A}}{B}S}} \\\\ &\qquad \qquad \qquad
                                          \inferenceg{^L|Sel}
                                          { 
                                          \begin{array}{c}
                                            \isIndex{\thread_A}{\ownerT kA}
                                            \qquad 
                                            \isIndex{\thread_{i}}{\ownerT
                                            k{B_i}}
                                            \qquad
                                            \irange{ i}{1}{|\VEC{B}|}
                                          \end{array}
                                          }
                                          { \isSType{
                                                             \choiceA{\thread_A[A]}{\VEC{\thread_r[B_r]}}{q}{k}{l}
                                                             } {k
                                                             \colon \selT{A}{\VEC{B}}{l}}} 
    \end{align*}
  \caption{\GCQ: Label Typing, $\isLType{\lambda}{k \colon \alpha}$}
  \label{fig:global:labeltyping}
  }
\end{figure}


We can proceed to establish the technical results of the session type discipline.
In the following, we write $\stypeEnv \action{k \colon \alpha}
\stypeEnv'$ to say that  $k \colon G \in \stypeEnv$, $G \action{\alpha}
G'$, and that $\stypeEnv'$ corresponds to the substitution   $\stypeEnv\subst{k \colon G'}{k \colon
  G}$. Figure \ref{fig:global:labeltyping} formalizes the
correspondence between labels in the choreography and their respective
global types ($\isLType{\lambda}{k \colon \alpha}$). We can now establish our type preservation theorem for
global types.

\begin{restatable}[Type
  Preservation for Global Types]{theorem}{theoremTypePreservationGlobal}
  \label{thm:global-type-preservation}

If 
$           \isType{\ltypeEnv}{\chor}{\stypeEnv}$, 
$\isState{\sigma}{\ltypeEnv}$,  and
$\conf{\sigma}{\chor}
     \action{\lambda} \conf{\sigma'}{\chor'}$, then there
     exists $\typeEnv', \stypeEnv'$
     s.t. 
     $\isType[\typeEnv']{\ltypeEnv'}{\chor'}{\stypeEnv'}$, $\isState{\sigma'}{\ltypeEnv'}$,
     and 
     \begin{itemize}
     \item if $\lambda = \initA{ \VEC{\thread_r[A_r]}
        }{ \VEC{\thread_s[B_s]}}{a}{k}$, then
       $\stypeEnv' = \stypeEnv$,
     \item otherwise $\stypeEnv \action{k \colon \alpha} \stypeEnv'$
       and $\isLType{\lambda}{k \colon \alpha}$.
     \end{itemize}
\end{restatable}
\begin{proof}
  The proof follows by rule induction on the transition rules of $\conf{\sigma}{\chor}
     \action{\lambda} \conf{\sigma'}{\chor'}$. Its details are
     presented in Appendix \ref{lem-type-progress-global-proof}.
\end{proof}

Type preservation for Global Types ensures that the transitions of a
well-typed choreography are still well typed, and, more importantly,
that choreography transitions performed corresponds to the protocol intented by
the global type, as evidenced by the type transitions in $\stypeEnv$.

Checking the decidability of type checking for session types depends
on the typing rules in Figure
\ref{fig:global:sessiontypes}.  The only rule of interest here is
$\Did{TG}$, that involves the type checking of the capability type
system.  The decidability of   $\isEntailed{\chor}$ is given in Theorem
\ref{thm:type-decidability}. All the other rules are syntax-directed. 

\begin{theorem}[Decidability]\label{thm:type-decidability-session-types}
For any $\typeEnv, \ltypeEnv, \chor $, the checking of $\isType{\ltypeEnv}{\chor}{\stypeEnv}$
is decidable.
\end{theorem}

\subsection{Linearity}

A conflict (race condition) may be generated when implementing
multiparty choreographies. While at the choreographic level one
imposes a sequence of interactions among participants, the
projection of a choreography into endpoints generate a set of participants acting
concurrently. A poorly defined choreography may lead to
implementations that do not follow the sequence imposed by the
choreography. Take the following choreography:

\begin{equation} \label{example:linearity}
  \initI{p[A]}{q[B]}{a(k)} \pfx \initI{r[D]}{s[E]}{a(k')} \pfx \chor'
\end{equation}

Here four processes start two different sessions using the same
service and same session key. Once projected, threads implementing session initiation constructs $  \initI{p[A]}{q[B]}{a(k)}$
and $\initI{r[D]}{s[E]}{a(k')} $ will compete. The race occurs when
the thread implementing $p[A]$ establishes a session with $s[E]$. Such
behavior will correspond to term
$\initI{p[A]}{s[E]}{a(k)}$, which do not 
exists in the  original choreography. Similar considerations apply for
the race between $r[D]$ and $q[B]$. We appeal to the use of linearity
conditions
\cite{honda2008multiparty,Carbone:2013:DMA:2429069.2429101}, that we
adapt to multiparty-collective interactions.

An interaction node, denoted by $n$, is an abstraction of a node in
the abstract syntax tree. Node $n$ can be 
\begin{enumerate}[i]
\item  $\initI{\VEC{p}}{\VEC{q}}{a}$ for
$(init)$, with $\fn(n) = \VEC{p}$, 
\item 
$\interactI{p}{\VEC{q}}$ for $(bcast)$ and  $(select)$, with
$\fn(n)=\{\VEC{p}, q\}$, or
\item $\interactI{\VEC{p}}{q}$ for $(reduce)$, with
$\fn(n)=\{p, \VEC{q}\}$.
\end{enumerate}
We say that $n_2$ \emph{depends on} $n_1$ in $\chor$, written $n_1 \prec n_2 \in \chor$, whenever $n_1$
precedes $n_2$ in $\chor$ (i.e.: $n_1$ and $n_2$ cannot appear in
different branches in  conditional and sum statements). An \emph{interaction dependency} $n_1
\prec_p n_2 \in \chor$
occurs whenever $n_1 \prec n_2 \in \chor$ and one of the following
conditions hold: 
\begin{itemize}
\item $n_1 = \initI{\VEC{p}}{\VEC{q}}{a} $ and
  $n_2 = \interactI{p}{\VEC{q}'} $ and $p \in \{\VEC{p},\VEC{q}\}$, or
\item $n_1 = \initI{\VEC{p}}{\VEC{q}}{a} $ and
  $n_2 = \interactI{\VEC{p}'}{q} $ and $p \in \VEC{p'}$ and 
  $p \in \{\VEC{p},\VEC{q}\}$,
  or
\item $n_1 = \initI{\VEC{p}}{\VEC{q}}{a} $ and
  $n_2 = \initI{\VEC{r}}{\VEC{s}}{b}$, where
  $p \in \{ \VEC{p}, \VEC{q}\}$, and $p \in \VEC{r}$, or
\item $n_1 = \interactI{\VEC{q}}{p}$ and $p \in \fn(n_2)$, 
  or
\item $n_1 = \interactI{q}{\VEC{r}}$ and $p \in \VEC{r}$ and $p \in \fn(n_2)$.
\end{itemize}
The interaction dependency  $n_1 \prec_p n_2 \in \chor$ says that the projection of
a process $p$ for the interaction node abstracted by $n_2$ cannot
occur before that for $n_1$.  Interaction dependencies are the basis
for establishing a linearity property.

\begin{definition}[Linearity \cite{M13:phd}]\label{def:linearity}
Let $\chor$ be a choreography. We say that $\chor$ is \emph{linear} if
for all nodes $n_1 = \initI{\VEC{p}}{\VEC{q}}{a}$ and $n_2 =
\initI{\VEC{r}}{\VEC{s}}{a}$ such that $n_1 \prec n_2 \in \chor$ we
have that $\forall r \in \VEC{r}. \exists p \in
\{\VEC{p},\VEC{q}\}. n_1 \prec_p\ldots\prec_r n_2$. 
\end{definition}

Intuitively, linearity checks that for dependent nodes $n_1, n_2$ such
that $n_1 \prec n_2$, if they both take the form of start nodes over
a common service name $a$, then all active processes used in  $n_2$ depend
on some process in $n_1$. In this way, the races between active
processes explained before are avoided.

In the following, we recall that  bound variables are renamed apart in
$\chor$;  That means that for two dependent $(start)$ nodes using the
same service name, the session keys used will be different. The use of
renaming for session keys proves useful by limiting additional races
where service processes compete with active processes in the
establishment of a new session.



\section{The Endpoint Quality Calculus (\EPQ), and Endpoint Projection }
\label{sec:epc}

The \EPQ calculus extends the Quality calculus
\cite{nielson2013calculus} with session-based communication and
input-output queues. In
addition to the syntactic categories defined in Section
\ref{sec:language}, $P, Q, \ldots$ denote processes, $k, \ldots, r,s$ denote names,  $p$ denotes an
  annotated thread $\thread[A]$, where   $\thread$ is a
thread.
We will use $\VEC{\thread}$ to denote  $\{\thread_1,
\ldots, \thread_j\}$ for a finite $j$.

\begin{definition}[\EPQ Syntax]
  \label{def:endpointsyntaxlable} { \begin{align*}
      &
                                                   & P,Q::=\ &
                                                               \phantom{{}\mid{}}
                                                               \initIn
                                                               {a[\VEC{A}]}{k}\pfx
                                                               P 
                                                   &\mid~&
                                                           \initOut{a[A]}{k}\pfx
                                                           P 
                                                   \\
      & & \mid~& \repInitIn {a[A]}{k}\pfx P 
                                                   &\mid~& P\pp
                                                           Q 
                                                   \\
      & &\mid~ & \bcast{k}{A}{\VEC{B}}{q}{e} P 
                                                   &\mid~&
                                                           \receive{k}{A}{B}{x}P 
                                                   \\
      & &\mid~ & \reduce{k}{\VEC{A}}{B}{q}{x}{op} \pfx P 
                                                   &\mid~ &
                                                            \send{k}{A}{B}{e}P  
                                                   \\
      & & \mid~& \colSelection kA{\VEC{B}}{q}l P 
                                                   &\mid~&
                                                        \colBranching kABlPI 
                                                   \\
      & & \mid~& \highlightBox{\waitInput k{\VEC{A}}B{op}x P} 
                                                   &\mid~& 
                                                           \highlightBox{\waitOutput
                                                           {k}{A}{\VEC{B}}{P} } 
                                                   \\
  &    
  &\mid~& \itn{e}{P}{Q} 
                                                   & \mid~  &  \INACT  \\
&  &\mid~&  \highlightBox{\new k P } 
                                                   & \mid~  &  
               \highlightBox{\queue kh } 
                                    \end{align*}
                                    \begin{align*}
&
  &    h::=\ & \phantom{{}\mid{}}
   \emptyset   &\mid~&m \cdot h 
  \\
&
  &    m::=\ & 
   (A, q: <. (\VEC{B: b}) : w .>) 
    &\mid~& (q: <. (\VEC{A : b : sb}) .>, B ) 
  \\
& &    sb::=\ & 
             \some{v}   &\mid~& \none \\ 
&
  &    w::=\ & 
               sb   &\mid~& l \\ 
& &    b::=\ & 
             \true   &\mid~& \false  
\end{align*}
}
\end{definition}

The first three terms
correspond to a session establishment phase. A process $ \initIn
{a[\VEC{A}]}{k}\pfx P$ acts as a requester for service $a$, with
roles $\VEC{A}$. It will interact with endpoint providers implementing
each of the behaviours in $\VEC{A}$, being those replicated services (i.e.:
$\repInitIn{a[A]}{k} \pfx P$), or one-time instances (i.e.:
$\initOut{a[A]}{k}\pfx P$).
In these cases, $P$ denotes the continuation process. The pair $\bcast{k}{A}{\VEC{B}}{q}{e} P$ and $\receive{k}{A}{B}{x}P$ models
one-to-many communications.  While the sender part of a broadcast is parameterised with a
quality predicate, the receiver does not require it, as receivers only
communicate with one sender. The pair $\reduce{k}{\VEC{A}}{B}{q}{x}{op} \pfx P$ and $\send{k}{A}{B}{e}P $ implements
many-to-one communication patterns. Dual to broadcast, here the
receiver process requires the quality predicate, while the sender
process does not. The pair $\colSelection kA{\VEC{B}}{q}l P$ and
$\colBranching kABlPI$ implements a one-to-many method selection (where $\{l_i\}$
should be pairwise distinct). 
Runtime processes $\waitInput k{\VEC{A}}B{op}x P$ and $\waitOutput
{k}{A}{\VEC{B}}{P}$ implement queue-synchronization processes, and
interact directly with input/output session queues $\queue kh$. Each queue contains messages with one sender and many recipients $   (A, q: <. \VEC{B: w}
.>) $ or many recipients and one sender $  (q: <. \VEC{A : sb} .>, B )$. 
Other process constructs, such as parallel composition, 
if-then
constructs, and restriction are standard. Boxed terms can only be used at runtime. The free session channels, free term variables and service
channels are defined as usual over processes  are denoted by $fsc(P), fv(P)$ and
$channels(P)$ respectively. 

\subsection{Semantics}

\EPQ is equipped with a
structural congruence relation over processes.
\begin{definition}[Structural Congruence in \EPQ] 
The structural congruence relation $\equiv$ in \EPQ is the least
congruence on processes supporting $\alpha$-renaming, such that 
$(P, \INACT, \pp)$ is
an abelian monoid, and the following rules are satisfied:
\begin{enumerate}[(i)] 
\item $\new r \INACT \equiv \INACT$,
\item $\new r\new s P \equiv \new s\new r P$,
\item $\new r(P \pp Q) \equiv \new r P \pp Q$ if $r \notin \fn(Q)$,
\item $\queue {k}{h \cdot (A, q: <. \VEC{B : w} .>) 
  \cdot (C, q': <. \VEC{D : w'} .>) \cdot h'} \equiv \queue{k}{h \cdot (C, q': <. \VEC{D : w'} .>)
  \cdot (A, q: <. \VEC{B : w} .>) \cdot h'}$ if $C \neq A$ or $\VEC{B}
\# \VEC{D}$,
\item $\queue {k}{h \cdot (q: <. \VEC{A
    : sb} .>, B)  \cdot (q': <. \VEC{C : sb'} .>, D) \cdot h'} \equiv
\queue{k}{h \cdot (q': <. \VEC{C : sb'}, D .>)
  \cdot (q: <. \VEC{A : sb}, B .>) \cdot h'}$ if $\VEC{A}
\# \VEC{C}$ or  $B \neq D$.
\end{enumerate}

\end{definition}

We give an operational semantics in terms of  labeled reductions $
\ltransition{P}{\emm}{P'}$, where 
{ \[
\emm ::=
\begin{array}{llll}
          \tau  &
          \mid~ \startAction{\VEC{A}}{\VEC{B}}{a}{k} &
          \mid~\aBroadcastO{A}{\VEC{B}}qkv \\
          \mid~\aBroadcastI{A}{\VEC{B}}{k}{v}&
          \mid~\aReduceO{\VEC{A}}{B}{k}{v} &
          \mid~\aReduceI{\VEC{A}}Bqkv\\
          \mid~\aSelectionO{A}{\VEC{B}}qkl&
          \mid~\aSelectionI{A}{\VEC{B}}kl &
          \mid~\downarrow_\tau          \quad\mid \quad\uparrow_\tau
        \end{array}
\]
 }

\begin{figure*}[t!]
{\small
  \begin{align*} & 
                   \inferenceg{^E|{Init}} 
                   { 
                   \VEC{C} = \VEC{A},\VEC{B} \quad 
                   \VEC{A} = A_1, \ldots, A_n \quad  
                   \VEC{B} = B_1, \ldots, B_m   \quad
                   R = \prod_{i \in [1, m]}
                   \repInitIn{a[B_i]}{k} \pfx Q_i
                   }
                   { 
                       \initIn{a[\VEC{C}]}{k} \pfx P 
                       \pp 
                       \prod_{i \in [2,n]} \initOut{a[A_i]}{k} \pfx P_i
                       \pp
                       R \action{\startAction {\VEC{A}}{\VEC{B}}ak } 
                       \new{k} (
                       P \pp 
                       \prod_{i \in [2, n]} P_i \pp
                       \prod_{i \in [1, m]} Q_i \pp
                       \queue k\emptyset 
                       )  \pp R
    }
    \\[\ruleskip]    & \qquad
                       \inferenceg{^E|{Bc.O}} 
                       { \evalOp ev 
                       }{
                       \begin{array}{l}
                         \bcast{k}{A}{\VEC{B}}{q}{e} P \pp 
                         \queue kh                        
                         \action{ 
                         \asynchEnqueueO }
                         \waitOutput{k}{A}{\VEC{B}} P \pp 
                         \queue k{h\cdot \qmsgI  {A}{\VEC{B : \false}}{q}{\some{v}}}
                       \end{array}
    }
    \\ & 
    \inferenceg{^E|{Bc.I}} 
    { 
    b_i \neq \true
    }{
    \begin{array}{l}
      \receive{k}{B_i}{A}{x} P \pp 
      \queue k{ \qmsgI  {A}{\ldots,
      (B_i : b_i), \ldots }{q}{sb} \cdot h} \\ \qquad \qquad \qquad \qquad \qquad \qquad \qquad \quad 
      \action{ \aBroadcastI{A}{B_i}{k}{sb} }
      P\subst{sb}{ x} \pp 
      \queue k{\qmsgI  A{\ldots,
      (B_i : \true), \ldots }q{sb} \cdot h} 
    \end{array}
    }
    \\[\ruleskip]    &                   
    \inferenceg{^E|{Rd.O}} 
    { b_i \neq \true \quad 
      \evalOp {e_i}{v_i} 
    }{
    \begin{array}{l}
      \send{k}{A_i}{B}{e_i} P \pp 
      \queue k{ (q : <. \ldots, (A_i:b_i: sb_i), \ldots .>, B) \cdot   h}\\ \qquad \qquad \qquad \qquad \qquad \qquad 
      \action{ \aReduceO{A_i}{B}{k}{\some{v_i}} }
      P \pp       \queue k{ (q : <. \ldots, (A_i:\true:\some{v_i}), \ldots .>,
      B) \cdot   h} 
    \end{array}
    }
    \\[\ruleskip] & \qquad
                       \inferenceg{^E|{Rd.I}} 
                       { 
    }{
    \begin{array}{l} 
      \reduce{k}{\VEC{A}}{B}{q}{x}{op} \pfx P \pp 
      \queue k{ h } 
      \action{ \asynchEnqueueI}
      { \waitInput{k}{\VEC{A}}{B}{op}{x}P \pp \queue k{h\cdot  \qmsgOInitial { A}{B}{q}{\none}{\false}    }}
    \end{array} 
    }
    \\[\ruleskip]    &         \qquad          \qquad
                   \inferenceg{^E|{Sel}} 
                   { 
                       }{
                       \begin{array}{l}
                       \colSelection{k}{A}{\VEC{B}}{q}{l} P \pp 
                       \queue kh 
                       \action{ 
                         \asynchEnqueueO}
                        \waitOutput{k}{A}{\VEC{B}} P \pp 
                       \queue k{h\cdot \qmsgI  A{(\VEC{B : \false})}ql}
                       \end{array} 
                       }
\\ & 
                   \inferenceg{^E|{Br}} 
                       { 
                       j \in I
                       \quad
                       b_i \neq \true
                       }{
                       \begin{array}{l}
                         \colBranching{k}{B}{A}{l}{P}{I}  \pp 
                         \queue k{ \qmsgI  {A}{ \ldots,
                         (B_i : b_i), \ldots  }{q}{l_j} \cdot h} \\ \qquad \qquad \qquad \qquad \qquad \qquad \qquad \qquad \quad
                         \action{ \aSelectionI{A}{B}{k}{l_j} }
                         P_j \pp 
                         \queue k{h\cdot \qmsgI  {A}{  \ldots,
                         (B_i : \true), \ldots  }{q}{l_j} \cdot h}
                       \end{array} 
                       }
    \\[\ruleskip]    &   
                   \inferenceg{^E|{Wait_B}} 
                       { 
                       \VEC{B} =\VEC{B'}, \VEC{B''}
                       \qquad 
                       q(\VEC{b'})
    }{                       \begin{array}{l}
                       {
                       \waitOutput{k}{A}{\VEC{B}} P  \pp 
                       \queue k{\qmsgI{A}{ \VEC{B' : b'}, \VEC{B'' : b''} }{q}{ sb  } \cdot h} \pp
                       \prod_{B_i \in \VEC{B''}}       \receive{k}{B_i}{A}{x_i} Q_i
                       }\\ \qquad \qquad \qquad \qquad \qquad \qquad \qquad \qquad \qquad 
                       \action{ \aBroadcastO{A}{\VEC{B}}{q}{k}{sb} }
                       { P \pp \queue k{h} \pp \prod_{B_i \in \VEC{B''}} Q_i\subst{\none}{x_i}} 
                             \end{array}
    }                       
    \\[\ruleskip]    &   
                       \inferenceg{^E|{Wait_S}} 
                       { 
                       \VEC{B} =\VEC{B'}, \VEC{B''}
                       \qquad 
                       q(\VEC{b})                        
                       }{ \begin{array}{l}
                            {
                            \waitOutput{k}{A}{\VEC{B}} P  \pp 
                            \queue k{\qmsgI{A}{ \VEC{B' : b'}, \VEC{B'' : b''} }{q}{ l  } \cdot h} 
                            \pp \prod_{B_i \in \VEC{B''}}  \colBranching{k}{B_i}{A}{l}{Q}{I}  
                            }
                            \\ \qquad \qquad \qquad \qquad \qquad \qquad \qquad \qquad \qquad 
                            \action{ \aSelectionO{A}{\VEC{B}}{q}{k}{l} }
                            { P \pp \queue k{h} }
                          \end{array} 
    }                      
\\ &
     \inferenceg{^E|{Wait_i}} 
     { 
     \VEC{A} =\VEC{A'}, \VEC{A''}
     \qquad 
     q(\VEC{b'}) 
     \qquad \evalOp {op(\VEC{sb'})}{\some{v}}
     }{
     \begin{array}{l}
       \waitInput{k}{\VEC{A}}{B}{op}{x}P  \pp 
       \queue k{(q: <. \VEC{A':b':sb'},\VEC{A'':b'':sb''} .> , B) \cdot h} \pp 
       \prod_{A_i \in \VEC{A''}}  \send{k}{A_i}{B}{e_i} Q_i
       \\ \qquad \qquad \qquad \qquad \qquad \qquad \qquad \qquad \qquad 
       \action{ \aReduceI{\VEC{A}}{B}{q}{k}{\some{v}}  }
       P\subst{\some{v}}{x} \pp 
       \queue k{h} \pp
       \prod_{A_i \in \VEC{A''}}  Q_i
     \end{array}
                       }
    \\[\ruleskip]    &                   
                       \qquad
                       \inferenceg{^E|{Res}}
                       {  P \action{\emm}  P' } 
                       {  \new{r} P  \action{ \emm} \new{r} P'}
                       \qquad
                       \inferenceg{^E|{If} }{
                       i =1 \text{ if } \evalOp{e}{\true} ~~i=2 \text{ o.w.} 
                       } 
                       {  \itn{e}{P_1}{P_2}
                       \action{\tau}
                       P_i} 
                       \qquad
                       \inferenceg{^E|{Par}}
                       {  P \action{\emm}  P' } 
                       {  P\pp Q  \action{\emm} P'\pp Q}
    \\ &  \qquad \qquad \qquad \qquad \qquad \qquad \qquad \qquad 
                       \inferenceg{^E|{Str}}
                       {  P \equiv P' \action{\emm}  Q' \equiv Q } 
                       {  P  \action{\emm} Q}
\end{align*}
}
  \caption{\EPQ: Operational Semantics }
  \label{figure:endpoint:sos}
\end{figure*}


The operational semantics for \EPQ is defined by the rules given in
Figure~\ref{figure:endpoint:sos}.
We give an intution of the most important rules. 
Rule $\Did{^E|_{Init}}$ describes 
session initiation: A requester  process $\initIn{a[\VEC{A},\VEC{B}]}{k} \pfx
P$ can establish a new session $k$, if it is in interaction with
active threads $\prod_{i \in \{2, \ldots, |\VEC{A}|\}} \initOut{a[A_i]}{k} \pfx P_i$,
and replicated services  $\prod_{i \in \{1, \ldots, |\VEC{B}|\}}
\repInitIn{a[B_i]}{k}.P$.

 Asynchronous queue-based communication is
implemented by the interplay of rules $\Did{^E|{Bc.O}}$,
$\Did{^E|{Bc.I}}$ and $\Did{^E|{Wait_o}}$.
Starting with a parallel composition of a sender process and a queue,
rule $\Did{^E|{Bc.O}}$  adds to the session queue the contents
resulting of evaluating expression $2$ at the sender side. In the
meantime, the sender process will move into a waiting state, denoted
by $\waitOutput{k}{A}{\VEC{B}} P$.
Rule $\Did{^E|{Bc.I}}$ captures the interplay between receivers and
the queue. Its transition updates the message on
top of  the session  queue, generating a substitution of the
communicated value on the receiver process. In order to avoid
performing the substitution multiple times over the same participant,
the queue will be modified to include information regarding the
identity of the receiver.
Finally, dequeueing occurs once the evaluation of the quality
predicate over the set of performed substitutions deems satisfiable
(Rule $\Did{^E|{Wait_B}}$). At this point, we will have the following concurrent processes:
\begin{itemize}
\item A sender in its waiting state, $\waitOutput{k}{A}{\VEC{B}} P$.
\item A queue tracking roles who have performed substitutions ($B’$),
and those who have not ($B’'$).
\item A parallel composition of all receiver processes who have not
  yet synchronised, $\prod_{B_i \in \VEC{B''}}       \receive{k}{B_i}{A}{x_i} Q_i$
\end{itemize} 
The consequence of this transition is the dequeueing of the top message from the queue, the activation of the sender process, and a none substitution on all receiver processes that did not synchronised with the queue. 
Similar considerations are given for the triad of label
selection rules 
$\Did{^E|{Sel}}$, $\Did{^E|{Br}}$, and $\Did{^E|{Wait_S}}$.  

Rules for reduce
act similarly as the ones for broadcast: A reduce process enqueues a message with
placeholders for each of the senders involved, as well as the quality
predicate, and blocks until enough senders have sent information
(c.f.: rule $\Did{^E|{Rd.I}}$). The presence of a sender will update
the queue, enclosing the new value $v_i$ as an optional datatype
(c.f.: rule $\Did{^E|{Rd.O}}$). The release of the reduce happens in
Rule $\Did{^E|{Wait_i}}$, once enough senders have contributed with
values, the substitution of $x$ with the result of the operation
$op(sb_1, \ldots, sb_n)$ is performed on the continuation of the
reduce, and the queue is updated.
The remaining rules are standard in
the session $\pi$ calculus.

\subsection{Endpoint Projection}

The projection function maps the behaviours described by a
choreography into endpoints. Special care must be payed when
constructing the endpoint projection. In particular, an endpoint may
implement different behaviours depending on the choices made in a
choreography. For instance, consider the following choreography:

\begin{equation}
  \ifthenelsek {e@p_A} {\left(\choiceA{p_A}{p_B}{}{k}{l_1} \right)} {\left(\choiceA{p_A}{p_B}{ }{k}{l_2} \right) }
\end{equation}

When  projecting the behavior of  thread $p_B$, it is not clear a priori whether the endpoint should implement the behaviour dictated by label $l_1$ or by $l_2$. We make  use of a merge operator \cite{Carbone:2012:SCP:2220365.2220367,DBLP:journals/corr/abs-1208-6483} to 
collect all such labels into a label branching operator. 

\begin{definition}[Merging]
$P \mergek Q$ is a partial commutative binary operator on processes that is well-defined iff $P \mergeable Q$ and it satisfies the following rules: 

\begin{align*}
  &\left(\colBranching{k}{A}{B}{l}{P}{I} \right) \mergek 
    \left(\colBranching{k}{A}{B}{l}{Q}{J} \right)  &&= k[A]?[B]\blacktriangleleft( \{l_i :
                                                      P_i\}_{i \in I \backslash J} \\ 
  & &&\qquad \qquad \quad \cup \{l_i:Q_i\}_{i\in J\backslash I} \\
  & &&\qquad \qquad \quad \cup \{l_i:(P_i \mergek Q_i)\}_{i \in I \cap J}) \\
  & P   \mergek Q    &&= P'   \mergek Q' \qquad ( P \equiv P', Q \equiv Q')   
\end{align*}
\end{definition}

Intuitively, the merge takes  branching processes with
the same roles and generates a single process with all their options.  Above, $P \mergeable Q$ denotes the smallest congruence relation over endpoints processes such that:

\begin{equation}
\myrule{}{
  \forall i\in (K \cup J ). P_i \mergeable Q_i \qquad \forall k \in (K \backslash J). \forall j \in (J \backslash K). l_k \neq l_j
}{\left(\colBranching{k}{A}{B}{l}{P}{K} \right) \mergeable 
    \left(\colBranching{k}{A}{B}{l}{Q}{J} \right)}
\end{equation}

\begin{definition}[Process Projection]\label{def:process-projection}
  $\encod{\chor}{p}$ is a partial function from
  choreographies to processes, defined on the structure of $\chor$
  according to the rules in Figure \ref{fig:process:projection}.

\begin{figure*}
  {
    \begin{align*} 
      &                     \encod{\initA{\VEC{\thread_r[A]}}{\VEC{\thread_s[B]}}{a}{k} \pfx \chor}{p} =
                                                                                            \left\{
                                                                                            \begin{array}{ll}
                                                                                              \initIn{a[\VEC{A},
                                                                                              \VEC{B}]}{k}
                                                                                              \pfx\encod{\chor}{p}
                                                                                              &
                                                                                              \text{if } p = \thread_1[A_1] \\
                                                                                              \initOut{a[A_i]}{k} \pfx \encod{\chor}{p}
                                                                                              & 
                                                                                              \text{if } p = \thread_i[A_i] \in \VEC{\thread_r}\backslash\thread_1\\
                                                                                              \repInitIn{a[B_i]}{k} \pfx \encod{\chor}{p}
                                                                                              &
                                                                                              \text{if } p = \thread_i[B_i] \in \VEC{\thread_s}\\
                                                                                              \encod{\chor}{p}
                                                                                              &
                                                                                              \text{otherwise} 
                                                                                            \end{array}
                                                                                            \right.
    \end{align*}
    \begin{align*} 
      &                     \encod{\bcastA{\thread_r[A]}{\VEC{\thread_s[B_s]:x}}{q}{e}{k} \pfx \chor}{p} =
                                                                                      \left\{
                                                                                      \begin{array}{ll}
                                                                                        \bcast{k}{A}{\VEC{B}}{q}{e} \encod{\chor}{p} 
                                                                                        &\text{if } p = \thread_r[A]\\
                                                                                        \receive{k}{B_i}{A}{x_i} \encod{\chor}{p} 
                                                                                        & \text{if } p = \thread_i[B_i] \in \VEC{\thread_s[B_s]}\\
                                                                                        \encod{\chor}{p}
                                                                                        &
                                                                                          \text{otherwise} 
                                                                                      \end{array}
                                                                                      \right.
    \end{align*}
    \begin{align*}\hspace{\bighorizontalskip}
      &\qquad                     \encod{\reduceA{\VEC{\thread_r[A_r].e_r}}{\thread_s[B]:x}{q}{k}{op} \pfx \chor}{p} =
                                                                                      \left\{
                                                                                      \begin{array}{ll}
                                                                                        \send{k}{A_i}{B}{e_i} \encod{\chor}{p} 
                                                                                        &\text{if } p = \thread_i[A_i] \in \VEC{\thread_r[A_r]}
                                                                                        \\
                                                                                        \reduce{k}{\VEC{A}}{B}{q}{x}{op} \pfx \encod{\chor}{p} 
                                                                                        &\text{if } p = \thread_s[B]\\
                                                                                        \encod{\chor}{p}
                                                                                        &
                                                                                          \text{otherwise} 
                                                                                      \end{array}
                                                                                      \right.
    \end{align*}
    \begin{align*}\hspace{\horizontalskip}
      &                     \encod{\choiceA{\thread_r[A]}{\VEC{\thread_s[B]}}{q}{k}{l}  \pfx \chor}{p}  = \left\{
                                                                                      \begin{array}{ll}
                                                                                        \colSelection{k}{A}{\VEC{B}}{q}{l} \encod{\chor}{p} 
                                                                                        &\text{if } p = \thread_r[A]\\
                                                                                        \branching{k}{B_i}{A}{l}{\encod{\chor}{p}}  
                                                                                        & \text{if } p = \thread_i[B_i] \in \VEC{\thread_s[B]}\\
                                                                                        \encod{\chor}{p}
                                                                                        &
                                                                                          \text{otherwise} 
                                                                                      \end{array}
                                                                                      \right.
    \end{align*}
    \begin{align*}
&                     \encod{\ifthenelsek {e@\thread_r} {\chor_1} {\chor_2}}{p} &=&
                                                                                     \left\{
                                                                                      \begin{array}{ll}
                                                                                        \itn{e}{\encod{\chor_1}{p}}{\encod{\chor_2}{p}}
                                                                                        &\text{if } p = \thread_r\\
                                                                                        \encod{\chor_1}{p} \mergek \encod{\chor_2}{p}
                                                                                        &
                                                                                          \text{otherwise} 
                                                                                      \end{array}
                                                                                      \right. \\
&                     \encod{\INACT}{p} &=&
                                                      \INACT 
    \end{align*}
    \caption{Process Projection, $\encod{\chor}{p}$, where
      $\VEC{\thread_r[A]}= \thread_1[A_1], \ldots, \thread_n[A_n]$ and       $\VEC{\thread_s[B]}= \thread_1[B_1], \ldots, \thread_m[B_m]$ }
    \label{fig:process:projection}
  }
\end{figure*}


\end{definition}

We provide some coments on the mechanics behind  $\encod{\chor}{p}$. 
Depending on the chosen thread $p$, a choreography term
$\initA{\VEC{\thread_r[A]}}{\VEC{\thread_s[B]}}{a}{k} \pfx \chor$ will
generate either (i) an initiating process
$\initIn{a[\VEC{A},\VEC{B}]}{k}\pfx\encod{\chor}{p}$, or (ii) an active
process $\initOut{a[A_i]}{k} \pfx \encod{\chor}{p}$, or (iii) a
service process $\repInitIn{a[B_i]}{k} \pfx \encod{\chor}{p}$, or (iv)
$\encod{\chor}{p}$ if $p$ was not one of the threads involved in
(start).
Collective communications are projected similarly, therefore
describing broadcast will suffice. The thread projection of
$\bcastA{\thread_r[A]}{\VEC{\thread_s[B_s]:x}}{q}{e}{k} \pfx \chor$
will generate either a quality broadcast $\bcast{k}{A}{\VEC{B}}{q}{e}
\encod{\chor}{p}$, or a receiver process $\receive{k}{B_i}{A}{x_i}
\encod{\chor}{p} $ for any of the roles $B_i \in B_s$. In any other
case,  $\encod{\chor}{p}$ will simply continue operating over the
continuation $\chor$.
Since the  conditional construct $\ifthenelsek {e@\thread_r} {\chor_1}
{\chor_2}$ depends solely on the guard of one given thread
$\thread_r$, its thread projection will generate a conditional
localized in such a thread. The projection of the conditional for any
other thread will merge both branches, in order to preserve the label
behaviors at each side. 

A service merge operator joins the behaviour of different service
processes started on the same public channel, playing the same
roles. Formally, the service merge operator, denoted
$\serviceMerge{\chor}{a}{R}$,  returns a set of annotated
threads, and it is defined below:
{
\begin{align*}
  & \serviceMerge{
    \initA{\VEC{\thread_r[A]}}{\VEC{\thread_s[B]}}{a}{k} \pfx
    \chor}{a}{R} &&=  \left\{ \begin{array}{ll}
                              s[R] \cup
                              \serviceMerge{\chor}{a}{R} & \text{if }  s[R] \in \VEC{\thread_s[B]}
                              \\
                              \serviceMerge{\chor}{a}{R} & \text{otherwise}
                            \end{array}
    \right.\\
  & \serviceMerge{ \ifthenelsek {e@\thread_r} {\chor_1} {\chor_2}}{a}{R} &&= \serviceMerge{\chor_1}aR \cup \serviceMerge{\chor_2}aR \\
  & \serviceMerge{
    \chorAct \pfx \chor }{a}{R} &&= \serviceMerge{\chor}aR \quad
                                   \text{if } \chorAct \neq (init) 
\end{align*}
}

We can finally provide a definition of the Endpoint Projection.

\begin{definition}[Endpoint Projection]\label{def:epp}
  Let $\chor = \new{\VEC{k}, \VEC{p}} \chor'$ with  a
  restriction-free  $\chor'$.
  The projection of $\chor$, denoted
  $\project{\chor}$, is defined as:
  \[\begin{array}{ll}
      \project{\chor}& =  \new{\VEC{k}} \left(\prod_{p \in \ft{\chor'}}
                       \encod{\chor'}{p} \pp
                       \prod_{k \in \fsc{\chor'} }
                       \queue{k}{\emptyset} \right)
                       \pp \prod_{a
                       \in \sv{\chor'},A \in \roles{\chor'}} \left(
                       \bigsqcup_{p \in \serviceMerge{\chor'}{a}{A}} \encod{\chor'}{p}
                       \right) 
    \end{array}
    \]
\end{definition}

Recall that $\fsc{\chor}$ and $\ft{\chor}$ contain the set of free session channels
and free threads in $\chor$, respectively.
Essentially, process $\new{\VEC{k}, \VEC{p}} \chor'$ contains active
sessions $\VEC{k}$ and the set of free threads ($\VEC{p}$). The
Endpoint projection of $\chor$ is the parallel
composition of all the active processes with associated empty queues, and the
parallel composition of 
replicated processes resulting from merging all service processes with
same service channel and same role. 

The persistent nature of service processes means that they will not
dissapear once engaged into a session initiation phase. Recalling the
definition of $\Did{^E|Init}$, a system will evolve into:
\begin{equation}
                       \new{k} (
                       P \pp 
                       \prod_{i \in [2, n]} P_i \pp
                       \prod_{i \in [1, m]} Q_i \pp
                       \queue k\emptyset 
                       )  \pp \prod_{i \in [1, m]}
                   \repInitIn{a[B_i]}{k} \pfx Q_i
\end{equation}
Processes in $\repInitIn{a[B_i]}{k} \pfx Q_i$ may not be used after
this interaction. The role of the prunning relation
\cite{Carbone:2012:SCP:2220365.2220367,Carbone:2013:DMA:2429069.2429101},
is to garbage-collect replicated services that are not in use.

\begin{definition}[Pruning] \label{def:pruning}
A pruning between endpoints $P$ and $Q$,
 written
$P \prune Q$, is the relation between $P$ and $Q$ such that $Q \equiv Q_0 \pp                    \prod_{i \in
  I}\repInitIn{a_i[A_i]}{k_i} \pfx R_i$ and it
satisfies the following conditions:
\begin{itemize}
  \item If $Q_0 \action{\mu} Q'_0 $, then $\exists P'$ s.t. $P
    \action{\mu} P'$ and $P' \prune Q'_0$,
    \item $\forall_{i \in I},  a_i \notin \fn(Q_0)$, and
      \item $P \mergek Q_0 = Q_0$.
\end{itemize}
\end{definition}

The first two conditions filter replicated services that are not used in
the evolution of the system, and the last one ensures that all the
labels used in $P$ exists on $Q_0$. 

\begin{lemma}[Pruning lemma]\label{lem:pruning}
  $\prune$ is a strong bisimulation, in the sense that:
  \begin{enumerate}
    \item if $P \prune Q$ and $P \action{\mu} P'$, then $Q
      \action{\mu} Q'$ and $P' \prune Q'$,
      \item if $P \prune Q$ and $Q \action{\mu} Q$, then $P
        \action{\mu} P'$ and $P' \prune Q'$,
        \item if $P \prune Q$ and $Q \prune R$, then $P \prune R$.
  \end{enumerate}
\end{lemma}
\begin{proof}
As in \cite[Lemma 5.29]{Carbone:2012:SCP:2220365.2220367}.
\end{proof}

\begin{figure*}
  {\small
    \begin{align*}
      &   \qquad \qquad         \qquad         \inferenceg{^L|init}
        { \isAProjection[\VEC{\ell}]{\VEC{\emm}}
        }
        { \isAProjection[ \initA{ \VEC{\thread_r[A_r]\{\Blue{X_r; Y_r}\}}
        }{ \VEC{\thread_s[B_s]\{\Blue{ Y_s}\}}}{a}{k}, \VEC{\ell}
        ]{\startAction {\VEC{A}}{\VEC{B}}ak, \VEC{\emm}}
        }
      \\\\      & \qquad \quad \inferenceg{^L|Bcast}
                  { 
                  \begin{array}{c}
                    J \subseteq \VEC{\thread_r[B_r]}
                    \quad
                    q(J)
                    \quad 
                    \VEC{\emm_1} =  \aBroadcastI{A}{B_1}{k}{sb}, \ldots,
                    \aBroadcastI{A}{B_j}{k}{sb}, \asynchEnqueueO 
                    \\
                    |\VEC{\emm_1}| = |J|+1
                    \quad
                    \isAProjection[\VEC{\ell}]{\VEC{\emm_2}}
                  \end{array}
      }
      { \isAProjection[         \bcastA{
      \thread_A[A]\{\Blue{ X_A;Y_A}\}
      }{\VEC{\thread_r[B_r]\{\Blue{X_r;Y_r}\}:sb}}{q}{v}{k}, \VEC{\ell}
      ]{ \aBroadcastO{A}{\VEC{B}}{q}{k}{sb}, \VEC{\emm_1},\VEC{\emm_2}  }
      }
      \\\\      & \inferenceg{^L|Red}
                  {
                  \begin{array}{c}
                    J \subseteq \VEC{\thread_r[A_r]}
                    \quad
                    q(J)
                    \quad 
                    \VEC{\emm_1} =  \aReduceO{A_1}{B}{k}{sb_1} , \ldots,
                    \aReduceO{A_j}{B}{k}{sb_j},\asynchEnqueueI
                    \\
                    |\VEC{\emm}| = |J|+1
                    \quad
                    \evalOp {op(sb_1, \ldots, sb_j)}{\some{v}}
                    \qquad
                    \isAProjection[\VEC{\ell}]{\VEC{\emm_2}}
                  \end{array}
      }
      { \isAProjection[         \reduceA{\VEC{\thread_r[A_r]\{\Blue{
      X_r;Y_r}\}.v_r}}{\thread_B[B]\{\Blue{
      X_B;Y_B}\}:\some{v}}{q}{k}{op}, \VEC{\ell}
      ]{ \aReduceI{\VEC{A}}{B}{q}{k}{\some{v}}, \VEC{\emm_1}, \VEC{\emm_2} }
      }
      \\\\      & \qquad \inferenceg{^L|Sel}
                  { 
                  J \subseteq \VEC{\thread_r[B_r]}
                  \quad
                  q(J)
                  \quad 
                  \VEC{\emm_1} = \aSelectionI{A}{B}{k}{l_j}, \ldots,
                  \aSelectionI{A}{B}{k}{l_j}, \asynchEnqueueO
                  \quad
                  |\VEC{\emm}| = |J|+1
                  \quad
                  \isAProjection[\VEC{\ell}]{\VEC{\emm_2}}
                  }
                  { \isAProjection[
                  \choiceA{\thread_A[A]\{\Blue{
                  X_A;Y_A}\}}{\VEC{\thread_r[B_r]\{\Blue{
                  X_r;Y_r}\}}}{q}{k}{l}, \VEC{\ell}
                  ]{ \aSelectionO{A}{\VEC{B}}{q}{k}{l}, \VEC{\emm_1}, \VEC{\emm_2} }
                  }
      \\\\      & \qquad  \qquad \qquad \inferenceg{^L|Comm}
                  { 
                  \isAProjection[\VEC{\ell}]{\emm', \emm, \VEC{\emm''}}
                  }
                  {                 \isAProjection[\VEC{\ell}]{\emm, \emm', \VEC{\emm''}}
                  }
                  \qquad
                  \inferenceg{^L|Tau}
                  { 
                  \isAProjection[\VEC{\ell}]{\VEC{\emm}}
                  }
                  {                 \isAProjection[\tau,\VEC{\ell}]{\tau, \VEC{\emm}}
                  }
                  \qquad
                  \inferenceg{^L|Empty}
                  { 
                  }
                  {                 \isAProjection[\emptyset]{\emptyset}
                  }
    \end{align*}
    \caption{Behavioural Implementation, $\isAProjection[\VEC{\ell}]{\VEC{\emm}}$}
    \label{fig:label:projection}
  }
\end{figure*}


Let $\action{\VEC{\ell}}$ (resp. $\action{\VEC{\emm}}$ ) be the finite  chain of labelled transitions such that
$\VEC{\ell} = \{\ell_1, \ldots, \ell_n\}$ and $P \action{\ell_1}
\ldots, \action{\ell_n} P'$ (resp. for $\emm$). We are ready to establish the correctness
of the Endpoint Projection. 
   \begin{restatable}[Correctness of the Endpoint
     Projection]{theorem}{theoremEPP} \label{thm:epp} 
     Let $\chor = \new{\VEC{k}, \VEC{p}} \chor_1$ with  a linear 
  restriction-free  $\chor_1$, and 
     $ \isType{\ltypeEnv}{\chor}{\stypeEnv}$, and
     $\isState{\sigma}{\ltypeEnv}$. Then
     \begin{itemize}
     \item{\bf (Soundness)} If
       $ \conf{\sigma}{\chor} \action{\ell} \conf{\sigma'}{\chor'}$
       and $\project{\chor} \prune P$, then $ \conf{\sigma'}{\chor'} \action{\VEC{\ell'}} \conf{\sigma''}{\chor''}$,
       $P \action{\VEC{\emm} } P'$, $\project{\chor''} \prune P'$ and
       $ \isAProjection[\ell,\VEC{\ell'}]{\VEC{\mu}} $.
     \item{\bf (Completeness)} If $\project{\chor} \action{\emm_1} P$
       then 
         there exists $P',\VEC{\ell'}$ such that
         $P \action{\VEC{\emm_2}} P'$,
         $\conf{\sigma}{\chor} \action{\ell} \conf{\sigma'}{\chor'} $,
         $\project{\chor'} \prune P'$, and
         $\isAProjection{\emm_1,\VEC{\emm_2}}$
     \end{itemize}
   \end{restatable}

The judgment $\isAProjection[\VEC{\ell}]{\VEC{\emm}}$ captures whether the labels
in $\VEC{\emm}$ correspond to the choreographic behavior in $\VEC{\ell}$,
and it is defined  as the minimal relation satisfying the   rules in
Figure \ref{fig:label:projection}.
\begin{proof} The proof proceeds by rule induction on the transition
  rules for  $\action{\ell}$ in the case of Soundness, and by induction on the structure of $
     \chor_1$ in the case of Completeness. The details on the proof are presented in Appendix \ref{epp-proof}.
\end{proof}

We can combine Theorem  \ref{thm:global-type-preservation} and Theorem \ref{thm:epp} to
derive that projections out of a well-typed choreography always progress.

\begin{theorem}[Availability By Design] \label{thm:availability-by-design}
     Let $\chor = \new{\VEC{k}, \VEC{p}} \chor'$ with  a
  restriction-free  $\chor'$ and $\chor$ is linear, and 
 $\isType{\ltypeEnv}{\chor}{\stypeEnv}$, 
 $\isState{\sigma}{\ltypeEnv}$,  then either there exists $P', \emm'$
 s.t.  $\project{\chor} \action{\VEC{\emm}} P$, and $P
 \action{\emm'} P'$, or $\project{\chor} \equiv \INACT$.
\end{theorem} 
\begin{proof}
  We proceed by structural induction on $\chor'$. If $\chor' = \INACT$,
  then the projection  $ \project{\new{\VEC{k}, \VEC{p}} \INACT} = \new{\VEC{k}} \left(\prod_{p \in \ft{\INACT}}
                       \encod{\INACT}{p} \pp
                       \prod_{k \in \fsc{\INACT} }
                       \queue{k}{\emptyset} \right)
                       \pp \prod_{a
                       \in \sv{\INACT},A \in \roles{\INACT}} \left(
                       \bigsqcup_{p \in \serviceMerge{\INACT}{a}{A}} \encod{\INACT}{p}
                       \right)  \equiv \INACT$ 
 and we are done. This reasoning also applies when $\chor' \equiv \INACT$.

If $\chor' \not\equiv \INACT$, then by applying inversion in   $\isType{\ltypeEnv}{\chor}{\stypeEnv}$, 
 we know that $\isEntailed{ \chor }$. By the application of Theorem
 \ref{thm:progress} along with assumption
 $\isState{\sigma}{\ltypeEnv}$, then we know that there exists
 $\lambda$ s.t. $\conf{\sigma}{\chor}
     \action{\lambda} \conf{\sigma'}{\chor''}$. From Lemma \ref{lem:pruning}, we know that there exists a $P$,
       s.t. $\project{\chor} \prune P$. Then by the application of
     Theorem \ref{thm:epp}, we know that $P \action{\VEC{\emm} } P'$
     and we are done.
\end{proof}







\section{Related Work} \label{sec:relwork}




Availability considerations in distributed systems have recently
spawned novel research strands in  regular languages
\cite{DBLP:conf/concur/HoenickeMO10,DBLP:conf/fsttcs/AbdullaAMS15},
continuous systems \cite{CPS-book}, and endpoint languages
\cite{nielson2013calculus}. To the best of our knowledge, this is the
first work considering availability from a choreographical perspective.

A closely related work is the Design-By-Contract approach for
multiparty interactions \cite{DBLP:conf/concur/BocchiHTY10}. In fact,
in both works communication actions are enriched with pre-/post-
conditions, similar to works in sequential programming
\cite{DBLP:journals/cacm/Hoare83}. The work on
\cite{DBLP:conf/concur/BocchiHTY10} enriches global types with
assertions, that are then projected to a session
$\pi-$calculus. Assertions may generate ill-specifications, and a
check for consistency is necessary. Our capability-based type system
guarantees temporal-satisfiability as in
\cite{DBLP:conf/concur/BocchiHTY10}, not requiring history-sensitivity
due to the simplicity of the preconditions used in our framework. The
most obvious difference with \cite{DBLP:conf/concur/BocchiHTY10} is the underlying semantics
used for communication, that allows progress despite some participants
are unavailable.

Other works have explored the behavior of communicating systems with
collective/broadcast primitives. In \cite{huttel2015broadcast}, the
expressivity of a calculus with bounded broadcast and collection is
studied. In \cite{Lopez2015OOPSLA}, the authors present a type theory
to check whether models for multicore programming behave according to
a protocol and do not deadlock. Our work differs from these approaches
in that our model focuses considers explicit considerations on
availability for the systems in consideration.  Also for multicore
programming, the work in \cite{DBLP:conf/coordination/CogumbreiroMV13}
presents a calculus with fork/join communication primitives, with a
flexible phaser mechanism that allows some threads to advance prior to
synchronization. The type system guarantees a node-centric progress
guarantee, ideal for multicore computing, but too coarse for
CPS. Finally, the work \cite{kouzapas2014session}, present endpoint
(session) types for the verification of communications using broadcast
in the $\Psi$-calculus. We do not observe similar considerations
regarding availability of components in this work.

The work \cite{castagna2011global} presented multiparty global types
with join and fork operators, capturing in this way some notions of
broadcast and reduce communications, which is similar to our
capability type-system. The difference with our approach is described
in Section  \ref{sec:language}. On the same branch
\cite{denielou2012multiparty} introduces multiparty global types with
recursion, fork, join and merge operations. The work does not provide a natural way of
encoding broadcast communication, but one could expect to be able to
encode it by composing fork and merge primitives.

The work by Kouzapas, Yoshida and Honda explore a session $\pi$
calculus with  an asynchronous
semantics based on input/output queues
\cite{DBLP:conf/forte/KouzapasYH11}. The language presented there
bears similarities with the Endpoint Calculus presented in Section
\ref{sec:epc}. The use of use of message queues and the use of
predicates to identify when a message has arrived to a local buffer
resembles the interplay between input/output queues and quality
predicates. In our model, collective operations such as broadcast and
reduce imply that there can be multiple orderings on the communication
events occurred (e.g.: we cannot guarantee when receivers of a
broadcast will consume the message). In future work, we would like to
explore how behavioral theories such as the one in
\cite{DBLP:conf/forte/KouzapasYH11} can be adapted for collective
communications. 

The current work is an extension of the Quality Choreographies work
presented at  \cite{Lopez2016Quality}. As mentioned in the
introduction, in this version we have provided a full methodology of
choreographic programming, where choreographies can project to a
novel asynchronous endpoint language. Moreover, implementations have
been proven to guarantee a novel availability-by-design property, the
corresponding deadlock-freedom property for failure-aware
communication protocols.  Technically, the choreographic
language presented in this version bears
differences in some of the language operators, as well as in the operational
semantics: the non-deterministic choice presented in
\cite{Lopez2016Quality} proved difficult to accommodate in an endpoint
projection that could respect the soundness and completeness
guarantees in Theorem \ref{thm:epp}.  Further restrictions involved
limitations on the quality predicates used for collective
selections. We would like to revisit such aspects in future works.


\section{Conclusions and Future Work} \label{sec:conclusions}

We have presented a process calculus aimed at studying 
protocols with variable availability conditions, as well as a type
system to ensure their progress. Paired with session types,
choreographies guide the correct implementation of distributed systems
with failure conditions, on a communication model based on synchronous
and collective communications. This constitutes the first
step towards a methodology for the
safe development of communication protocols in CPS. 
Some important considerations have been left out for future
work. First, linearity considerations require each participant to
implement one unique behavior. This is not natural in failure-aware
communication, that requires several copies of the same component to
be deployed, all of them implementing the same behavior. A possible
extension will be to integrate parameterized or index-based multiparty
session types in our analysis, taking inspiration from the works of
\cite{DBLP:journals/corr/abs-1208-6483,Lopez2015OOPSLA}.  Other
possible efforts 
include the modification of the type theory to
cater for recursive behavior, non-determinism, and considerations of
compensating \cite{DBLP:journals/entcs/Carbone09,DBLP:conf/concur/CarboneHY08,Lopez2011Integrating-Tim} and time \cite{DBLP:conf/concur/BocchiYY14,DBLP:conf/concur/BocchiLY15}.
Type checking is
computationally expensive, because for each collective interaction one must
perform the analysis on each subset of participants involved. The situation will be critical
once recursion is considered. We believe that 
the efficiency of type checking can be improved by modifying the theory so it
generates one formulae for all subsets.

Traditional design mechanisms (including sequence charts of UML and
choreographies) usually focus on the desired behavior of systems. In
order to deal with the challenges from security and safety in CPS it
becomes paramount to cater for failures and how to recover from
them. This was the motivation behind the development of the Quality
Calculus that not only extended a $\pi$-calculus with quality predicates
and optional data types, but also with mechanisms for programming the
continuation such that both desired and undesired behavior was
adequately handled. In this work we have incorporated the quality
predicates into choreographies and thereby facilitate dealing with
systems in a failure-aware fashion. However, it remains a challenge to
incorporate the consideration of both desired and undesired behavior
that is less programming oriented (or EndPoint Projection oriented)
than the solution presented by the Quality Calculus. This may require
further extensions of the calculus with \emph{fault-tolerance} considerations.



\subsection*{Acknowledgments.}
Part of the research leading to these results  has received funding
from the Danish
Foundation for Basic Research, project \emph{IDEA4CPS}
(DNRF86-10), and the European Union Seventh Framework Programme (FP7/2007-2013)
under grant agreement no.~318003 (TRE$_\mathrm{S}$PASS). This
publication reflects only the authors' views and the Union is not
liable for any use that may be made of the information contained
herein.
López has benefitted from travel support by the EU COST
Action IC1201: \emph{Behavioural Types for Reliable Large-Scale
  Software Systems}~(BETTY).



 \bibliographystyle{plainurl}
\bibliography{biblio}

 \appendix
\onecolumn 
\allowdisplaybreaks 
 \section{Proofs}
\label{appendix:proofs}
\subsection{Results related to states}
\begin{lemma}[Validity: States] \label{lem:states:validity}
  If $\isState{\sigma}{\ltypeEnv}$,
  then $\isAFormula$ and $\isAState$.
\end{lemma}
\proof
  By rule induction on the first hypothesis.
\qed

\begin{lemma}[Weakening: States] \label{lem:states:weakening}
  If $\isState{\sigma}{\ltypeEnv}$,
  then $\isStateEx{\sigma, (\thread \left[A\right], k, X)}{\ltypeEnv, \thread \colon \ownerTP{k}{A}{X}}$
\end{lemma}
\proof
  It follows by induction on the hypothesis.
\qed
\begin{lemma}[Update: States] \label{lem:states:update}
 If $\isState{\sigma}{\ltypeEnv}$ and $\isState{\sigma'}{\ltypeEnv'}$,
  then $\isState{\updateState{\sigma'} }{(\ltypeEnv \backslash
    \delta), \ltypeEnv'}$, where $    \delta = \{ (\thread, k, X) ~|~ (\thread, k, X) \in \sigma \land
    (\thread, k, Y) \in \sigma'\}$.
\end{lemma}
\proof
It follows directly from the hypotheses, the definition of $\updateState{\sigma'}$
and Definition \ref{app:state-satisfaction}. 
\qed
\subsection{Results related to choreographies}
\begin{lemma}[Weakening: choreographies] \label{lem:choreo:weakening}
  Let $\isAFormula[\psi]$. If $\isEntailed{\chor}$,
  then $\isEntailed[\ltypeEnv, \psi]{\chor}$.
\end{lemma}
\proof
 By rule induction on the hypothesis.
\qed
\begin{lemma}[Strengthening: choreographies] \label{lem:choreo:strength}
  If $\isEntailed[\ltypeEnv, \thread \colon \ownerTP{k}{A}{X} ]{\chor}$  and $X \notin \fforms{\chor}$, 
  then $\isEntailed{\chor}$.
\end{lemma}
\proof
  By rule induction on the first hypothesis.
\qed
\begin{lemma}[Substitution] \label{lem:chor-substitution}
 Let $t$ be a term, and $x@p \in \vars{\chor}$.
  If   $\isEntailed{\chor}$, then   $\isEntailed{\theta(\chor)}$.
\end{lemma}
\proof
  By rule  induction on the hypothesis. Note that $x \notin \ltypeEnv$. 
\qed

\begin{lemma}[Subject Congruence]\label{lem:subject-cong}
If $\chor \equiv \chor'$ and $\isEntailed{\chor}$, then $\isEntailed{\chor'}$.
\end{lemma}
\proof
 It proceeds by induction on the depth of the first
 premise. 
\qed

\begin{lemma}[Inversion Lemma] \label{app:inv-lemma-typing-formation}
Let $\isEntailed{ \chor}$ then either:
{
\begin{itemize}
   \item $\chor = \chorAct \pfx \chor'$, and:
     \begin{itemize}
     \item
       $\chorAct = \initA{\VEC{\thread_r[A_r]\Blue{\{Y_r\}}}}{
         \VEC{\thread_s[B_s]\Blue{\{Y_s\}}}}{a}{k}$
       and
       $ \isEntailed[\ltypeEnv, \isInit{
         \VEC{\thread_r[A_r]\Blue{\{Y_r\}}},
         \VEC{\thread_s[B_s]\Blue{\{Y_s\}}}, k}]{\chor'}$,
       and
       $ \{\VEC{\thread_s},k\} \# (\threads{\ltypeEnv} \cup
       \keys{\ltypeEnv}) $, or
     \item
       $\chorAct = \bcastA{\thread_A[A]\Blue{\{X_A;Y_A\}}}{\VEC{\thread_r[B_r]\Blue{\{X_r;Y_r\}}:x_r}}{q}{e}{k}$
       and \\
       $ \forall^{\ge 1} J.\ s.t. \big(J \subseteq \widetilde{B} ~\land~ q(J) $,
       $ \ltypeEnv = \psi_A, (\psi_j) _{j \in
                        J},  \ltypeEnv'  ~\land~ $
       $ \phi = \thread_A \colon \ownerTP{k}{A}{X_A} \bigotimes_{j \in
         J} (\thread_j \colon \ownerTP{k}{B_j}{X_j} )   ~\land~$
       $ \phi' = \thread_A \colon \ownerTP{k}{A}{Y_A} \bigotimes_{j
         \in J} \left( \thread_j \colon \ownerTP{k}{B_j}{Y_j} \right)  ~\land~
       $
       $ \typerule {\psi_A, \left( \psi_j\right)_{j \in J}, \phi
         \lolli \phi' }{}{ \phi' }\big) : $ 
       $ \isEntailed[ \thread_A \colon \ownerTP{k}{A}{Y_A}, \left(\thread_j \colon
         \ownerTP{k}{B_j}{Y_j} \right) _{j \in J}, \ltypeEnv']{\chor'} $,
       and $ \isOptData[e@\thread_A]$, and
       $(\isOptData[x_i@\thread_i])_{i \in \VEC{\thread_r}} $, or
     \item
       $\chorAct =
       \reduceA{\VEC{\thread_r[A_r]\Blue{\{X_r;Y_r\}}.e_r}}{\thread_B[B]\Blue{\{X_B;Y_B\}}:x}{q}{k}{op}
       $ and\\
       $ \forall^{\ge 1} J.\ s.t. \big( J \subseteq \widetilde{A}
       ~\land~ q(J)  ~\land~$
       $ \ltypeEnv = \psi_B, (\psi_j)_{j \in J}, \ltypeEnv'   ~\land~$
       $ \phi =\thread_B \colon \ownerTP{k}{B}{X_B} \bigotimes_{j \in
         J} (\thread_j \colon \ownerTP{k}{A_j}{X_j} )   ~\land~$
       $ \phi' = \thread_B \colon \ownerTP{k}{B}{Y_B} \bigotimes_{j
         \in J} \left( \thread_j \colon \ownerTP{k}{A_j}{Y_j} \right)  ~\land~
       $ 
       $ \typerule {\psi_B, \left( \psi_j \right)_{j \in J},
         \phi \lolli \phi' }{}{ \phi' } \big) :$ 
       $ \isEntailed[ \thread_B \colon \ownerTP{k}{B}{Y_B},
       \left(\thread_j \colon
         \ownerTP{k}{A_j}{Y_j} \right)_{j \in J}, \ltypeEnv']{\chor'}$,
       and $(\isOptData[e_i@\thread_i])_{i \in \VEC{\thread_r}}$,
       and $\isOptData[x@\thread_B]$, or
     \item
       $\chorAct =
       \choiceA{\thread_A[A]\Blue{\{X_A;Y_A\}}}{\VEC{\thread_r[B_r]\Blue{\{X_r;Y_r\}}}}{q}{k}{l_h}$
       and\\
       $ \forall^{\ge 1} J.\ s.t. \big(J \subseteq \widetilde{B}
       ~\land~ q(J) \land$
       $ \ltypeEnv = \psi_A, (\psi_j)_{j \in J}, \ltypeEnv'   ~\land~$ 
       $ \phi = \thread_A \colon \ownerTP{k}{A}{X_A} \bigotimes_{j \in
         J} (\thread_j \colon \ownerTP{k}{B_j}{X_j} )  ~\land~$
       $ \phi' = \thread_A \colon \ownerTP{k}{A}{Y_A} \bigotimes_{j
         \in J} \left( \thread_j \colon \ownerTP{k}{B_j}{Y_j} \right)
        ~\land~$
       $ \typerule {
         \psi_A, (\psi_j)_{j \in J},  \phi \lolli \phi' }{}{ \phi' } \big):$
       $ \isEntailed[ \thread_A \colon \ownerTP{k}{A}{Y_A} ,
       (\thread_j \colon \ownerTP{k}{B_j}{Y_j}) _{j \in J}, \ltypeEnv']{\chor'} $,
       or
     \end{itemize}
\item $\chor = \ifthenelsek{e@\thread}{\chor_1}{\chor_2}$, and $\isEntailed[\ltypeEnv]{\chor_1}$, and
  $\isEntailed[\ltypeEnv]{\chor_2}$, or 
\item $\chor = \INACT$.
\end{itemize}
}
\end{lemma}
\proof
  By case analysis on the type formation rules. 
\qed

\begin{lemma}[Subject Swap] \label{lem:subject-swap}
If $\chor \swaps \chor'$ and $\isEntailed{\chor}$, then $\isEntailed{\chor'}$.
\end{lemma}
\proof
 It proceeds by induction on the depth of the first premise. Most of
 the cases are straightforward except $       \chorAct \pfx (\chorAct' \pfx \chor) \swaps \chorAct'
   \pfx (\chorAct \pfx \chor)$, which requires the application of Lemma
 \ref{app:inv-lemma-typing-formation} and case analysis.
\qed

\theoremTypePreservation*
\label{lem-type-preservation-proof}
\proof
  It follows by rule induction on the first hypothesis. We have eight
  cases.

 \Case{} $\Did{BCast},\Did{Sel}$ rules: Standard Inversion/formation
 rules. Process typing requires substitution (Lemma
 \ref{lem:chor-substitution}) and  state typing requires validity
 (Lemma \ref{lem:states:validity}) and state update
 (Lemma \ref{lem:states:update}). We proceed to show the case for $\Did{BCast}$.


\setcounter{equation}{0}
\begin{align}
  & \text{Hypothesis} && \begin{array}{l}
                           \conf{\sigma}{    \left(
                           \bcastA{
                           \thread_A[A]\{\Blue{ X_A;Y_A}\}
                           }{\VEC{\thread_r[B_r]\{\Blue{ X_r;Y_r}\}:x_r}}{q}{e}{k} \right) \pfx \chor
                           } \\
                           \qquad \qquad \qquad \qquad \qquad \qquad \qquad \qquad \qquad \qquad \action{    
                           \theta(\chorAct)
                           }\ 
                           \conf{\sigma[\sigma']}{\theta(\chor) }
                         \end{array}
  \\
  & \text{Hypothesis} &&       \isEntailed{ \left(
    \bcastA{
    \thread_A[A]\{\Blue{ X_A;Y_A}\}
    }{\VEC{\thread_r[B_r]\{\Blue{ X_r;Y_r}\}:x_r}}{q}{e}{k} \right) \pfx \chor }  \\
  & \text{Hypothesis} && \isState{\sigma}{\ltypeEnv} \\
  & 3, \text{lem. \ref{lem:states:validity}} &&         \isAFormula \\
  & 3, \text{lem. \ref{lem:states:validity}} &&         \isAState\\
  & 1, \text{inversion} &&  J' \subseteq \VEC{\thread_r} \\
  & 1, \text{inversion} &&  q(J')\\
  & 1, \text{inversion} && \forall {\iset i{\{A\} \cup J'}}: \Blue{X_i} \subseteq
        \sigma(\thread_i, k) \land \sigma'(t_i, k) = \exchange{X_i}{Y_i}(\sigma(t_i,k))        \label{Eq-sigmaprime-bcast}\\
  & 1, \text{inversion} &&  \evalOp{e@\thread_A}{ v } \\
  & 1, \text{inversion} &&   \forall {\iset {i}{\VEC{\thread_r}}}: \theta(x_{i}) =
        \left\{
        \begin{array}{l} 
          \some{v} \quad i \in J'  \\
          \none \quad \text{otherwise}
        \end{array}
        \right. \\
  & 2, \text{inversion} && 
\begin{array}{c}
                        \forall^{\ge 1} J.\ s.t. \left(
                        \begin{array}{c}
                          J \subseteq \VEC{\thread_r}
                        ~\land~
                        q(J) 
                        ~\land~
                        \ltypeEnv = \psi_A, (\psi_j)_{j \in
                        J},  \ltypeEnv' 
                        \\[\ruleskip] ~\land~
                        \phi = \thread_A \colon
                        \ownerTP{k}{A}{X_A} \bigotimes_{j \in J} (\thread_j \colon
                        \ownerTP{k}{B_j}{X_j}  ) 
                        \\[\ruleskip] 
                        ~\land~
                        \phi' = \thread_A \colon
                        \ownerTP{k}{A}{Y_A}      \bigotimes_{j \in J} \left( \thread_j \colon
                        \ownerTP{k}{B_j}{Y_j} \right) 
                        \\[\ruleskip]                         ~\land~
                          \typerule
                        {
                        \psi_A, (\psi_j)_{j \in
                        J},
                        \phi \lolli \phi'
                        }{}{
                        \phi'
                        } 
                        \end{array}
                        \right): 
                        \\
                        \isEntailed[      \thread_A \colon
                        \ownerTP{k}{A}{Y_A},  \left(\thread_j 
                        \colon \ownerTP{k}{B_j}{Y_j} \right)_{j \in
                        J},
                        \ltypeEnv']{\chor} 
                      \end{array} \\
  & 2, \text{inversion} &&     \isOptData[e@\thread_A] \\
  & 2, \text{inversion} &&     \isOptData[x_i@\thread_i] \quad \irange{i}{1}{|\VEC{\thread_r}|} 
\end{align}
From the definition of $\fv{\cdot}$ we know that
$x_i@\thread_i \in \fv{ \bcastA{    \thread_A[A] \{\Blue{ X_A;Y_A}\}
    }{\\ \VEC{\thread_r[B_r] \{\Blue{ X_r;Y_r}\}:x_r}}{q}{e}{k}  \pfx
    \chor} $,  ${\irange{i}{1}{|\thread_r|}}$. From this, Equations 9 and 11 and the  application of
  Lemma \ref{lem:chor-substitution} we can
  conclude 
\begin{equation}
  \begin{array}{c}
    \forall^{\ge 1} J.\ s.t. \left(
    \begin{array}{c}
      J \subseteq \VEC{\thread_r}
      ~\land~
      q(J) 
      ~\land~
      \ltypeEnv = \psi_A, (\psi_j)_{j \in
      J},  \ltypeEnv' 
      \\[\ruleskip] ~\land~
      \phi = \thread_A \colon
      \ownerTP{k}{A}{X_A} \bigotimes_{j \in J} (\thread_j \colon
      \ownerTP{k}{B_j}{X_j}  ) 
      \\[\ruleskip] 
      ~\land~
      \phi' = \thread_A \colon
      \ownerTP{k}{A}{Y_A}      \bigotimes_{j \in J} \left( \thread_j \colon
      \ownerTP{k}{B_j}{Y_j} \right) 
      \\[\ruleskip]                         ~\land~
      \typerule
      {
      \psi_A, (\psi_j)_{j \in
      J},
      \phi \lolli \phi'
      }{}{
      \phi'
      } 
    \end{array}
    \right): 
    \\
    \isEntailed[      \thread_A \colon
    \ownerTP{k}{A}{Y_A},  \left(\thread_j 
    \colon \ownerTP{k}{B_j}{Y_j} \right)_{j \in
    J},
    \ltypeEnv']{\theta(\chor)} 
  \end{array} 
\end{equation}

For state typing we need to show that 
\[
\begin{array}{c}
  \forall^{\ge 1} J.\ s.t. \left(
  \begin{array}{c}
    J \subseteq \VEC{\thread_r}
    ~\land~
    q(J) 
    ~\land~
    \ltypeEnv = \psi_A, (\psi_j)_{j \in
    J},  \ltypeEnv' 
    \\[\ruleskip] ~\land~
    \phi = \thread_A \colon
    \ownerTP{k}{A}{X_A} \bigotimes_{j \in J} (\thread_j \colon
    \ownerTP{k}{B_j}{X_j}  ) 
    \\[\ruleskip] 
    ~\land~
    \phi' = \thread_A \colon
    \ownerTP{k}{A}{Y_A}      \bigotimes_{j \in J} \left( \thread_j \colon
    \ownerTP{k}{B_j}{Y_j} \right) 
    \\[\ruleskip]                         ~\land~
      \typerule{ \psi_A, (\psi_j)_{j \in
      J},
      \phi \lolli \phi'
      }{}{
      \phi'
      } 
  \end{array}
  \right): 
  \\
  \isState{\updateState{\sigma'} }{\thread_A \colon
  \ownerTP{k}{A}{Y_A},  \left(\thread_j 
  \colon \ownerTP{k}{B_j}{Y_j} \right)_{j \in
  J},
  \ltypeEnv'}
\end{array}
\]
By the application of state update rule and state satisfaction we
get:

\begin{align}
 & 6, 7, 11 &&      J' \subseteq J 
\end{align}
From Eq. \ref{Eq-sigmaprime-bcast} we know that $\sigma $ contains all
triples $(\thread_i, k, \Blue{X_i})_{i\in \{A\}\cup J'}$ and
possibly more. We denote with $\sigma''$ the set of additional tuples
in $\sigma$. From Eq. 3 we know that
\begin{equation}
  \isState{\sigma''}{\ltypeEnv'}
\end{equation}
\begin{align}
& \ref{Eq-sigmaprime-bcast}, \text{state satisfaction} &&
                                                           \exists R.~
                                                           \isState{\sigma}{(\thread_i
                                                           \colon
                                                           \ownerTP{k}{R}{X_i})_{i
                                                           \in \{A\}
                                                           \cup J'}}
  \\
 & \ref{Eq-sigmaprime-bcast}, \text{state satisfaction} &&
                                                           \exists R.~
                                                           \isState{\sigma'}{(\thread_i
                                                           \colon
                                                           \ownerTP{k}{R}{Y_i})_{i
                                                           \in \{A\}
                                                           \cup J'}}
  \\
 & 3, 16, \text{lem. \ref{lem:states:update}} &&
                                                           \exists R.~ \isState{\updateState{\sigma'}  }{(\ltypeEnv \backslash
    \delta), (\thread_i \colon \ownerTP{k}{R}{Y_i})_{i
                                                           \in \{A\}
                                                           \cup J'}} \\
 & 3, 16, \text{lem. \ref{lem:states:update}} && \delta = \{ (\thread, k, X) ~|~ (\thread, k, X) \in \sigma \land
    (\thread, k, Y) \in \sigma'\}
\end{align}
Moreover, we know that $\delta \cup \sigma'' = \sigma$, and from the
definition of store update, we know that $  \updateState{\sigma'} = (\sigma\backslash \delta), \sigma' $.  We can rewrite
Eq. 19 as 
\begin{equation}
                                                           \exists R.~
                                                           \isState{
                                                             (\sigma'',\delta)\backslash
                                                           \delta, \sigma'}{(\ltypeEnv \backslash
    \delta), (\thread_i \colon \ownerTP{k}{R}{Y_i})_{i
                                                           \in \{A\}
                                                           \cup J'}} 
\end{equation}
By replacing $\ltypeEnv \backslash \delta$ by $\ltypeEnv'$ in Eq. 21,
the definition of state update,  and by
applying simple formula exchange, we get:
\begin{equation}
                                                           \exists R.~
                                                           \isState{
                                                            \updateState{\sigma'}}{ (\thread_i \colon \ownerTP{k}{R}{Y_i})_{i
                                                           \in \{A\}
                                                           \cup J'}, \ltypeEnv'} 
\end{equation}



 \Case{} $\Did{^G|Red}$: It corresponds to the same equivalence class as
 the case for $\Did{^G|Bcast}$. Its proof is analogous.

 \Case{} $\Did{Init}$ rule: Standard inversion/formation
 rules. It requires state validity (Lemma \ref{lem:states:validity}),
 state weakening (Lemma \ref{lem:states:weakening}) and state update lemma (Lemma \ref{lem:states:update}).

\setcounter{equation}{0}
\begin{align}
  & \text{Hypothesis} &&
     \new{\VEC {m}}  \conf{\sigma}{
    \initA{ \VEC{\thread_r[A_r]\{\Blue{ Y_r}\}}
    }{ \VEC{\thread_s[B_s]\{\Blue{ Y_s}\}}}{a}{k}  \pfx \chor
    }
      \action{\chorAct  } 
      \new{\VEC{m},\VEC{n}}\conf{\updateState{\sigma'}}{\chor} \\
  & \text{Hypothesis} &&       \isEntailed{ \initA{\widetilde{\thread_r[A_r]\{\Blue{Y_r}\}}}{
    \widetilde{\thread_s[B_s]\{\Blue{Y_s}\}}}{a}{k} \pfx \chor }  \\
  & \text{Hypothesis} && \isState{\sigma}{\ltypeEnv} \\
  & 3, \text{lem. \ref{lem:states:validity}} &&         \isAFormula \\
  & 3, \text{lem. \ref{lem:states:validity}} &&         \isAState\\
  & 1, \text{inversion} && \sigma' = [(\thread_i,k) |-> \Blue{ Y_i}
                           ]^{|\VEC{\thread_r}|+|\VEC{\thread_s}|}_{i=1}  \\
  & 1, \text{inversion} &&        \VEC{n} = \VEC{\thread_s}, \{k\} \\
  & 1, \text{inversion} &&         \VEC{n} ~\#~ \VEC{m} \\
  & 2, \text{inversion} &&        
    \{\VEC{\thread_s},k\} ~\#~ (\threads{\ltypeEnv} \cup
    \keys{\ltypeEnv}) \\
  & 2, \text{inversion} &&        \isEntailed[\ltypeEnv, \isInit{
                           \VEC{\thread_r[A_r]\Blue{\{Y_r\}}},
                           \VEC{\thread_s[B_s]\Blue{\{Y_s\}}},
                           k}]{\chor}  \label{proof-init-proc-typing}
\end{align}
Process typing is derived already in Eq.
\ref{proof-init-proc-typing}. Moreover, by the application of the
definition of  $\isInit{}$ to Eq. \ref{proof-init-proc-typing} we know
that $\isEntailed[\ltypeEnv, \VEC{\thread_r \colon
\ownerTP{k}{A_r}{\Blue{Y_r}}}, \VEC{\thread_s \colon
\ownerTP{k}{B_s}{\Blue{Y_s}}}]{\chor} $.  We now proceed with state
typing. 

\begin{align}
  & 6, \text{def. \ref{app:state-satisfaction}} && \exists
                                                   A. (\isState{\sigma'}{\thread_i
                                                   \colon
                                                   \ownerTP{k}{A}{Y_i}})_{i
                                                   =
                                                   1}^{|\VEC{\thread_r}|+|\VEC{\thread_s}|}\\
  & 11, \text{def. of $\isInit{}$} &&\exists
                                                   A.(\isState{\sigma'}{\isInit{
         \VEC{\thread_r[A_r]\Blue{\{Y_r\}}},
         \VEC{\thread_s[B_s]\Blue{\{Y_s\}}}, k}})\\
  & 3, 12, \text{lemma \ref{lem:states:update}} && \isState{\updateState{\sigma'} }{(\ltypeEnv \backslash
    \delta), \isInit{
         \VEC{\thread_r[A_r]\Blue{\{Y_r\}}},
         \VEC{\thread_s[B_s]\Blue{\{Y_s\}}}, k}}\\
  & 3, 12, \text{lemma \ref{lem:states:update}} && \delta = \{ (\thread, k, X) ~|~ (\thread, k, X) \in \sigma \land
    (\thread, k, Y) \in \sigma'\}
\end{align}
However, from Eq. 9 we know that $k \notin \keys{\ltypeEnv}$,
therefore $\delta = \emptyset$.  Using this fact we can write Eq. 13
as 
\begin{equation}
\isState{\updateState{\sigma'} }{\ltypeEnv, \isInit{
         \VEC{\thread_r[A_r]\Blue{\{Y_r\}}},
         \VEC{\thread_s[B_s]\Blue{\{Y_s\}}}, k}}
   \end{equation}
   Which is what we wanted to show.

 \Case{} $\Did{Cong}$ rule: One case for each
 congruence relation.  $\equiv$ requires subject congruence (Lemma
 \ref{lem:subject-cong}), and $\swaps$ requires subject swap (Lemma \ref{lem:subject-swap}).

 \Case{} $\Did{If}$ rule: it follows a standard induction.
\qed

\theoremProgressChor*
\label{lem-type-progress-proof}
\proof
Proof by contradiction. Let us assume that $\isEntailed \chor$, $\isState{\sigma}{\ltypeEnv}$
  and $\chor \not\equiv \INACT$ and $\conf{\sigma}{\chor}
  \not\action{\lambda}$. We proceed by
  induction on the structure of  $\chor$ to show that such $\chor$
  does not exists.
\qed

\subsection{Results related to session types}

\begin{lemma}[Substitution] \label{lem:global-substitution}
If $\isType[\typeEnv, x@p \colon S]{\ltypeEnv}{\chor}{\stypeEnv}$
and $\isIndex{v}{S}$ then   
$\isType{\ltypeEnv}{\theta(\chor)}{\stypeEnv}$.
\end{lemma}
\begin{proof}
  It follows by rule induction on the first hypothesis. 
\end{proof}

\begin{lemma}[Subject Congruence]\label{lem:global-congruence}
If $\isType{\ltypeEnv}{\chor}{\stypeEnv}$ and
$\chor \equiv \chor'$, then   $\isType{\ltypeEnv}{\chor'}{\stypeEnv}$.
\end{lemma}

\begin{proof}
It follows by induction on the depth of the
    premise $\chor \equiv \chor'$.
\end{proof}

We write $\stypeEnv \swapsT \stypeEnv'$ to denote that $dom(\stypeEnv)
= dom(\stypeEnv')$ and for all $k \in dom(\stypeEnv)$, $\stypeEnv(k)
\swapsT \stypeEnv'(k)$. Similarly, we say $\stypeEnv' \subseteq \stypeEnv$ when  $ k\colon G'
\in \stypeEnv'$, implies that $\exists \alpha,k$; $k\colon G
\in \stypeEnv$ and $\ltransition{G}{\alpha}{G'}$, $\forall k\colon G'
\in \stypeEnv'$.

\begin{lemma}[Subject Swap]\label{lem:global-swap}
If $\isType{\ltypeEnv}{\chor}{\stypeEnv}$ and
$\chor \swaps \chor'$, then  there exists $ \stypeEnv'$ s.t.
$\isType{\ltypeEnv}{\chor'}{\stypeEnv'}$ and $\stypeEnv \swapsT \stypeEnv'$.
\end{lemma}
\begin{proof}
  It follows by induction on the depth of the
    premise $\chor \swaps \chor'$, as well as the swap relation rules
    for global types in Figure \ref{fig:swap-rel-types}.
\end{proof}

\theoremTypePreservationGlobal*
\label{lem-type-progress-global-proof}
\proof
The proof follows by rule induction on $\conf{\sigma}{\chor}
     \action{\lambda} \conf{\sigma'}{\chor'}$, using Theorem
     \ref{thm:type-preservation}  to guarantee the type
     preservation of judgments $\isEntailed{\chor}$. We have seven
     cases:

\Case{} Rule $\Did{^G|Init}$:

  \setcounter{equation}{0}
  \begin{align}
    & \text{Hypothesis} &&      
                           \begin{array}{l}
                             \conf{\sigma}{\initA{
                             \VEC{\thread_r[A_r]\{\Blue{Y_r}\}}}{
                             \VEC{\thread_s[B_s]\{\Blue{
                             Y_s}\}}}{a}{k}  \pfx \chor'} \\
                             \quad \action{\initA{ \VEC{\thread_r[A_r]\{\Blue{Y_r}\}}
        }{ \VEC{\thread_s[B_s]\{\Blue{ Y_s}\}}}{a}{k}} 
                             \conf{\sigma[\sigma'[\sigma'']]}{\new{\VEC{\thread_s},k}\chor'} 
                           \end{array}
    \\
    & \text{Hypothesis} &&
                           \isType{\ltypeEnv}{ \initA{\VEC{\thread_r[A_r]\Blue{\{Y_r\}}}}{
      \VEC{\thread_s[B_s]\Blue{\{Y_s\}}}}{a}{k} \pfx
                           \chor'}{\stypeEnv}\\
    & \text{Hypothesis} && \isState{\sigma}{\ltypeEnv} \\
    & \text{A.2, inversion} && \isSType{ \initA{\VEC{\thread_r[A_r]\Blue{\{Y_r\}}}}{
      \VEC{\thread_s[B_s]\Blue{\{Y_s\}}}}{a}{k} \pfx \chor' }
                               {\stypeEnv} \\
    & \text{A.4, inversion} &&          \isIndex{a}{\serviceT GAB} \\
    & \text{A.4, inversion} &&
                               \isSType[\Gamma,\isInit{\{
                               \VEC{\thread_r[A_r] },
                               \VEC{\thread_s[B_s]}\},
                               k}]{\chor'}{\stypeEnv, k \colon G}  \\ 
    & \text{A.4, inversion} &&          
          \VEC{\thread_s}\ \#\ \typeEnv \\
    & \text{A.2, inversion} && \isEntailed{ \initA{\VEC{\thread_r[A_r]\Blue{\{Y_r\}}}}{
      \VEC{\thread_s[B_s]\Blue{\{Y_s\}}}}{a}{k} \pfx \chor' }\\
    & \text{A.1, A.3, A.8, thm. }\ref{thm:type-preservation} &&
                                                                 \isState{\sigma[\sigma'[\sigma'']]}{\ltypeEnv'}\\
    & \text{A.1, A.3, A.8, thm. }\ref{thm:type-preservation} &&
                                                                 \isEntailed[\ltypeEnv']{
                                                                \new{\VEC{\thread_s},k}
                                                                \chor'
                                                                }
  \end{align}
  Moreover, we know that $\VEC{\thread_s} \# \stypeEnv$ since
  $\stypeEnv$ only contain information regarding session
  variables. Also, recall that function
  $\isInit{\{\VEC{\thread_r[A_r]},\VEC{\thread_s[B_s]}\} , k}$ returns a
  list  of     ownership types $\VEC{ \thread_p \colon \ownerT kA }$
  where $\forall \thread_p \in \VEC{\thread_p}, \thread_p \in
  \{\VEC{\thread_r},\VEC{\thread_s}\}$.  We can conclude by the
  sequence of applications of $\Did{TGres}$ to type the redex. Let
  $\typeEnv' = \typeEnv,\isInit{\{
                               \VEC{\thread_r[A_r] },
                               \VEC{\thread_s[B_s]}\},
                               k}\backslash \VEC{\thread_s}$, then
\begin{align}
  & 
      \text{A.6}, \VEC{\thread_s} \# \stypeEnv, 
      \text{ rule }
    \Did{TGres} \times |\VEC{\thread_s}| \text{ times}
  &&
     \isSType[\typeEnv']{\new{\VEC{\thread_s} }\chor'}{\stypeEnv, k
     \colon G} \\
  & 
      \text{A.11, rule }    \Did{TGres} 
  &&
     \isSType[\typeEnv'\backslash k]{\new{\VEC{\thread_s},k
     }\chor'}{\stypeEnv } \\
    & \text{A.10, A.12, Rule } \Did{TG} &&
                                \isType[\typeEnv'\backslash k]{\ltypeEnv'}{\new{\VEC{\thread_s},k
     }\chor'}{\stypeEnv }
  \end{align}
  
\Case{} Rule $\Did{^G|Bcast}$:
  \setcounter{equation}{0}
  \begin{align}
    & \text{Hypothesis} &&          \begin{array}{l}
                                      \conf{\sigma}{    \left(
                                      \bcastA{
                                      \thread_A[A]\{\Blue{ X_A;\!Y_A}\}
                                      }{\VEC{\thread_r[B_r]\{\Blue{
                                      X_r;\!Y_r}\}\!:\! x_r}}{q}{e}{k}
                                      \right) \!\pfx\! \chor' 
                                      } \\ \qquad \qquad \qquad 
                                      \action{    
                                      \theta(        \bcastA{
                                      \thread_A[A]\{\Blue{ X_A;Y_A}\}
                                      }{\VEC{\thread_r[B_r]\{\Blue{X_r;Y_r}\}:x_r}}{q}{v}{k})
                                      }\ 
                                      \conf{\sigma[\sigma']}{\theta(\chor') }
                                    \end{array}
    \\
    & \text{Hypothesis} &&
                           \isType{\ltypeEnv}{  \left( \bcastA{\thread_A[A]\Blue{\{X_A;Y_A
                           \}}}{\VEC{\thread_r[B_r]\Blue{\{X_r;Y_r\} }:x_r}}{q}{e}{k}
                           \right) \pfx \chor'}{\stypeEnv}\\
    & \text{Hypothesis} && \isState{\sigma}{\ltypeEnv} \\
    & \text{A.2, inversion} && \isSType{
                               \left( \bcastA{\thread_A[A]\Blue{\{X_A;Y_A
                               \}}}{\VEC{\thread_r[B_r]\Blue{\{X_r;Y_r\} }:x_r}}{q}{e}{k}
                               \right) \pfx \chor'  }{\stypeEnv', k \colon \left(
                               \bcastT{A}{\VEC B}{S} \pfx G \right)
                               }
  \end{align}
  \begin{align}
    & \text{A.1, inversion} &&         J \subseteq \VEC{\thread_r}\\
    & \text{A.1, inversion} &&                 q(J)\\
    & \text{A.1, inversion} &&                 \forall {\iset
                               i{\{\thread_A\} \cup J}}: \Blue{ X_i}
                               \subseteq        \sigma(\thread_i, k)
                               \land        \sigma'(t_i, k) =
                               \exchange{\Blue{ X_i}}{\Blue{
                               Y_i}}(\sigma(t_i,k)) \\
    & \text{A.1, inversion} &&                 \evalOp{e@\thread_A}{ v }\\
    & \text{A.1, inversion} &&                 \forall {\irange
                               {i}{1,}{,|\VEC{\thread_r}|}}:
                               \theta(x_{i}) = 
                               \left\{
                               \begin{array}{l} 
                                 \some{v} ~~ \thread_i \in J  \\
                                 \none \qquad \text{o.w.}
                               \end{array}
    \right.\\
    & \text{A.4, inversion} &&
                               \isIndex[\typeEnv]{ \thread_A} {
                               \ownerT{k}{A} }  \\
    & \text{A.4, inversion} &&                 \isIndex[\typeEnv]{
                               \thread_i} { \ownerT {k}{B_i} }  \quad \irange{i}{1,}{,|\VEC{\thread_r}|}\\
    & \text{A.4, inversion} &&
                               \isIndex{e@\thread_A}{S} \\
    & \text{A.4, inversion} &&                 \isSType[\typeEnv,
                               \VEC{x_r@\thread_r[B_r]} \colon
                               S]{\chor'}{\stypeEnv', k \colon G } \\
    & \text{A.8, A.9, A.12, A.13, lem.}\ref{lem:global-substitution}
                            &&
                               \isSType[\typeEnv]{\theta(\chor')}{\stypeEnv',
                               k \colon G } \\
    & \text{A.4, A.14, rule } \Did{^G|Bcast}
                            &&\stypeEnv', k \colon \left(\bcastT{A}{\VEC
                               B}{S} \pfx G \right) \action{k \colon
                               \alpha } \stypeEnv', k \colon G \quad 
                               {\small \alpha = \bcastT{A}{\VEC{B}}{S}}\\
    & \text{A.1, A.9, A.11, A.12, rule } \Did{^L|Bcast}
                            && \isLType{
                               \bcastA{
                               \thread_A[A]
                               }{\VEC{\thread_r[B_r]:sb_r}}{q}{v}{k}
                               }
                               {k \colon \bcastT {A}{\VEC{B}}S }
  \end{align}

\Case{} Rule $\Did{^G|Sel}$: The case is analogous to the one above
(excluding substitutions)

\Case{} Rule $\Did{^G|Red}$: The case is analogous to the $\Did{^G|Bcast}$.

\Case{} Rule $\Did{^G|Res}$: This case follows straightforwardly after
application of the induction hypothesis.

\Case{} Rule $\Did{^G|If}$:
  \setcounter{equation}{0}
  \begin{align}
    & \text{Hypothesis} &&                                  \conf{\sigma}{\ifthenelsek{e@\thread}{\chor_1}{\chor_2}} 
                            \action{\tau} 
                            \conf{\sigma}{\chor_i } \\
    & \text{Hypothesis} &&
                           \isType{\ltypeEnv}{ \ifthenelsek
                           {e@\thread} {\chor_1}
                           {\chor_2}}{\stypeEnv}\\
    & \text{Hypothesis} && \isState{\sigma}{\ltypeEnv} \\
    & \text{A.2, inversion} && \isSType{ \ifthenelsek {e@\thread} {\chor_1} {\chor_2}
        }{\stypeEnv}\\
    & \text{A.2, inversion} && \isEntailed{ \ifthenelsek {e@\thread} {\chor_1} {\chor_2}
        }\\
    & \text{A.4, inversion} && \isSType{ \chor_1}{\stypeEnv}\\
    & \text{A.4, inversion} && \isSType{ \chor_2}{\stypeEnv}
  \end{align}
Assume $\evalOp{e@\thread}{\true}$ (the other case is analogous). 
\begin{align}
  & \text{case} &&
                         \conf{\sigma}{\ifthenelsek{e@\thread}{\chor_1}{\chor_2}}
                         \action{\tau} \conf{\sigma}{\chor_1 } \\ 
  & \text{A.3, A.5, A.8, thm. } \ref{thm:type-preservation} &&
                                                               \isEntailed[\ltypeEnv']{\chor_1}\\ 
  & \text{A.3, A.5, A.8, thm. } \ref{thm:type-preservation} &&
                                                               \isState{\sigma'}{\ltypeEnv'} \\
    & \text{A.6, A.9, rule } \Did{TG} && \isType[\typeEnv]{\ltypeEnv'}{ \chor_1 }{\stypeEnv}
\end{align}

\Case{} Rule $\Did{^G|Cong}$:
  \setcounter{equation}{0}
  \begin{align}
    & \text{Hypothesis} &&      \conf{\sigma}{\chor}
                     \action{\lambda}
                     \conf{\sigma'}{\chor'''} \\
    & \text{Hypothesis} &&
                           \isType{\ltypeEnv}{ \chor}{\stypeEnv}\\
    & \text{Hypothesis} && \isState{\sigma}{\ltypeEnv} \\
    & \text{A.1, inversion} && \chor \,\mathcal R\, \chor' \\
    & \text{A.1, inversion} && \conf{\sigma}{\chor'} \action{\lambda}
                               \conf{\sigma'}{\chor''} \\ 
    & \text{A.1, inversion} && \chor'' \,\mathcal R\, \chor''' \\
    & \text{A.1, inversion} && \mathcal R \in \{ \equiv, \swaps\} 
  \end{align}
 From eq. A.7 we have two sub-cases. The case for $\equiv$ follows from
 application of Lemma \ref{lem:global-congruence}, and the case for
 $\swaps$ follows from the application of Lemma \ref{lem:global-swap}.

\qed

\subsection{Results Related to the Endpoint Projection}

\begin{lemma}[Substitution: Process Projection] \label{lem:subst-process-proj}
  $\encod{\chor\subst{v}{x@p}}{p} = \encod{\chor}{p}\subst{v}{x@p}$.
\end{lemma}
\begin{proof}
It follows directly from Definition \ref{def:process-projection}.
\end{proof}

\begin{lemma}[Substitution: Projection Locality]\label{lem:subst-locality}
Let $\chor = \new{\VEC{p},\VEC{k}}\chor'$,  $q \in \ft{\chor'}$, and\\ 
$\project{\chor} = \new{\VEC{k}} \left(\prod_{p \in \ft{\chor'}}
                       \encod{\chor'}{p} \pp
                       \prod_{k \in \fsc{\chor'} }
                       \queue{k}{\emptyset} \right)
                       \pp \prod_{a
                       \in \sv{\chor'},A \in \roles{\chor'}} \left(
                       \bigsqcup_{p \in \serviceMerge{\chor'}{a}{A}} \encod{\chor'}{p}
                       \right)  $. Then
                       \\$\begin{array}{ll}\project{\chor\subst{v}{x@q}}
                            =& \new{\VEC{k}}  \left( \encod{\chor'\subst{v}{x@q}}{q} \pp \prod_{p \in \fn(\chor)\backslash q}
                       \encod{\chor'}{p} \pp
                       \prod_{k \in \fn(\chor') }
                       \queue{k}{\emptyset} \right) \\ &
                       \pp \prod_{a
                       \in \sv{C'},A \in \roles{\chor'}} \left(
                       \bigsqcup_{p \in \serviceMerge{\chor'}{a}{A}} \encod{\chor'}{p}
                       \right) .\end{array} $
\end{lemma}
\begin{proof}
From Definition \ref{def:epp} we know that there is only one thread
projection for $q$ in $\project{\chor}$. The rest follows directly
from Lemma
\ref{lem:subst-process-proj}.
\end{proof}

\begin{lemma}[$\equiv$ Preserves Linearity]
  If $\chor \equiv \chor'$ and $\chor$ is linear, then $\chor'$ is linear.
\end{lemma}
\begin{proof}
  Follows by rule induction on the derivation of the hypothesis.
\end{proof}


\begin{lemma}[Projection Congruence] \label{lem:projection-congruence}
  If $\chor \equiv \chor'$ then $\project{\chor} \equiv \project{\chor'}$.
\end{lemma}
\begin{proof}
  It follows by rule induction on the rules for $\chor \equiv \chor'$.
\end{proof}

\begin{lemma}[Session Linearity] \label{lem:session-linearity}
  If $\isType{\ltypeEnv}{\chor}{\stypeEnv}$, and there exists $P, Q$,
  s.t.  $\project{\chor}
  \equiv \new{\VEC{k}} (P \pp Q)$. We have either
  \begin{itemize}
    \item $P =
  \bcast{k}{A}{\VEC{B}}{q}{e} P'$, and $Q$ does not contain actions
  with free subject $ k[A]![\VEC{B}]$, or
    \item $P =
  \colSelection{k}{A}{\VEC{B}}{q}{l} P'$, and $Q$ does not contain actions
  with free subject $ k[A]![\VEC{B}]$, or
  \item $P =
      \reduce{k}{\VEC{A}}{B}{q}{x}{op} \pfx P'$, and $Q$ does not contain actions
  with free subject $ k[\VEC{A}]?[B]$.
\end{itemize}
\end{lemma}
\begin{proof}
  We analyze each of the cases separately. From
  $\isType{\ltypeEnv}{\chor}{\stypeEnv}$ and $P =
  \bcast{k}{A}{\VEC{B}}{q}{e} P'$ we know that $\chor$ corresponds to
  $\bcastA{
    \thread_A[A]\{\Blue{ X_A;\!Y_A}\}
    }{\VEC{\thread_r[B_r]\{\Blue{ X_r;\!Y_r}\}\!:\! x_r}}{q}{e}{k} \!\pfx\! \chor'$. Moreover, performing inversion in the
  typing rule for broadcast allow us to conclude that there is only
  one judgment for $\isIndex[\typeEnv]{ \thread_A} { \ownerT{k}{A}
  }$. A similar reasoning is performed for Reduce and Collective selections.
\end{proof}

\begin{lemma}[$\swaps$ Preserves Linearity]
  If $\chor \swaps \chor'$ and $\chor$ is linear, then $\chor'$ is linear.
\end{lemma}
\begin{proof}
  It follows by rule induction on the rules for $\chor \swaps \chor'$.
\end{proof}

We say that $a[A]$ \emph{is enabled in} $P$ if $P$ contains a sub-term
$\initOut {a}{k}\pfx P$ or  $ \repInitIn {a}{k}\pfx P$.

\begin{lemma}[Projections of linear choreographies do not introduce races]\label{lem:linearity}
  Let $\chor$ be a linear choreography, and $\project{\chor}
  \action{\VEC{\emm}} P$ for a finite $\VEC{\emm}$. If $P$ contains a sub-term $\initIn{a}{k}\pfx
  Q$, then there exists at most one sub-term $P'$ in $P$ s.t. $a[A]$ is
  enabled in $P'$ for any $A$.
\end{lemma}
\begin{proof}
Since $P$ contains a
  sub-term $\initIn{a}{k}\pfx
  Q$ then we know that $\chor$ contains a sub-term $\initA{
    \VEC{p}}{ \VEC{p}}{a}{k}$.   We proceed by induction on the length
  of $\VEC{\emm}$. 

\Case{} $|\VEC{\emm}| = 0$: then $\project{\chor} = P$. The interaction
  dependencies imposed by the linearity of $\chor$ say that 
   $\forall r \in \VEC{r}$ there is an interaction dependency with 
    threads in $\VEC{p}, \VEC{q}$.  $n_1 = \initA{
    \VEC{p}}{ \VEC{q}}{a}{k}$, $n_2 = \initA{
    \VEC{r}}{ \VEC{s}}{a}{k}$ in $\chor$. Then the thread projection
  of $r \in \VEC{r}$ is indeed a continuation of one of the projections of
  threads in $\VEC{p}, \VEC{q}$. Finally, the sub-term $\initIn{a}{k}\pfx
  Q$ corresponding to the thread projection of $r \in \VEC{r}$ are 
  disabled until the terms generated by the projection of $\initA{
    \VEC{p}}{ \VEC{q}}{a}{k}$ evolve. 

\Case{} $|\VEC{\emm}| \ge  0$: We assume that  $\project{\chor}
  \action{\VEC{\emm}} P' \action{\emm'} P$, and that there exists at
  most one sub-term $P''$ in $P'$ s.t. $a[A]$ is 
  enabled in $P''$ for any $A$. We check now all the possible
  derivation sequences for $P' \action{\emm'} P$. The interesting one
  refers to rule $\Did{^E|{Init}}$, that will consume  term $\initIn{a}{k}\pfx
  Q$ when reducing to $P$. The existence of at most one sub-term $P''$
  in $P$
  s.t. $a[A]$ is enabled in $P''$ from the same argument regarding
  linearity conditions as the case for $|\VEC{\emm}| =0 $.
\end{proof}

\begin{lemma}[Swapping: Endpoint Invariance]\label{lem:swapping-invariance-endpoint}
  If $\chor \swaps \chor'$, then $\project{\chor} = \project{\chor'}$.
\end{lemma}
\begin{proof}
The proof proceeds by rule induction on the swapping relation rules
in Figure \ref{fig:swap-rel}, followed by case analysis on the shape that $\chorAct$ can
  take. Here we present the case where $\chorAct =
  \bcastA{p_r}{\VEC{p_s:x_s}}{q}{e}{k}$. The other cases are similar.

\Case{} $\chorAct =
  \bcastA{p_r}{\VEC{p_s:x_s}}{q}{e_1}{k}$.

  \setcounter{equation}{0}
  \begin{align}
    & \text{Hypothesis} && \begin{array}{l}           \ifthenelsek {e_1@p} { 
  \bcastA{p_r}{\VEC{p_s:x_s}}{q}{e_1}{k} \pfx \chor_1 }
      {   \bcastA{p_r}{\VEC{p_s:x_s}}{q}{e_1}{k} \pfx \chor_2} \\ \qquad \swaps
        \bcastA{p_r}{\VEC{p_s:x_s}}{q}{e_1}{k} \pfx\ifthenelsek {e_1@p} { \chor_1} { \chor_2} 
                           \end{array}
    \\
    & \text{A.1, inversion} &&             p \notin \threads{  \bcastA{p_r}{\VEC{p_s:x_s}}{q}{e_1}{k}}
    \\
    & \text{A.1, def. } \project{\cdot} &&  \project{\chor}
                                           = \begin{array}{l}
                                               \new{\VEC{k}} \left(\prod_{p \in \ft{\chor'}}
                                               \encod{\chor'}{p} \pp
                                               \prod_{k \in \fsc{\chor'} }
                                               \queue{k}{\emptyset}
                                               \right) \\
                                               \pp \prod_{a
                                               \in \sv{\chor'},A \in \roles{\chor'}} \left(
                                               \bigsqcup_{p \in \serviceMerge{\chor'}{a}{A}} \encod{\chor'}{p}
                                               \right) 
                                             \end{array}\\
    & \text{A.3, def. } \encod{\cdot}{p} && \project{\chor} = \begin{array}{l}
                                                    \new{\VEC{k}} 
                                                    \left(              
                                                    \begin{array}{l}          
                                                      \itn{e}{\encod{\bcastA{p_r}{\VEC{p_s:x_s}}{q}{e_1}{k}
                                                      \pfx
                                                      \chor_1}{p}\\\quad}{
                                                      \encod{\bcastA{p_r}{\VEC{p_s:x_s}}{q}{e_1}{k}
                                                      \pfx
                                                      \chor_2}{p}} \\
                                                      \pp 
                                                      \prod_{p' \in
                                                      (\ft{\chor_1},\ft{\chor_2})\backslash
                                                      p}
                                                      \encod{\bcastA{p_r}{\VEC{p_s:x_s}}{q}{e_1}{k}
                                                      \pfx (
                                                      \chor_1 \mergek \chor_2)}{p'} 
                                                      \\
                                                      \pp
                                                      \prod_{k \in \fsc{\chor} }
                                                      \queue{k}{\emptyset}
                                                    \end{array}
                                                                \right) \\\qquad \quad
                                                    \pp \prod_{ a,A } \left(
                                                    \bigsqcup_{p
                                                    \in
                                                    \serviceMerge{\chor}{a}{A}}
                                                      \encod{\chor}{p}
                                                    \right)                             
                                                  \end{array}  
  \end{align}

  \begin{align}
  & \text{A.3, def. } \encod{\cdot}{p} && \project{\chor'} = \begin{array}{l}  \left(              
  \begin{array}{l}          
    \bcast{k}{A}{\VEC{B}}{q}{e} (
    \encod{\chor_1}{\thread_1} \mergek  \encod{\chor_2}{\thread_1})\\
    \pp 
    \prod_{i
    \in
    [1,|\VEC{\thread_r}|]}
    \receive{k}{B_i}{A}{x_i} (
    \encod{\chor_1}{\thread_i} \mergek  \encod{\chor_2}{\thread_i}) 
    \\
    \pp
    \prod_{\thread_h
    \in
    \ft{\chor'}\backslash
    {\VEC{\thread_r},\VEC{\thread_s}}}
    \encod{\ifthenelsek {e_1@p} {\chor_1}{\chor_2}}{\thread_h}\\
    \pp
    \prod_{k' \in \fsc{\chor'} }
    \queue{k'}{\emptyset}
    \pp 
    \queue{k}{\emptyset}
  \end{array} \right) \\
  \pp \prod_{ a,A } \left(
  \bigsqcup_{p
  \in
  \serviceMerge{\chor'}{a}{A}}
  \encod{\chor'}{p}                 \right)                             \end{array} 
  \end{align}
From A.4 and A.5 it suffices to observe that for $\thread_h \in
\ft{\chor'}\backslash\{\VEC{\thread_r},\VEC{\thread_s}\}$ the
projection do not change, and that for all the other processes, the
merge between continuations $\chor_1$ and $\chor_2$ is maintained. 
\end{proof}

\begin{lemma}[Passive Process Pruning Invariance] \label{lem:passive-invariance}
Assume a restriction-free $\chor$. If
$\ltransition{\conf{\sigma}{\chor}}{\lambda}{\conf{\sigma'}{\chor'}}$
then $\encod{\chor'}{p}
\prune \encod{\chor}{p}$, $\forall p\in \fn(\chor)\backslash \fn(\ell)$.
\end{lemma}
\begin{proof}
  The proof proceeds by rule induction on
  $\ltransition{\conf{\sigma}{\chor}}{\ell}{\conf{\sigma'}{\chor'}}$. Cases
 $\Did{^G|Init}, \Did{^G|Bcast}, \Did{^G|Sel}$, $\Did{^G|Red}$ are
  straightforward. In the conditional case, we have that $\conf{\sigma}{\ifthenelsek {e_1@p} { \chor_1} { \chor_2}}
  \action{\tau} \conf{\sigma}{\chor_1}$ (the other case is
  analogous). Then 
  \[
\begin{array}{l}
  \begin{array}{l}
    \left(\prod_{p \in \ft{\chor_1}}
    \encod{\chor_1}{p} \pp
    \prod_{k \in \fsc{\chor_1} }
    \queue{k}{\emptyset}
    \right) \\
    \pp \prod_{a
    \in \sv{\chor_1},A \in \roles{\chor_1}} \left(
    \bigsqcup_{p \in \serviceMerge{\chor_1}{a}{A}} \encod{\chor_1}{p}
    \right) 
  \end{array} \\ \qquad 
  \prune 
  \begin{array}{l}
    \left(\begin{array}{l}
            \itn{e}{\encod{\chor_1}{p}}{\encod{\chor_2}{p}} 
            \pp 
            \prod_{p' \in
            \{\ft{\chor_1},\ft{\chor_2}\}\backslash p}
            (\encod{  \chor_1}{p'} \mergek \encod{\chor_2}{p'} ) 
            \pp
            \prod_{k \in \fsc{\chor} }
            \queue{k}{\emptyset}
          \end{array}
    \right) \\
    \pp \prod_{a
    \in \{\sv{\chor_1},\sv{\chor_2}\},A \in \{\roles{\chor_1},\roles{\chor_2},p\}} \left(
    \bigsqcup_{p \in \serviceMerge{\chor'}{a}{A}}
    \encod{\ifthenelsek {e_1@p} { \chor_1} { \chor_2}}{p}
    \right) 
  \end{array}
\end{array}
\] Follows directly from the definition of pruning in Def. \ref{def:pruning}. The case for $\Did{^G|Cong}$ follows from application of Lemma
  \ref{lem:swapping-invariance-endpoint} for the case of swapping, and
  of Lemma \ref{lem:projection-congruence} for structural congruence.
\end{proof}

\theoremEPP* \label{epp-proof}
\begin{proof}
  {\bf On Soundness:}\\
  The proof proceeds by rule induction on the transition rules for
  $\action{\ell}$ in Figure \ref{fig:global:semantics}. We have seven
  cases:

  \Case{} Rule $\Did{^G|Init}$:
  \setcounter{equation}{0}
  \begin{align}
    & \text{Hypothesis} &&     \chor = \new{\VEC{k}, \VEC{p}}
                           \initA{\VEC{\thread_r[A_r]}}{\VEC{\thread_{s}[B_s]}}{a}{k} \pfx \chor_1\\
    & \text{Hypothesis} &&     \initA{\VEC{\thread_r[A_r]}}{\VEC{\thread_{s}[B_s]}}{a}{k} \pfx
                           \chor_1 \text{ is restriction-free} \\
    & \text{Hypothesis} &&     \initA{\VEC{\thread_r[A_r]}}{\VEC{\thread_s[B_s]}}{a}{k} \pfx
                           \chor_1 \text{ is linear}\\
    & \text{Hypothesis} &&     \isType{\ltypeEnv}{\new{\VEC{k},
                           \VEC{p}}
                           \initA{\VEC{\thread_r[A_r]}}{\VEC{\thread_s[B_s]}}{a}{k}
                           \pfx   \chor_1}{\stypeEnv}\\
    & \text{Hypothesis} &&     \isState{\sigma}{\ltypeEnv}\\
    & \text{Hypothesis} &&                             \begin{array}{l}       \conf{\sigma}{\new{\VEC{k},
                                                         \VEC{p}} (\initA{
                                                         \VEC{\thread_r[A_r]\{\Blue{Y_r}\}}}{
                                                         \VEC{\thread_s[B_s]\{\Blue{ Y_s}\}}}{a}{k}
                                                         \pfx \chor_1)}
                                                         \action{\ell} \\ \qquad \qquad \qquad \qquad \qquad \qquad 
                                                         \conf{\sigma[\sigma'[\sigma'']]}{\chor''}
                                                       \end{array}\\
    & \text{A.6} &&                \chor'' = \new{\VEC{k},\VEC{p}}(\new{k,\VEC{\thread_s}}(\chor_1)) \\
    & \text{A.6, inversion} &&                \ell =\initA{ \VEC{\thread_r[A_r]\{\Blue{Y_r}\}}
                               }{ \VEC{\thread_s[B_s]\{\Blue{ Y_s}\}}}{a}{k}  \\
    & \text{A.6, inversion} &&                        \sigma' = [(\thread_i,k)  |-> \Blue{ Y_i} ]_{i = 1}^{|\VEC{\thread_r}|} \\
    & \text{A.6, inversion} &&                        \sigma'' =
                               [(\thread_i,k)  |-> \Blue{ Y_i} ]_{i =
                               1}^{|\VEC{\thread_s}|} \\
    & \text{A.1, def. } \project{\cdot} &&  \project{\chor}
                                           = \begin{array}{l}
                                               \new{\VEC{k}} \left(\prod_{p \in \ft{\chor'}}
                                               \encod{\chor'}{p} \pp
                                               \prod_{k \in \fsc{\chor'} }
                                               \queue{k}{\emptyset}
                                               \right) \\
                                               \pp \prod_{a
                                               \in \sv{\chor'},A \in \roles{\chor'}} \left(
                                               \bigsqcup_{p \in \serviceMerge{\chor'}{a}{A}} \encod{\chor'}{p}
                                               \right) 
                                             \end{array}\\
    & \text{A.10, exp. } &&  
                            \begin{array}{l}\project{\chor} = \new{\VEC{k}}  \left( \begin{array}{l}
                                                      \initIn{a[\VEC{A},\VEC{B}]}{k} \pfx\encod{\chor_1}{\thread_1}
                                                      \pp 
                                                      \prod_{i \in [2,
                                                      |\VEC{\thread_r}|\,]
                                                      }
                                                      \initOut{a[A_i]}{k}
                                                      \pfx
                                                      \encod{\chor_1}{\thread_i} 
                                                      \\
                                                      \pp 
                                                      \prod_{j \in
                                                      [1,|\VEC{\thread_s}|\,]}
                                                      \repInitIn{a[B_j]}{k}
                                                      \pfx \encod{\chor_1}{\thread_j}
                                                      \pp
                                                      \prod_{\thread_k \in \ft{\chor_1}
                                                      \backslash \{\VEC{\thread_r},
                                                      \VEC{\thread_s}} \encod{\chor_1}{\thread_k}
                                                      \\\pp
                                                      \prod_{k \in \fsc{\chor'} }
                                                      \queue{k}{\emptyset} 
                                                    \end{array}
                              \right)
                              \\ \qquad \quad 
                              \pp \prod_{a
                              \in \sv{\chor'},A \in \roles{\chor'}} \left(
                              \bigsqcup_{p \in \serviceMerge{\chor'}{a}{A}} \encod{\chor'}{p}
                              \right) \\\qquad \quad = P_1
                            \end{array} \\
    & \text{A.11, exp. } \serviceMerge{\cdot}{a}{A} && 
                                                       \begin{array}{l}  P_1 = \new{\VEC{k}} \left( \begin{array}{l}
                                                                                                      \initIn{a[\VEC{A},\VEC{B}]}{k}
                                                                                                      \pfx\encod{\chor_1}{\thread_1} 
                                                                                                      \pp 
                                                                                                      \prod_{i \in [2,
                                                                                                      |\VEC{\thread_r}|\,]}
                                                                                                      \initOut{a[A_i]}{k}
                                                                                                      \pfx
                                                                                                      \encod{\chor_1}{\thread_i} 
                                                                                                      \\
                                                                                                      \pp 
                                                                                                      \prod_{j \in
                                                                                                      [1,|\VEC{\thread_s}|\,]}
                                                                                                      \repInitIn{a[B_j]}{k}
                                                                                                      \pfx  \left(
                                                                                                      \bigsqcup_{p
                                                                                                      \in
                                                                                                      \serviceMerge{\chor'}{a}{A}}
                                                                                                      \encod{\chor_1}{p} 
                                                                                                      \right)  \\
                                                                                                      \pp
                                                                                                      \prod_{\thread_k \in \ft{\chor_1}
                                                                                                      \backslash \{\VEC{\thread_r},
                                                                                                      \VEC{\thread_s\}}} \encod{\chor_1}{\thread_k}
                                                                                                      \pp
                                                                                                      \prod_{k \in \fsc{\chor'} }
                                                                                                      \queue{k}{\emptyset} 
                                                                                                    \end{array}
                              \right)
                              \\\qquad \quad
                              \pp \prod_{a'
                              \in \sv{\chor'}\backslash a,A \in \roles{\chor'}} \left(
                              \bigsqcup_{p \in \serviceMerge{\chor'}{a'}{A}} \encod{\chor'}{p}
                              \right)  \\ \qquad \quad = P_2
                                                       \end{array} \\
    &  \begin{array}{l}\text{A.13, rules  }\\ \Did{^E|Init},
         \Did{^E|Res},  \end{array}   &&   \begin{array}{l}  P_2
                                           \action{\startAction
                                           {\VEC{A}}{\VEC{B}}ak }\\  
                                           \new{\VEC{k}} 
                                           \left( \begin{array}{l} \new{k'}\left(
                                                    \prod_{\thread_i \in
                                                    \VEC{\thread_r}, \VEC{\thread_s}}
                                                    \encod{\chor_1}{\thread_i}
                                                    \pp
                                                    \queue{k'}{\emptyset}
                                                    \right)
                                                    \\
                                                    \pp
                                                    \prod_{\thread_k \in \ft{\chor_1}
                                                    \backslash \{\VEC{\thread_r},
                                                    \VEC{\thread_s\}}} \encod{\chor_1}{\thread_k}
                                                    \pp
                                                    \prod_{k \in \fsc{\chor'} }(
                                                    \queue{k}{\emptyset})
                                                  \end{array}
                                           \right)
                                           \\
                                           \pp 
                                           \prod_{j \in
                                           [1,|\VEC{\thread_s}|\,]}
                                           \repInitIn{a[B_j]}{k}
                                           \pfx  \left(
                                           \bigsqcup_{p
                                           \in
                                           \serviceMerge{\chor'}{a}{A}}
                                           \encod{\chor_1}{p} 
                                           \right) \\
                                           \pp \prod_{a'
                                           \in \sv{\chor'}\backslash a,A \in \roles{\chor'}} \left(
                                           \bigsqcup_{p \in \serviceMerge{\chor'}{a'}{A}} \encod{\chor'}{p}
                                           \right) 
                                         \end{array} = P_3\\
    &  \text{A.14,  } \equiv  &&   \begin{array}{l}  P_3 \equiv  
                                           \new{\VEC{k},k'} 
                                           \left( \begin{array}{l} 
                                                    \prod_{\thread_i \in
                                                    \VEC{\thread_r}, \VEC{\thread_s}}
                                                    \encod{\chor_1}{\thread_i}
                                                    \pp
                                                    \prod_{\thread_k \in \ft{\chor_1}
                                                    \backslash \{\VEC{\thread_r},
                                                    \VEC{\thread_s\}}}
                                                    \encod{\chor_1}{\thread_k}
                                                    \\
                                                    \pp
                                                    \prod_{k \in \fsc{\chor'},k' }(
                                                    \queue{k}{\emptyset})
                                                  \end{array}
                                           \right)
                                           \\
                                           \pp 
                                           \prod_{j \in
                                           [1,|\VEC{\thread_s}|\,]}
                                           \repInitIn{a[B_j]}{k}
                                           \pfx  \left(
                                           \bigsqcup_{p
                                           \in
                                           \serviceMerge{\chor'}{a}{A}}
                                           \encod{\chor_1}{p} 
                                           \right) \\
                                           \pp \prod_{a'
                                           \in \sv{\chor'}\backslash a,A \in \roles{\chor'}} \left(
                                           \bigsqcup_{p \in \serviceMerge{\chor'}{a'}{A}} \encod{\chor'}{p}
                                           \right) 
                                         \end{array}
  \end{align}
Moreover, from Lemma
                                           \ref{lem:passive-invariance}
                                           we know that $\forall p \in
                                           \fn(\initA{\VEC{\thread_r[A_r]}}{\VEC{\thread_{s}[B_s]}}{a}{k}
                                             \pfx
                                             \chor_1)\backslash\{ \thread_r,\VEC{\thread_s},k\} $,
                                           $ \encod{\chor_1}{p}\prune \encod{\initA{\VEC{\thread_r[A_r]}}{\VEC{\thread_{s}[B_s]}}{a}{k} \pfx \chor_1}{p}$. The projection of $\chor''$  up to reordering and alpha
conversion in eq. A.7 is 
\begin{equation}
\project{\new{\VEC{k} k',\VEC{p} \VEC{\thread_s}}(\chor_1)}
                                           = \begin{array}{l}
                                               \new{\VEC{k},k'} \left(\prod_{p \in \ft{\chor_1}}
                                               \encod{\chor_1}{p} \pp
                                               \prod_{k \in \fsc{\chor_1} }
                                               \queue{k}{\emptyset}
                                               \right) \\
                                               \pp \prod_{a
                                               \in \sv{\chor_1},A \in \roles{\chor_1}} \left(
                                               \bigsqcup_{p \in \serviceMerge{\chor_1}{a}{A}} \encod{\chor_1}{p}
                                               \right) 
                                             \end{array}
                                           \end{equation} 
                                           That corresponds to
                                           eq. A.15. It is easy to see
                                           that $\isAProjection[\initA{ \VEC{\thread_r[A_r]\{\Blue{Y_r}\}}
                               }{ \VEC{\thread_s[B_s]\{\Blue{ Y_s}\}}}{a}{k}]{\startAction
                                           {\VEC{A}}{\VEC{B}}ak }$
                                         from the rules in Figure
                                         \ref{fig:label:projection}.

  \Case{} Rule $\Did{^G|Red}$:
  \setcounter{equation}{0}
  \begin{align}
    & \text{Hypothesis} &&     \chor = \new{\VEC{k}, \VEC{p}}
                           \reduceA{\VEC{\thread_r[A_r]\{\Blue{
    X_r;Y_r}\}.e_r}}{\thread_B[B]\{\Blue{ X_B;Y_B}\}:x}{q}{k}{op} \pfx \chor_1\\
    & \text{Hypothesis} &&     \reduceA{\VEC{\thread_r[A_r]\{\Blue{
    X_r;Y_r}\}.e_r}}{\thread_B[B]\{\Blue{
    X_B;Y_B}\}:x}{q}{k}{op} \pfx \chor_1 \text{ is restriction-free} \\
    & \text{Hypothesis} &&     \reduceA{\VEC{\thread_r[A_r]\{\Blue{
    X_r;Y_r}\}.e_r}}{\thread_B[B]\{\Blue{
    X_B;Y_B}\}:x}{q}{k}{op} \pfx \chor_1 \text{ is linear}\\
    & \text{Hypothesis} &&     \isType{\ltypeEnv}{\new{\VEC{k},
                           \VEC{p}}
                           \reduceA{\VEC{\thread_r[A_r]\{\Blue{
                           X_r;Y_r}\}.e_r}}{\thread_B[B]\{\Blue{
                           X_B;Y_B}\}:x}{q}{k}{op} \pfx \chor_1}{\stypeEnv}\\
    & \text{Hypothesis} &&     \isState{\sigma}{\ltypeEnv}\\
    & \text{Hypothesis} &&             \begin{array}{l}
    \conf{\sigma}{  \reduceA{\VEC{\thread_r[A_r]\{\Blue{
    X_r;Y_r}\}.e_r}}{\thread_B[B]\{\Blue{
    X_B;Y_B}\}:x}{q}{k}{op} \pfx \chor_1\ }
    \action{ \ell}\ \\ \qquad \qquad \qquad \qquad \qquad \qquad \qquad \qquad \qquad \qquad \qquad \qquad \qquad 
    \conf{\sigma[\sigma']}{\chor''}
    \end{array} \\
    & \text{A.6, inversion} &&                  J \subseteq \VEC{t_r} \\
    & \text{A.6, inversion} &&                q(J) \\
    & \text{A.6, inversion} &&                (\evalOp{e_i@\thread_i}{ v_i })_{\iset{\thread_i}{J}} \\
    & \text{A.6, inversion} &&                                 \forall {\iset {\thread_i}{\{\thread_B\} \cup J}}: \Blue{ X_i} \subseteq
                 \sigma(\thread_i, k) \land
                 \sigma'(\thread_i, k) = \exchange{\Blue{ X_i}}{\Blue{
                 Y_i}}(\sigma(\thread_i,k)) \\
    & \text{A.6, inversion} &&                                {\forall {\irange {i}{1,}{,|\VEC{\thread_r}|}}: \theta(x_i) =
                 \left\{
                 \begin{array}{l} 
                   \some{v_i} ~~ \thread_i \in J  \\
                   \none \qquad \text{o.w.}
                 \end{array}
                 \right.}
  \\
    & \text{A.6, inversion} &&
                               \evalOp{\mathsf{op}(\theta)}{ \some{v}
                               }         \\
    & \text{A.6, inversion} &&                \chor'' = \new{\VEC{k},\VEC{p}}\chor_1 \subst{\some{v}}{x@\thread_{B}} \\
    & \text{A.6, inversion} &&                \ell =\reduceA{\VEC{\thread_r[A_r]\{\Blue{
                 X_r;Y_r}\}.v_r}}{\thread_B[B]\{\Blue{
                 X_B;Y_B}\}: \some{v}}{q}{k}{op}   \\
    & \text{A.1, def. } \project{\cdot} &&  \project{\chor}
                                           = \begin{array}{l}
                                               \new{\VEC{k},k} \left(\prod_{p \in \ft{\chor'}}
                                               \encod{\chor'}{p} \pp
                                               \prod_{k \in \fsc{\chor'} }
                                               \queue{k}{\emptyset}
                                               \right) \\
                                               \pp \prod_{a
                                               \in \sv{\chor'},A \in \roles{\chor'}} \left(
                                               \bigsqcup_{p \in \serviceMerge{\chor'}{a}{A}} \encod{\chor'}{p}
                                               \right) = P_1
                                             \end{array}\\
    & \text{A.15, exp. } &&  P_1
                                           = \begin{array}{l}
                                               \new{\VEC{k},k}
                                               \left(
                                               \begin{array}{l}
                                                 \prod_{i \in [1,                                                 |\VEC{\thread_r}| ]}
                                                 \send{k}{A_i}{B}{e_i}
                                                 \encod{\chor_1}{\thread_i} 
                                               \pp
                                               \reduce{k}{\VEC{A}}{B}{q}{x}{op}
                                               \pfx
                                               \encod{\chor_1}{\thread_B}\\
                                               \pp
                                               \queue{k}{\emptyset}
                                               \pp \prod_{p \in
                                                 \ft{\chor'}\backslash
                                                 \VEC{\thread_r},\thread_B}
                                               \encod{\chor_1}{p} 
                                               \pp
                                               \prod_{k' \in
                                                 \fsc{\chor'}
                                                 \backslash k}
                                               \queue{k'}{\emptyset}
                                               \end{array}
                                               \right) \\
                                               \pp \prod_{a
                                               \in \sv{\chor'},A \in \roles{\chor'}} \left(
                                               \bigsqcup_{p \in \serviceMerge{\chor'}{a}{A}} \encod{\chor'}{p}
                                               \right) = P_2
                                             \end{array} 
  \end{align}
Then we can proceed by moving the reduce into a blocking state,
updating the session queue with information regarding expected senders.
\begin{align}
  & \begin{array}{l} \text{A.16, rules } \\\Did{^E|Rd.I}, 
        \Did{^E|Res}
      \end{array}
    &&  \begin{array}{l}P_2       \action{ \asynchEnqueueI} \\
                                              \new{\VEC{k},k}
                                              \left(
                                              \begin{array}{l}
                                                \prod_{i \in [1,                                                 |\VEC{\thread_r}| ]}
                                                \send{k}{A_i}{B}{e_i}
                                                \encod{\chor_1}{\thread_i} 
                                                \pp
                                                \waitInput{k}{\VEC{A}}{B}{op}{x}
                                                \encod{\chor_1}{\thread_B} \\
                                                \pp
                                                \queue{k}{  \qmsgOInitial { A}{B}{q}{\none}{\false}    }\\
                                                \pp \prod_{p \in
                                                \ft{\chor'}\backslash
                                                \VEC{\thread_r},\thread_B}
                                                \encod{\chor_1}{p} 
                                                \pp
                                                \prod_{k' \in
                                                \fsc{\chor'}
                                                \backslash k}
                                                \queue{k'}{\emptyset}
                                              \end{array}
                                              \right) \\
                                              \pp \prod_{a
                                              \in \sv{\chor'},A \in \roles{\chor'}} \left(
                                              \bigsqcup_{p \in \serviceMerge{\chor'}{a}{A}} \encod{\chor'}{p}
                                              \right) = P_3
                                            \end{array} 
  \end{align}

From eq. A.7 we know that for a subset $J$ of threads the quality
predicate $q$ is deemed to be satisfied. We proceed by firing enough
sender to guarantee $q(J)$:
\begin{align}
  & \begin{array}{l} \text{A.9, A.17, rules } \\\Did{^E|Rd.I}, 
        \Did{^E|Res}~J \text{ times }
      \end{array}
    &&  \begin{array}{l}P_3             \action{
          \aReduceO{A_1}{B}{k}{\some{v_1}} } \ldots       \action{ \aReduceO{A_j}{B}{k}{\some{v_j}} }\\
                                              \new{\VEC{k},k}
                                              \left(
                                              \begin{array}{l}
                                                \prod_{i \in [1, |J| ]}
                                                \encod{\chor_1}{\thread_i} 
                                                \pp
                                                \prod_{\thread_i \in
                                                \VEC{\thread_r}\backslash
                                                J}
                                                \send{k}{A_i}{B}{e_i}
                                                \encod{\chor_1}{\thread_i} \\
                                                \pp
                                                \waitInput{k}{\VEC{A}}{B}{op}{x}
                                                \encod{\chor_1}{\thread_B} \\
                                                \pp        \queue
                                                k{(q:
                                                <. \VEC{A':\true:sb},\VEC{A'':\false:\none}
                                                .> , B) }\\ 
                                                \pp \prod_{p \in
                                                \ft{\chor'}\backslash
                                                \VEC{\thread_r},\thread_B}
                                                \encod{\chor_1}{p} 
                                                \pp
                                                \prod_{k' \in
                                                \fsc{\chor'}
                                                \backslash k}
                                                \queue{k'}{\emptyset}
                                              \end{array}
                                              \right) \\
                                              \pp \prod_{a
                                              \in \sv{\chor'},A \in \roles{\chor'}} \left(
                                              \bigsqcup_{p \in \serviceMerge{\chor'}{a}{A}} \encod{\chor'}{p}
                                              \right) = P_s
                                            \end{array} 
  \end{align} 
Where $\VEC{A} = \VEC{A'},\VEC{A''}$ and $\forall <.A: \true:sb
.> \in \VEC{<.A': \true:sb
.>}$ has the form of $\some{v_i}$. Now we are in position to
unpause the reduce process and apply the continuation to senders not used
in the reduce. 

\begin{align}
  & \begin{array}{l} \text{A.9, A.12, A.18, rules } \\\Did{^E|Wait_i}, 
        \Did{^E|Res}
      \end{array}
    &&  \begin{array}{l}P_s                    \action{ \aReduceI{\VEC{A}}{B}{q}{k}{\some{v}}  }\\
                                              \new{\VEC{k},k}
                                              \left(
                                              \begin{array}{l}
                                                \prod_{i \in [1, |J| ]}
                                                \encod{\chor_1}{\thread_i} 
                                                \pp
                                                \prod_{\thread_i \in
                                                \VEC{\thread_r}\backslash
                                                J}
                                                \encod{\chor_1}{\thread_i} \\
                                                \pp
                                                \encod{\chor_1\subst{\some{v}}{x}}{\thread_B} 
                                                \pp        \queue{k}{\emptyset}\\ 
                                                \pp \prod_{p \in
                                                \ft{\chor'}\backslash
                                                \VEC{\thread_r},\thread_B}
                                                \encod{\chor_1}{p} 
                                                \pp
                                                \prod_{k' \in
                                                \fsc{\chor'}
                                                \backslash k}
                                                \queue{k'}{\emptyset}
                                              \end{array}
                                              \right) \\
                                              \pp \prod_{a
                                              \in \sv{\chor'},A \in \roles{\chor'}} \left(
                                              \bigsqcup_{p \in \serviceMerge{\chor'}{a}{A}} \encod{\chor'}{p}
                                              \right) 
                                            \end{array} 
  \end{align} 
   By applying the endpoint projection function on the redex of
  eq. A.6.:

  \begin{align}
    & \begin{array}{l}\text{A.6, def. } \project{\cdot}, \text{Lemmas}\\
                         \ref{lem:subst-process-proj}, \ref{lem:subst-locality} 
                       \end{array}
    && 
       \begin{array}{l}
         \project{\new{\VEC{k}\VEC{p}}
         \chor_1\subst{\some{v}}{x@\thread_B}} =\\ 
         \qquad 
         \new{\VEC{k}} \left(\begin{array}{l}
                         \encod{\chor_1\subst{\some{v}}{x}}{\thread_B}
                         \pp
                         \prod_{p \in \ft{\chor'}\backslash
                               \thread_B} \encod{\chor_1}{p} 
                               \\
                         \pp
                         \prod_{k \in \fsc{\chor_1} }
                         \queue{k}{\emptyset}
                       \end{array}
         \right) \\          \qquad 
         \pp \prod_{a
         \in \sv{\chor'},A \in \roles{\chor'}} \left(
         \bigsqcup_{p \in \serviceMerge{\chor'}{a}{A}} \encod{\chor'}{p}
         \right) 
       \end{array}
  \end{align}
That corresponds to Eq. A. 19 up - to pruning. To prove behavioral
implementation, we just need to check that the labels generated in
eq. A.17, A.18 and A.19 correspond to the $\ell$ generated in
eq. A.6. This follows after application of rule $\Did{^L|Red}$.

  \Case{} Rule $\Did{^G|Bcast}$: This case corresponds to the same
  equivalent class as the one from  $\Did{^G|Red}$. Its proof is shown
  above. 

  \Case{} Rule $\Did{^G|Sel}$: This case is analogous to
  broadcast. Recall that $q$ for collective selection has been
  restricted to $\forall$. Such restriction is fundamental to
  guarantee that after all endpoints have chosen their branch, the
  selector can continue. 

  \Case{} Rule $\Did{^G|If}$:
  \setcounter{equation}{0}
  \begin{align}
    & \text{Hypothesis} &&     \chor = \new{\VEC{k}, \VEC{p}}
                           \ifthenelsek{e@\thread}{\chor_1}{\chor_2}\\
    & \text{Hypothesis} &&     \ifthenelsek{e@\thread}{\chor_1}{\chor_2} \text{ is restriction-free} \\
    & \text{Hypothesis} &&     \ifthenelsek{e@\thread}{\chor_1}{\chor_2} \text{ is linear}\\
    & \text{Hypothesis} &&     \isType{\ltypeEnv}{\new{\VEC{k},
                           \VEC{p}} \ifthenelsek{e@\thread}{\chor_1}{\chor_2}}{\stypeEnv}\\
    & \text{Hypothesis} &&     \isState{\sigma}{\ltypeEnv}\\
    & \text{Hypothesis} &&  \conf{\sigma}{\ifthenelsek{e@\thread}{\chor_1}{\chor_2}} 
                     \action{\tau} 
                     \conf{\sigma}{\chor_i } \\
    & \text{Hypothesis} &&  \project{\new{\VEC{k},
                           \VEC{p}}\ifthenelsek{e@\thread}{\chor_1}{\chor_2}}
                           \prune P
  \end{align}
  Let us assume $\evalOp{e@\thread}{\true} $ (the other case is
  analogous).
  \begin{align}
    & \text{A.1, }\evalOp{e@\thread}{\true} &&     \conf{\sigma}{\new{\VEC{k}, \VEC{p}}
                           \ifthenelsek{e@\thread}{\chor_1}{\chor_2}} 
                                               \action{}
                                               \conf{\sigma}{\new{\VEC{k},
                                               \VEC{p}}\chor_1} \\
    & \text{A.7, def. } \project{\cdot} &&     
                                                 \project{\chor}
                                           = \begin{array}{l}
                                               \new{\VEC{k}} \left(\prod_{p \in \ft{\chor'}}
                                               \encod{\chor'}{p} \pp
                                               \prod_{k \in \fsc{\chor'} }
                                               \queue{k}{\emptyset}
                                               \right) \\
                                               \pp \prod_{a
                                               \in \sv{\chor'},A \in \roles{\chor'}} \left(
                                               \bigsqcup_{p \in \serviceMerge{\chor'}{a}{A}} \encod{\chor'}{p}
                                               \right) 
                                             \end{array} = P_1\\
    & \text{A.9, def. } \encod{\cdot}{p} && P_1 = \begin{array}{l}
                                                    \new{\VEC{k}} 
                                                    \left(              
                                                    \begin{array}{l}          
                                                      \itn{e}{\encod{\chor_1}{p}}{\encod{\chor_2}{p}}
                                                      \pp 
                                                      \prod_{p' \in
                                                      (\ft{\chor_1},\ft{\chor_2})\backslash
                                                      p}
                                                      \encod{\chor_1}{p'} \mergek \encod{\chor_2}{p'}
                                                      \\
                                                      \pp
                                                      \prod_{k \in \fsc{\chor'} }
                                                      \queue{k}{\emptyset}
                                                    \end{array} \right) \\
                                                    \pp \prod_{ a,A } \left(
                                                    \bigsqcup_{p
                                                    \in
                                                    \serviceMerge{\ifthenelsek {e@p} {\chor_1} {\chor_2}}{a}{A}}
                                                    \encod{\chor_1}{p} \mergek \encod{\chor_2}{p}
                                                    \right)     = P_2                        
                                                  \end{array}  \\
    & \text{A.10, rules } \Did{^E|if}, \Did{^E|Res} && P_2 \action{\tau} \begin{array}{l}
                                                    \new{\VEC{k}} 
                                                    \left(              
                                                    \begin{array}{l}          
                                                      \encod{\chor_1}{p}
                                                      \pp 
                                                      \prod_{p \in
                                                      (\ft{\chor_1},\ft{\chor_2})\backslash
                                                      p}
                                                      \encod{\chor_1}{p} \mergek \encod{\chor_2}{p}
                                                      \\
                                                      \pp
                                                      \prod_{k \in \fsc{\chor'} }
                                                      \queue{k}{\emptyset}
                                                    \end{array} \right) \\
                                                    \pp \prod_{ a,A } \left(
                                                    \bigsqcup_{p
                                                    \in
                                                    \serviceMerge{\ifthenelsek {e@p} {\chor_1} {\chor_2}}{a}{A}}
                                                    \encod{\chor_1}{p} \mergek \encod{\chor_2}{p}
                                                    \right)                   
                                                  \end{array}
  \end{align}
From Lemma \ref{lem:passive-invariance} we know that $\forall
p \in \fn(\chor) $, $ \project{\new{\VEC{k}, \VEC{p}}\chor_1} 
                       \prune \project{\new{\VEC{k},
                           \VEC{p}}\ifthenelsek{e@\thread}{\chor_1}{\chor_2}}$,
                       hence, by the application of the pruning Lemma
                       (lemma \ref{lem:pruning})  along hypothesis A.7
                       then we can conclude $ \project{\new{\VEC{k}, \VEC{p}}\chor_1} 
                       \prune P$. $\isAProjection[\tau]{\tau}$ follows
                       after application of Rule $\Did{^L|Tau}$.

  \Case{} Rule $\Did{^G|Cong}$: The case has two sub-cases, one for
  structural congruence and another for the swapping congruence. The
  first sub-case follows after application of Lemma
  \ref{lem:projection-congruence} and the second after
  application of Lemma \ref{lem:swapping-invariance-endpoint}

  {\bf On Completeness:}\\
  The proof proceeds by induction on the structure of $ \chor'$. We
  have six cases:\\
  \Case{}
  $\chor' = \initA{\VEC{p_r}}{\VEC{p_{s}}}{a}{k} \pfx \chor_1$:

  \setcounter{equation}{0}
  \begin{align}
    & \text{Hypothesis} &&     \chor = \new{\VEC{k}, \VEC{p}}
                           \initA{\VEC{\thread_r[A_r]}}{\VEC{\thread_{s}[B_s]}}{a}{k} \pfx \chor_1\\
    & \text{Hypothesis} &&     \initA{\VEC{\thread_r[A_r]}}{\VEC{\thread_{s}[B_s]}}{a}{k} \pfx
                           \chor_1 \text{ is restriction-free} \\
    & \text{Hypothesis} &&     \initA{\VEC{\thread_r[A_r]}}{\VEC{\thread_s[B_s]}}{a}{k} \pfx
                           \chor_1 \text{ is linear}\\
    & \text{Hypothesis} &&     \isType{\ltypeEnv}{\new{\VEC{k},
                           \VEC{p}}
                           \initA{\VEC{\thread_r[A_r]}}{\VEC{\thread_s[B_s]}}{a}{k}
                           \pfx   \chor_1}{\stypeEnv}\\
    & \text{Hypothesis} &&     \isState{\sigma}{\ltypeEnv}\\
    & \text{Hypothesis} &&    \project{\new{\VEC{k}, \VEC{p}}
                           \initA{\VEC{\thread_r[A_r]}}{\VEC{\thread_s[B_s]}}{a}{k} \pfx
                           \chor_1} \action{\emm_1} P\\
    & \text{6, def. } \project{\cdot} &&    \begin{array}{l}
                                              \project{\new{\VEC{k},
                                              \VEC{p}}
                                              \initA{\VEC{\thread_r[A_r]}}{\VEC{\thread_s[B_s]}}{a}{k}
                                              \pfx \chor_1}  =
                                              \\\new{\VEC{k}}
                                              \left(\prod_{p \in
                                              \ft{\chor_1},\VEC{\thread_r},\VEC{\thread_s}} 
                                              \encod{\initA{\VEC{\thread_r[A_r]}}{\VEC{\thread_s[B_r]
                                              }}{a}{k} \pfx 
                                              \chor_1}{p} \pp
                                              \prod_{k \in \fsc{\chor_1} }
                                              \queue{k}{\emptyset} \right) \\
                                              \pp \prod_{ a,A } \left(
                                              \bigsqcup_{p \in
                                              \serviceMerge{\initA{\VEC{p_r}}{\VEC{p_{s}}}{a}{k}
                                              \pfx 
                                              \chor_1}{a}{A}}
                                              \encod{\initA{\VEC{p_r}}{\VEC{p_{s}}}{a}{k}
                                              \pfx 
                                              \chor_1}{p}
                                              \right) 
                                            \end{array}
    \\
    & \text{7, def. } \encod{\cdot}{} && \label{eq.a7}   = \begin{array}{l}
                                                             \new{\VEC{k}} 
                                                             \left(              
                                                             \begin{array}{l}          
                                                               \initIn{a[\VEC{A},\VEC{B}]}{k}
                                                               \pfx
                                                               \encod{\chor_1}{\thread_1} 
                                                               \pp 
                                                               \prod_{i
                                                               \in
                                                               [2,|\VEC{\thread_r}|]}
                                                               \initOut{a[A_i]}{k}
                                                               \pfx
                                                               \encod{\chor_1}{\thread_i} 
                                                               \\
                                                               \pp \prod_{j \in [1, |\VEC{\thread_s}|]}
                                                               \repInitIn{a[B_j]}{k} \pfx
                                                               \encod{\chor_1}{\thread_j}
                                                               \pp
                                                               \prod_{\thread_h
                                                               \in
                                                               \ft{\chor'}\backslash
                                                               {\VEC{\thread_r},\VEC{\thread_s}}}
                                                               \encod{\chor_1}{\thread_h}
                                                               \\
                                                               \pp
                                                               \prod_{k \in \fsc{\chor'} }
                                                               \queue{k}{\emptyset}
                                                             \end{array} \right) \\
                                                             \pp \prod_{ a,A } \left(
                                                             \bigsqcup_{p
                                                             \in
                                                             \serviceMerge{\initA{\VEC{p_r}}{\VEC{p_{s}}}{a}{k}
                                                             \pfx 
                                                             \chor_1}{a}{A}}
                                                             \encod{\initA{\VEC{p_r}}{\VEC{p_{s}}}{a}{k}
                                                             \pfx 
                                                             \chor_1}{p}                 \right)                             \end{array} 
    \\
    & \text{A.1, A.5, rules} \Did{^G|Init}, \Did{^G|Res}
                        &&     \begin{array}{l}
                                 \conf{\sigma}{\chor}
                                 \action{\initA{ \VEC{\thread_r[A_r]\{\Blue{Y_r}\}}
                                 }{ \VEC{\thread_h[B_s]\{\Blue{
                                 Y_s}\}}}{a}{k}  }\\ \qquad \qquad \qquad \qquad \quad
                                 \conf{\sigma[\sigma'[\sigma'']]}{\new{\VEC{k},\VEC{p}}\new{k,\VEC{\thread_s}}\chor_1}
                               \end{array}
    \\
    & \text{A.1, A.5, rules} \Did{^G|Init}, \Did{^G|Res} &&             \sigma' = [(\thread_i,k)  |-> \Blue{ Y_i} ]_{i = 1}^{|\VEC{\thread_r}|}    \\
    & \text{A.1, A.5, rules} \Did{^G|Init}, \Did{^G|Res} &&
                                                            \sigma'' =
                                                            [(\thread_i,k)
                                                            |-> \Blue{
                                                            Y_i} ]_{i =
                                                            1}^{|\VEC{\thread_s}|} \\
    & \text{I.H. } && \forall \emm'_1. \project{\new{\VEC{k},
                      \VEC{p}}\chor_1} \action{\emm'_1}
                      P_1\\
    & \text{I.H. } &&  P_1 \action{\VEC{\emm'_2}}
                      P'_1\\
    & \text{I.H. } && \conf{\sigma_1}{\new{\VEC{k},
                      \VEC{p}}\chor_1} \action{\VEC{\ell'}} \conf{\sigma_2}{\chor_1'} \\
    & \text{I.H. } && \project{\chor_1'} \prune P'_1\\
    & \text{I.H. } &&
                      \isAProjection[\VEC{\ell'}]{\emm'_1,\VEC{\emm'_2}}
  \end{align}
  We have two cases:
  $\conf{\sigma}{\chor} \action{\initA{
      \VEC{\thread_r[A_r]\{\Blue{Y_r}\}} }{
      \VEC{\thread_s[B_s]\{\Blue{ Y_s}\}}}{a}{k} , \VEC{\ell'}}
  \conf{\sigma_2}{\chor_1'} $,
  or there exists $\chor_s$ s.t. $\chor \swaps \chor_s$, and
  $\conf{\sigma}{\chor_s} \action{\VEC{\ell'_1},\initA{
      \VEC{\thread_r[A_r]\{\Blue{Y_r}\}} }{
      \VEC{\thread_s[B_s]\{\Blue{ Y_s}\}}}{a}{k},\VEC{\ell'_2} }
  \conf{\sigma_2}{\chor_1'}$,
  and $\vec{\ell'} = \VEC{\ell'_1},\VEC{\ell'_2}$. From Lemma
  \ref{lem:swapping-invariance-endpoint} we know that
  $\project{\chor}$ and $\project{\chor_s}$ are the same, so we
  consider only one of them. We perform case analysis on the labels
  $\emm_1$ that our $\project{\chor}$ in eq. \ref{eq.a7} can generate.

  \subcase{ $\emm_1 = \startAction{\VEC{A}}{\VEC{B}}ak $}
  \begin{align}
    & \begin{array}{l}\text{A.8, rules }  \Did{^E|{Init}}, \\ \Did{^E|{Par}},
        \Did{^E|{Struct}}
      \end{array}
    &&     \begin{array}{l}\project{\chor} \action{\startAction
             {\VEC{A}}{\VEC{B}}ak }  \\ 
             \new{\VEC{k},
             \VEC{p}} \left(                  
             \begin{array}{l} \new{k} (
               \encod{\chor_1}{\thread_1}
               \pp 
               \prod_{i \in [2,|\VEC{\thread_r}|]} \encod{\chor_1}{\thread_i}
               \pp \prod_{j \in [1, |\VEC{\thread_s}|]}
               \encod{\chor_1}{\thread_j}\pp
               \queue k\emptyset 
               )  \\ 
               \pp
               \prod_{\thread_h
               \in
               \ft{\chor_1}\backslash{\VEC{\thread_r}, \VEC{\thread_s}}}
               \encod{\chor_1}{\thread_h}
               \pp \prod_{j \in [1, |\VEC{\thread_s}|]}
               \repInitIn{a[B_j]}{k} \pfx
               \encod{\chor_1}{\thread_j}
               \\
               \pp
               \prod_{k' \in \fsc{\chor_1} }
               \queue{k'}{\emptyset}
             \end{array}
             \right) \\                 
             \pp \prod_{ a,A } \left(
             \bigsqcup_{p \in \serviceMerge{\initA{\VEC{p_r}}{\VEC{p_{s}}}{a}{k} \pfx
             \chor_1}{a}{A}} \encod{\initA{\VEC{p_r}}{\VEC{p_{s}}}{a}{k} \pfx
             \chor_1}{p}\right)
           \end{array}\\
    & \begin{array}{l}\text{A.18, scope extr.,}\\\text{lemma } \ref{lem:linearity} 
      \end{array}
    &&     \begin{array}{l}\project{\chor} \action{\startAction
             {\VEC{A}}{\VEC{B}}ak }  \\ 
             \new{\VEC{k},
             \VEC{p}} \left(                  
             \begin{array}{l} \new{k} (
               \prod_{\thread_h
               \in
               \ft{\chor_1},\VEC{\thread_r}, \VEC{\thread_s}}
               \encod{\chor_1}{\thread_h}             
               \pp
               \prod_{k' \in \fsc{\chor_1},k }
               \queue{k'}{\emptyset}
               )
             \end{array}
             \right) \\ 
             \pp \prod_{ a,A } \left(
             \bigsqcup_{p \in \serviceMerge{\initA{\VEC{p_r}}{\VEC{p_{s}}}{a}{k} \pfx
             \chor_1}{a}{A}} \encod{\initA{\VEC{p_r}}{\VEC{p_{s}}}{a}{k} \pfx
             \chor_1}{p} \right) \\
             \pp \prod_{j \in [1, |\VEC{\thread_s}|]}
             \repInitIn{a[B_j]}{k} \pfx
             \encod{\chor_1}{\thread_j} \qquad = P_{11}
           \end{array}
    \\   & \text{A.15, def. } \project{\cdot} &&    \begin{array}{l}
                                                      \project{\new{\VEC{k},
                                                      \VEC{p}}\chor_1}  =
                                                      \new{\VEC{k}}
                                                      \left(\prod_{p \in
                                                      \ft{\chor_1}} 
                                                      \encod{\chor_1}{p} \pp
                                                      \prod_{k \in \fsc{\chor_1} }
                                                      \queue{k}{\emptyset} \right) \\
                                                      \pp \prod_{ a,A } \left(
                                                      \bigsqcup_{p \in
                                                      \serviceMerge{\chor_1}{a}{A}}
                                                      \encod{\chor_1}{p} 
                                                      \right) =  P_{12}
                                                    \end{array}\\
    & \text{A. 9, A.14} 
    &&  \conf{\sigma}{\new{\VEC{k},\VEC{p}}\chor'}
       \action{\initA{
       \VEC{\thread_r[A_r]\{\Blue{Y_r}\}} 
       }{ \VEC{\thread_s[B_s]\{\Blue{
       Y_s}\}}}{a}{k}}
       \conf{\sigma'}{\new{\VEC{k},\VEC{p}}\new{k}\chor_1}\action{\VEC{\ell'}}
       \conf{\sigma''}{\chor'_1}\\
    & \text{A.16, A.18} 
    &&  \isAProjection[\initA{
       \VEC{\thread_r[A_r]\{\Blue{Y_r}\}}, 
       }{ \VEC{\thread_s[B_s]\{\Blue{
       Y_s}\}}}{a}{k},\VEC{\ell'} ]{\startAction{\VEC{A}}{\VEC{B}}ak,
       \emm'_1,\VEC{\emm'_2}}
  \end{align}

  \subcase{ $\emm_1 \neq \startAction{\VEC{A}}{\VEC{B}}ak $}: From
  this case we know that the reduction has been performed by threads
  outside the session establishment phase.  \setcounter{equation}{17}
  \begin{align}
    & \begin{array}{l}\text{A.8, rules }  \Did{^E|{Par}},\\\Did{^E|{Res}}
      \end{array}
    &&     \begin{array}{l}\project{\chor} \action{\emm_1}  \\ 
             \new{\VEC{k},
             \VEC{p}} \left(                  
             \begin{array}{l} \new{k} (
               \encod{\chor_1}{\thread_1}
               \pp 
               \prod_{i \in [2,|\VEC{\thread_r}|]} \encod{\chor_1}{\thread_i}
               \pp \prod_{j \in [1, |\VEC{\thread_s}|]}
               \encod{\chor_1}{\thread_j}\pp
               \queue k\emptyset 
               )  \\ 
               \pp
               P_s
               \pp \prod_{j \in [1, |\VEC{\thread_s}|]}
               \repInitIn{a[B_j]}{k} \pfx
               \encod{\chor_1}{\thread_j} 
             \end{array}
             \right) \\                 
             \pp \prod_{ a,A } \left(
             \bigsqcup_{p \in \serviceMerge{\initA{\VEC{p_r}}{\VEC{p_{s}}}{a}{k} \pfx
             \chor_1}{a}{A}} \encod{\initA{\VEC{p_r}}{\VEC{p_{s}}}{a}{k} \pfx
             \chor_1}{p}\right) =P_{11}
           \end{array}
  \end{align}

  Where $P_s $ captures the evolution of
  $\new{\VEC{k}, \VEC{p}} \left(\prod_{\thread_h \in
      \ft{\chor_1}\backslash{\VEC{\thread_r}, \VEC{\thread_s}}}
    \encod{\chor_1}{\thread_h} \pp \prod_{k' \in \fsc{\chor_1} }
    \queue{k'}{\emptyset} \right)$.
  We can split the reduction chain to one that considers the
  evolutions of $P_{11}$, followed by the execution of the start
  action.

\begin{align}
  & \begin{array}{l}\text{A.18, sub-case }  
    \end{array} && \begin{array}{l}\project{\chor} \action{\emm_1}  
                     P_{11} \action{\VEC{\emm_2}} P_{12} \action{{\startAction
                     {\VEC{A}}{\VEC{B}}ak }} P_{13} \action{\VEC{\emm_4}} P_{14}
                   \end{array}
\end{align}
This thesis now follows from the induction hypothesis.

\Case{} $\chor' = \bcastA{p_r}{\VEC{p_s:x_s}}{q}{e}{k} \pfx \chor_1$:
\setcounter{equation}{0}
\begin{align}
  & \text{Hypothesis} &&     \chor = \new{\VEC{k}, \VEC{p}} \bcastA{p_r}{\VEC{p_s:x_s}}{q}{e}{k} \pfx \chor_1\\
  & \text{Hypothesis} &&     \bcastA{p_r}{\VEC{p_s:x_s}}{q}{e}{k} \pfx \chor_1 \text{ is restriction-free} \\
  & \text{Hypothesis} &&     \bcastA{p_r}{\VEC{p_s:x_s}}{q}{e}{k} \pfx \chor_1 \text{ is linear}\\
  & \text{Hypothesis} &&     \isType{\ltypeEnv}{\new{\VEC{k}, \VEC{p}}\bcastA{p_r}{\VEC{p_s:x_s}}{q}{e}{k} \pfx \chor_1}{\stypeEnv}\\
  & \text{Hypothesis} &&     \isState{\sigma}{\ltypeEnv}\\
  & \text{Hypothesis} &&    \project{\new{\VEC{k},
                         \VEC{p}}\bcastA{p_r}{\VEC{p_s:x_s}}{q}{e}{k}
                         \pfx \chor_1} \action{\emm_1} P\\
  & \text{I.H. } && \forall \emm'_1. \project{\new{\VEC{k},
                    \VEC{p}}\chor_1} \action{\emm'_1}
                    P_1\\
  & \text{I.H. } &&  P_1 \action{\VEC{\emm'_2}}
                    P'_1\\
  & \text{I.H. } && \conf{\sigma_1}{\new{\VEC{k},
                    \VEC{p}}\chor_1} \action{\VEC{\ell'}} \conf{\sigma_2}{\chor_1'} \\
  & \text{I.H. } && \project{\chor_1'} \prune P'_1\\
  & \text{I.H. } &&
                    \isAProjection[\VEC{\ell'}]{\emm'_1,\VEC{\emm'_2}}
\end{align}
The case is similar to the reasoning for the start case. Notice that
although there cannot be in-session linearity (receivers may implement
the same role), its impact only increases size and partitions of the
reduction chain needed to execute the dequeuing of an action in the
session queue. In particular, the projection
$\project{\new{\VEC{k},
    \VEC{p}}(\bcastA{\thread_r[A]}{\VEC{\thread_s[B_s]:x_s}}{q}{e}{k}
  \pfx \chor_1)}$ generates the following endpoints
\[
\begin{array}{l}
  \new{\VEC{k}} 
  \left(              
  \begin{array}{l}          
    \bcast{k}{A}{\VEC{B}}{q}{e} 
    \encod{\chor_1}{\thread_1} 
    \pp 
    \prod_{i
    \in
    [1,|\VEC{\thread_r}|]}
    \receive{k}{B_i}{A}{x_i} 
    \encod{\chor_1}{\thread_i} 
    \\
    \pp
    \prod_{\thread_h
    \in
    \ft{\chor'}\backslash
    {\VEC{\thread_r},\VEC{\thread_s}}}
    \encod{\chor_1}{\thread_h}
    \pp
    \prod_{k' \in \fsc{\chor'} }
    \queue{k'}{\emptyset}
    \pp 
    \queue{k}{\emptyset}
  \end{array} \right) \\
  \pp \prod_{ a,A } \left(
  \bigsqcup_{p
  \in
  \serviceMerge{\bcastA{\thread_r[A]}{\VEC{\thread_s[B_s]:x_s}}{q}{e}{k} \pfx \chor_1}{a}{A}}
  \encod{\bcastA{\thread_r[A]}{\VEC{\thread_s[B_s]:x_s}}{q}{e}{k} \pfx \chor_1}{p}                 \right)                             \end{array} 
\]

The cases related to the reduction chains generated by processes in
$ \prod_{\thread_h \in \ft{\chor'}\backslash
  {\VEC{\thread_r},\VEC{\thread_s}}} \encod{\chor_1}{\thread_h} \pp $
$ \prod_{k' \in \fsc{\chor'} } \queue{k'}{\emptyset} $, as well as the
transitions generated by the swapping congruence relation are similar
to the analysis in the start case. We will focus then on the behavior
generated by remaining processes. Applying rules
$\Did{^E|{Bc.O}}, \Did{^E|{Par}} $ with $e\downarrow v$ to the
projected process leads to
\[
\project{\chor} \action{ \asynchEnqueueO }
\begin{array}{l}
  \new{\VEC{k}} 
  \left(              
  \begin{array}{l}          
    \waitOutput{k}{A}{\VEC{B}} 
    \encod{\chor_1}{\thread_1} 
    \pp 
    \prod_{i
    \in
    [1,|\VEC{\thread_r}|]}
    \receive{k}{B_i}{A}{x_i} 
    \encod{\chor_1}{\thread_i} 
    \\
    \pp
    \prod_{\thread_h
    \in
    \ft{\chor'}\backslash
    {\VEC{\thread_r},\VEC{\thread_s}}}
    \encod{\chor_1}{\thread_h}\\
    \pp
    \prod_{k' \in \fsc{\chor'} }
    \queue{k'}{\emptyset}
    \pp 
    \queue{k}{h\cdot \qmsgI{A}{\VEC{B : \false}}{q}{\some{v}}}    
  \end{array} \right) \\
  \pp \prod_{ a,A } \left(
  \bigsqcup_{p
  \in
  \serviceMerge{\bcastA{\thread_r[A]}{\VEC{\thread_s[B_s]:x_s}}{q}{e}{k} \pfx \chor_1}{a}{A}}
  \encod{\bcastA{\thread_r[A]}{\VEC{\thread_s[B_s]:x_s}}{q}{e}{k} \pfx \chor_1}{p}                 \right)                             \end{array} = P_{1}
\]
With a blocked output, and receivers ready to interact. Their
interaction cannot be assumed to happen in a given order. In general,
each of the receive actions can be preceded or succeeded by a
sequence of actions $\VEC{\emm'}$ generated from the interaction of
processes outside session $k$.  After a finite sequence of reductions
$ P_{1} \action{\VEC{\emm_1}} P_{2} \action{
  \aBroadcastI{A}{B_1}{k}{\some{v}} } P_{3} \action{ \VEC{\emm_{2}}}
\ldots \action{\VEC{\emm_{j-1}}} P_{j} \action{
  \aBroadcastI{A}{B_j}{k}{\some{v}} } P_{j+1}
\action{\VEC{\emm_{j+1}}}_\equiv P_{n}$,
with a given $j = |J|$. $P_n$ has now the form:
\[
P_n =
\begin{array}{l}
  \new{\VEC{k}} 
  \left(              
  \begin{array}{l}          
    \waitOutput{k}{A}{\VEC{B}} 
    \encod{\chor_1}{\thread_1} 
    \pp 
    \prod_{j
    \in
    [1,|J|]}
    \encod{\chor_1}{\thread_i}\subst{\some{v}}{x_j}
    \\\pp 
    \prod_{i
    \in
    [1,|\VEC{\thread_r}\backslash J|]}
    \receive{k}{B_i}{A}{x_i} 
    \encod{\chor_1}{\thread_i} 
    \pp P_s\\
    \pp 
    \queue{k}{h_1 \cdot \qmsgI{A}{ \VEC{B' : \true}, \VEC{B'' :
    \false}}}{q}{\some{v}} \cdot h_2    
  \end{array} \right) \\
  \pp \prod_{ a,A } \left(
  \bigsqcup_{p
  \in
  \serviceMerge{\bcastA{\thread_r[A]}{\VEC{\thread_s[B_s]:x_s}}{q}{e}{k} \pfx \chor_1}{a}{A}}
  \encod{\bcastA{\thread_r[A]}{\VEC{\thread_s[B_s]:x_s}}{q}{e}{k} \pfx \chor_1}{p}                 \right)                             \end{array} 
\]

With $P_s$ the result of the interactions of processes and queues not
involved in session $k$, $\VEC{B} = \VEC{B'},\VEC{B''}$, and $h_1,h_2$
the result of messages on the same session. From Eq. A.4 and Lemma
\ref{lem:session-linearity} we know that
$ \qmsgI{A}{\VEC{B : b}}{q}{sb} \not\in h_1,h_2$ for any
$sb$. Inversion on Eq. A.4 guarantees that predicate $q(\VEC{B'})$ is
satisfiable. From the application of structural congruence rules to
rearrange $h_1, h_2$, the application of rules $\Did{^E|Wait_B}$,
$\Did{^E|Res}$ and $\Did{^E|Par}$ on $P_n$, and Lemma
\ref{lem:subst-process-proj} we get:

\[
P_n \action{ \aBroadcastO{A}{\VEC{B}}{q}{k}{sb} }
\begin{array}{l}
  \new{\VEC{k}} \left(
  \begin{array}{l}          
    \encod{\chor_1}{\thread_1} 
    \pp 
    \prod_{j
    \in
    [1,|J|]}
    \encod{\chor_1 \subst{\some{v}}{x_j}}{\thread_i} \pp 
    \prod_{i
    \in
    [1,|\VEC{\thread_r}\backslash J|]}
    \encod{\chor_1 \subst{\none}{x_i}}{\thread_i}
    \pp P_s\\
    \pp 
    \queue{k}{h' }   
  \end{array} \right) \\
  \pp \prod_{ a,A } \left(
  \bigsqcup_{p
  \in
  \serviceMerge{\bcastA{\thread_r[A]}{\VEC{\thread_s[B_s]:x_s}}{q}{e}{k} \pfx \chor_1}{a}{A}}
  \encod{\bcastA{\thread_r[A]}{\VEC{\thread_s[B_s]:x_s}}{q}{e}{k} \pfx
  \chor_1}{p}                 \right)                             
\end{array} 
\]

With $h' = h_1 \cdot h_2$. The thesis now follows from the application of the induction hypothesis.

                                                \Case{}
                                                $\chor'
                                                =\reduceA{\VEC{p_r.e_r}}{p_{s}:x}{q}{k}{op}
                                                \pfx \chor_1$:
                                                This case corresponds
                                                to the same
                                                equivalence class as
                                                the one for
                                                broadcast. They differ
                                                on the fact that we
                                                must account that for
                                                each asynchronous
                                                output, a subsequence
                                                of actions due to
                                                concurrent sessions
                                                can occur. Moreover,
                                                we must guarantee a
                                                session linearity
                                                condition requiring that there
                                                are no other reduce
                                                operations under the
                                                same roles.  This is
                                                guaranteed by Lemma
                                                \ref{lem:session-linearity}.

                                                \Case{}
                                                $\chor' =
                                                \choiceA{p_r}{\VEC{p_s}}{q}{k}{l}
                                                \pfx \chor_1$:
                                                Recall that according
                                                to the syntactic
                                                restrictions for
                                                collective selections
                                                introduced in Section
                                                \ref{sec:language},
                                                predicate $q$ must
                                                correspond to a
                                                $\forall$
                                                operator. The case
                                                follows in a similar
                                                way as the case for
                                                broadcast.

                                                \Case{}
                                                $\chor' =
                                                \ifthenelsek{e@p}{\chor_1}{\chor_2}$:
                                                \setcounter{equation}{0}
                                                \begin{align}
                                                  & \text{Hypothesis} &&     \chor = \new{\VEC{k}, \VEC{p}}\ifthenelsek {e@p} {\chor_1} {\chor_2}\\
                                                  & \text{Hypothesis} &&     \ifthenelsek {e@p} {\chor_1} {\chor_2} \text{ is restriction-free} \\
                                                  & \text{Hypothesis} &&     \ifthenelsek {e@p} {\chor_1} {\chor_2} \text{ is linear}\\
                                                  & \text{Hypothesis} &&     \isType{\ltypeEnv}{\new{\VEC{k}, \VEC{p}}\ifthenelsek {e@p} {\chor_1} {\chor_2}}{\stypeEnv}\\
                                                  & \text{Hypothesis} &&     \isState{\sigma}{\ltypeEnv}\\
                                                  & \text{Hypothesis} &&    \project{\new{\VEC{k}, \VEC{p}}
                                                                         \ifthenelsek {e@p} {\chor_1} {\chor_2}}
                                                                         \action{\emm_1} P\\
                                                  & \text{A.6, def. }
                                                    \project{\cdot} &&    
                                                                       \begin{array}{l}
                                                                         \project{\new{\VEC{k},
                                                                         \VEC{p}}
                                                                         \ifthenelsek {e@p} {\chor_1} {\chor_2}}  =
                                                                         \\\new{\VEC{k}}
                                                                         \left(\prod_{p \in
                                                                         \ft{\chor_1},\ft{\chor_2}} 
                                                                         \encod{\ifthenelsek {e@p} {\chor_1} {\chor_2}}{p} \pp
                                                                         \prod_{k \in \fsc{\chor_1},\fsc{\chor_2} }
                                                                         \queue{k}{\emptyset} \right) \\
                                                                         \pp \prod_{ a,A } \left(
                                                                         \bigsqcup_{p \in
                                                                         \serviceMerge{\ifthenelsek {e@p} {\chor_1} {\chor_2}}{a}{A}}
                                                                         \encod{\chor_1}{p} \mergek \encod{\chor_2}{p}
                                                                         \right) 
                                                                       \end{array}  \\
                                                  & \text{A.7, def. } \encod{\cdot}{} &&   = \begin{array}{l}
                                                                                               \new{\VEC{k}} 
                                                                                               \left(              
                                                                                               \begin{array}{l}          
                                                                                                 \itn{e}{\encod{\chor_1}{p}}{\encod{\chor_2}{p}}
                                                                                                 \pp 
                                                                                                 \prod_{p \in
                                                                                                 (\ft{\chor_1},\ft{\chor_2})\backslash
                                                                                                 p}
                                                                                                 \encod{\chor_1}{p} \mergek \encod{\chor_2}{p}
                                                                                                 \\
                                                                                                 \pp
                                                                                                 \prod_{k \in \fsc{\chor'} }
                                                                                                 \queue{k}{\emptyset}
                                                                                               \end{array} \right) \\
                                                                                               \pp \prod_{ a,A } \left(
                                                                                               \bigsqcup_{p
                                                                                               \in
                                                                                               \serviceMerge{\ifthenelsek {e@p} {\chor_1} {\chor_2}}{a}{A}}
                                                                                               \encod{\chor_1}{p} \mergek \encod{\chor_2}{p}
                                                                                               \right)                             
                                                                                             \end{array} 
                                                \end{align}
                                                Assume
                                                $e \downarrow \true$
                                                (the opposite case is
                                                analogous). We have
                                                the following
                                                induction hypotheses:
                                                \begin{align}
                                                  & \text{I.H. } && \forall \emm'_1. \project{\new{\VEC{k},
                                                                    \VEC{p}}\chor_1} \action{\emm'_1}
                                                                    P_1\\
                                                  & \text{I.H. } &&  P_1 \action{\VEC{\emm'_2}}
                                                                    P'_1\\
                                                  & \text{I.H. } && \conf{\sigma}{\new{\VEC{k},
                                                                    \VEC{p}}\chor_1} \action{\VEC{\ell'}} \conf{\sigma'}{\chor_1'} \\
                                                  & \text{I.H. } && \project{\chor_1'} \prune P'_1\\
                                                  & \text{I.H. } &&
                                                                    \isAProjection[\VEC{\ell'}]{\emm'_1,\VEC{\emm'_2}}
                                                \end{align}
                                                By the application of
                                                $\Did{^G|If}$ along
                                                with eq.  $A.11$, we
                                                form the following
                                                reduction chain:
                                                \begin{equation}
                                                  \conf{\sigma}{\new{\VEC{k}, \VEC{p}}\ifthenelsek {e@p} {\chor_1} {\chor_2}} \action{\VEC{\ell''}} \conf{\sigma'}{\chor_1'}
                                                \end{equation}
                                                With
                                                $\VEC{\ell''}=
                                                \VEC{\ell'_1},\tau,
                                                \VEC{\ell'_2} $
                                                and
                                                $\VEC{\ell'} =
                                                \VEC{\ell'_1},\VEC{\ell'_2}$.

                                                According to the use
                                                of the swap
                                                congruence, we must
                                                consider whether in
                                                the reduction chain
                                                $\project{\chor}
                                                \action{\VEC{\emm}} P$
                                                the projection of $p$
                                                executes or not the
                                                tau action associated
                                                to the conditional.

                                                \subcase{}
                                                $\project{\chor}
                                                \action{\tau} P$:

\begin{align}
  & \text{A.8, rule } \Did{^E|If} && \project{\chor}  \action{\tau} \begin{array}{l}
                                                                      \new{\VEC{k'}} 
                                                                      \left(              
                                                                      \begin{array}{l}          
                                                                        \encod{\chor_1}{p}
                                                                        \pp 
                                                                        P_s
                                                                      \end{array} \right) \\
                                                                      \pp \prod_{ a,A } \left(
                                                                      \bigsqcup_{p
                                                                      \in
                                                                      \serviceMerge{\ifthenelsek {e@p} {\chor_1} {\chor_2}}{a}{A}}
                                                                      \encod{\chor_1}{p} \mergek \encod{\chor_2}{p}
                                                                      \right)                             
                                                                    \end{array} 
\end{align}
Where $P_s$ denotes processes in
$\new{\VEC{k}}( \prod_{p \in
  (\ft{\chor_1},\ft{\chor_2})\backslash p}
\encod{\chor_1}{p} \mergek \encod{\chor_2}{p} \pp \prod_{k \in
  \fsc{\chor'} } \queue{k}{\emptyset}) $.
We proceed by the application of induction hypotheses in
eq. $A.9$, $A.10$ for $P_s$.
\begin{align}
  & \text{A.15, eq. A.9, A.10} && \project{\chor}  \action{\tau} \begin{array}{l}
                                                                   \new{\VEC{k'}} 
                                                                   \left(              
                                                                   \begin{array}{l}          
                                                                     \encod{\chor_1}{p}
                                                                     \pp 
                                                                     P_s
                                                                   \end{array} \right) \\
                                                                   \pp \prod_{ a,A } \left(
                                                                   \bigsqcup_{p
                                                                   \in
                                                                   \serviceMerge{\ifthenelsek {e@p} {\chor_1} {\chor_2}}{a}{A}}
                                                                   \encod{\chor_1}{p} \mergek \encod{\chor_2}{p}
                                                                   \right)                             
                                                                 \end{array}
  \action{\VEC{\emm_2}} P'
\end{align}
The thesis now follows from the application of induction hypotheses in
$A.12$ and $A.13$, where $\emm_1 = \tau$ and
$\VEC{\ell} = \VEC{\ell''}$.

\subcase{} $\project{\chor} \action{\emm_1} P$, $\emm_1 \neq \tau$:
\setcounter{equation}{14}
\begin{align}
  & \text{A.9, } \emm_1 = \emm'_1 && \project{\chor}  \action{\emm_1} \begin{array}{l}
                                                                        \new{\VEC{k'}} 
                                                                        \left(              
                                                                        \begin{array}{l}          
                                                                          \itn{e}{\encod{\chor_1}{p}}{\encod{\chor_2}{p}}
                                                                          \pp 
                                                                          P_s
                                                                        \end{array} \right) \\
                                                                        \pp \prod_{ a,A } \left(
                                                                        \bigsqcup_{p
                                                                        \in
                                                                        \serviceMerge{\ifthenelsek {e@p} {\chor_1} {\chor_2}}{a}{A}}
                                                                        \encod{\chor_1}{p} \mergek \encod{\chor_2}{p}
                                                                        \right)                             
                                                                      \end{array} = P
\end{align}
Where $P_s$ denotes new processes and session queues generated from
the evolution of processes in
$\new{\VEC{k}}( \prod_{p \in
  (\ft{\chor_1},\ft{\chor_2})\backslash p}
\encod{\chor_1}{p} \mergek \encod{\chor_2}{p} \pp \prod_{k \in
  \fsc{\chor'} } \queue{k}{\emptyset}) $.
\begin{align}
  & \text{A.15, rule  } \Did{^E|if},  && P  \action{\tau} \begin{array}{l}
                                                             \new{\VEC{k'}} 
                                                             \left(\begin{array}{l}          
                                                                     \encod{\chor_1}{p}
                                                                     \pp 
                                                                     P_s
                                                                   \end{array}\right) \\
                                                             \pp \prod_{ a,A } \left(
                                                             \bigsqcup_{p
                                                             \in
                                                             \serviceMerge{\ifthenelsek {e@p} {\chor_1} {\chor_2}}{a}{A}}
                                                             \encod{\chor_1}{p} \mergek \encod{\chor_2}{p}
                                                             \right)                             
                                                           \end{array} = P'
\end{align}
The thesis now follows by the application of induction hypotheses in
eq. $A.12$, $A.13$

\begin{align}
  & \text{A.15, eq. A.9, A.10} && \project{\chor}  \action{\emm_1} P
                                  \action{\tau} P' \action{\VEC{\emm'_2}} P''
\end{align}
 where $\VEC{\emm_2} = \tau,\VEC{\emm'_2}$.

\Case{} $\chor' = {\chor_1} + {\chor_2}$:  This case is essentially
the same as the deterministic choice explained above. When proving
$\conf{\sigma}{\new{\VEC{k},\VEC{p}} \chor_1 + \chor_2} \action{\VEC{\ell}}
\conf{\sigma'}{\chor'}$, we are reminded that the reduction chain
$\action{\VEC{\ell}}$ can be given from the $\conf{\sigma}{\new{\VEC{k},\VEC{p}} \chor_1 } \action{\VEC{\ell_1}}
\conf{\sigma'}{\chor'_1}$ or from $\conf{\sigma}{\new{\VEC{k},\VEC{p}}  \chor_2} \action{\VEC{\ell_2}}
\conf{\sigma'}{\chor'_2}$. They correspond to the possible evolutions
in
\[ \project{ \new{\VEC{k},\VEC{p}} (\chor_1 + \chor_2)} =
\begin{array}{l} \new{\VEC{k}}
  \left(\prod_{p \in
  \ft{\chor_1},\ft{\chor_2}} 
   (\encod{\chor_1}{p} + \encod{\chor_2}{p}) \pp
  \prod_{k \in \fsc{\chor_1},\fsc{\chor_2} }
  \queue{k}{\emptyset} \right) \\
  \pp \prod_{ a,A } \left(
  \bigsqcup_{p \in
  \serviceMerge{ {\chor_1} + {\chor_2}}{a}{A}}
  \encod{\chor_1 +\chor_2}{p}
  \right) 
\end{array}  
\]

\Case{} $\chor' = \INACT$: This case is vacuously true.

\end{proof}


\end{document}